\documentclass[12pt]{article}
\usepackage{jheppub}
\usepackage{youngtab}
\usepackage{graphicx}
\usepackage{subcaption}
\usepackage{comment}
\newcommand{\m}{\mathfrak{v}}
\newcommand{\fb}{\mathfrak{b}}
\newcommand{\cA}{\mathcal{A}}

\newcommand{\cM}{\mathcal{M}}
\newcommand{\cL}{\mathcal{L}}
\newcommand{\cB}{\mathcal{B}}
\newcommand{\cY}{\mathcal{Y}}
\newcommand{\cV}{\mathcal{V}}
\newcommand{\cW}{\mathcal{W}}

 \Yboxdim{8pt}

\title{Miura operators, degenerate fields and the M2-M5 intersection}
\author[a]{Davide Gaiotto,}
\author[b]{Miroslav Rap\v{c}\'{a}k}
\affiliation[a]{Perimeter Institute for Theoretical Physics, Waterloo, Ontario, Canada N2L 2Y5}
\affiliation[b]{Center for Theoretical Physics, University of California, Berkeley, USA}

\abstract{We determine the mathematical structures which govern the $\Omega$ deformation of supersymmetric intersections of M2 and M5 branes. We find that the supersymmetric intersections govern many aspects of the theory of W-algebras, including degenerate modules, the Miura transform and Coulomb gas constructions. We give an algebraic interpretation of the Pandharipande-Thomas box counting in $\mathbb{C}^3$.}

\begin{document}
\maketitle
\section{Introduction}
M-theory is described at low energy by an eleven-dimensional supergravity theory endowed with very specific effective action \cite{Witten:1995ex}. The form of the effective action is constrained to some (unknown) degree by supersymmetry and dualities \cite{Horava:1996ma,Green:1997as,Russo:1997mk,Green:2005ba}. Holography also provides strong constraints, as the flat space S-matrix can be extracted from the correlation functions of the world-volume SCFTs for stacks of M2 or M5 branes \cite{Chester:2018aca,Alday:2020tgi}. Besides the ambitious possibility of a full conformal bootstrap of either theories 
\cite{Agmon:2017xes,Beem:2015aoa}, exact localization results can at least determine certain protected coefficients 
in the supergravity effective action \cite{Chester:2018aca}. Amusingly, one may even discover information about one of the two worldvolume theories by applying holographic considerations to the defects which arise from the intersection of M2 and M5 branes
\cite{Drukker:2020swu}.

There is an alternative way to employ M2 and M5 branes to constrain the supergravity effective action. At low energy, a stack of branes is described by some specific collection of irrelevant couplings between the supergravity fields and the worldvolume SCFT. Supersymmetry and dualities will impose constraints on these couplings, which involve both the bulk couplings and the OPE of the SCFT. Further constraints will arise from the existence of various supersymmetric intersections between stacks of branes. It would be interesting to know to which degree these constraints fix the bulk couplings or the SCFT OPEs. As these constraints arise even from the coupling to a single brane, they may contain information which is distinct from the holographic considerations and may not require one to study directly strongly-coupled SCFTs.

At least in protected subsectors, this idea is known to work in string theory: the effective action of some topological string theories is 
determined uniquely by the requirement of a consistent coupling to space-filling branes \cite{Costello:2015xsa}. Furthermore, 
two beautiful works \cite{Costello:2016nkh,Costello:2017fbo} determined protected OPEs of M2 and M5 brane SCFTs 
from similar consideration in a twisted, $\Omega$-deformed background which we will refer to as ``twisted M-theory''. 

The same structure also encodes certain ``higher operations'' in twisted M-theory, which should be in principle computable by a descent procedure in the physical theory \cite{Beem:2018fng} and thus constrain some protected couplings in the supergravity effective action. Unfortunately, 
such a translation is not currently available. Even the precise mathematical relation between the higher operations involved in the M2 and M5 brane analyses has not been worked out, because of the holomorphic-topological nature of the setup \cite{Costello:2020ndc}.

One of the objectives of this paper is to spell out such a relation by studying the supersymmetric intersection points of stacks of M2 and M5 branes. Ultimately, we will find that the whole structure, including the protected OPEs of stacks of branes, can be reconstructed from the knowledge of the simplest such intersection: a single M2 brane completely transversal to a single M5 brane.  

\subsection{The basic setup}

The ``twisted M-theory'' setup introduced in \cite{Costello:2016nkh,Costello:2017fbo} simplifies drastically the low energy theory:
\begin{itemize}
\item The bulk theory becomes five-dimensional and locally trivial. It can be described in terms of a gauge field which descends from the three-form field in the physical theory. The data of the metric collapses locally to an identification of space-time with 
$\mathbb{R} \times \mathbb{C}^2$, equipped with the canonical complex symplectic form on the $\mathbb{C}^2$ factor. 
The theory is topological along $\mathbb{R}$ and holomorphic along $\mathbb{C}^2$. At leading order, it can be described 
by a Chern-Simons-like action employing the Poisson bracket on $\mathbb{C}^2$.
\item M2 branes become one-dimensional objects extended along the topological direction, 
endowed with a world-line algebra of local operators which arises as an $\Omega$-deformation 
of the physical worldvolume theory as in \cite{Yagi:2014toa}. 
By a slight abuse of notation, we will call these objects ``line defects''. 
\item M5 branes become two-dimensional objects extended along an holomorphic direction, endowed with a world-sheet chiral algebra of local operators which arise as an $\Omega$-deformation of the physical worldvolume theory as in \cite{Yagi:2012xa}. See also \cite{Nekrasov:2010ka,Gaiotto:2011nm}. By a slight abuse of notation, we will call these objects ``surface defects''.  
\item Intersections of M2 and M5 branes become zero-dimensional objects, supporting spaces of local operators equipped with actions of the algebras and chiral algebras of the intersecting branes \cite{Gaiotto:2019wcc} which generalize the module structures in \cite{Bullimore:2016nji,Cordova:2017mhb}. By a slight abuse of notation, we will call these objects ``junctions''.
\end{itemize}
The simplified setting retains some traces of the intricate collection of couplings of the physical theory. Costello identified 
the constraints on gauge-invariant couplings defining line defects \cite{Costello:2017fbo} 
and surface defects \cite{Costello:2016nkh}.\footnote{The natural mathematical language used to describe these gauge-invariant couplings is that of Koszul duality. In this paper we employ a more intuitive physical language. We will attempt to provide a translation between the two languages in a few footnotes, leaving a fully rigorous mathematical formulation to future work.}
In both cases, the constraints fix the OPE of an infinite tower of operators involved in the couplings, which are inherited from protected OPE coefficients of an infinite tower of BPS operators in the physical theories. 

\begin{figure}[h]
    \centering
        \includegraphics[width=5.5cm]{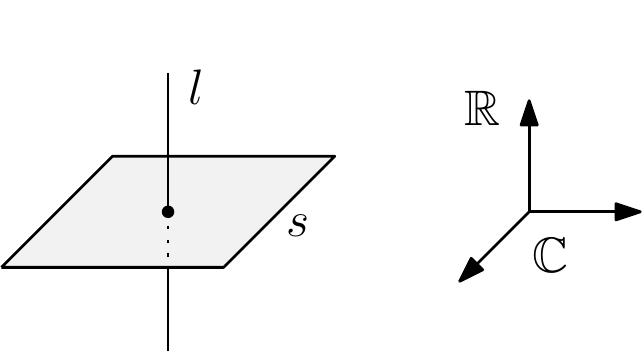}
 \caption{We place line defects along $\mathbb{R}$ and surface defects along one of the $\mathbb{C}$'s inside $\mathbb{C}^2\times \mathbb{R}$ and study their mutual junctions.}
\label{pic29}
\end{figure}

Our main objective is to identify the gauge-invariance constraints on junctions and the solutions to these constraints which arise in specific physical intersections, completing the program initiated in \cite{Gaiotto:2019wcc}. 

We should stress that we only consider junctions between line defects and surface defects. The intersection of transverse 
surface defects is also potentially interesting, but we will not consider it in this paper. In other words, we focus on line and surface defects which lie within a $\mathbb{R} \times \mathbb{C}$ factor in the local five-dimensional geometry. See figure \ref{pic29}.

\subsection{Gauge constraints}
The OPE constraints identified by Costello take the following form: 
\begin{itemize}
\item The couplings defining gauge-invariant line defects involve a collection of elements $t_{n,m}$ in the world-line theory, 
with non-negative integer $n$ and $m$. They must satisfy the commutation relations of the algebra $A$, deforming the universal enveloping algebra of $\mathfrak{u}(1)\otimes \mathbb{C}[z,w]$, known as the 1-shifted affine Yangian of $\mathfrak{u}(1)$.\footnote{The algebra $A$ was introduced first under the name deformed double-current algebra in \cite{Guay1} and studied further by \cite{Guay2,Guay3}. The deformed double-current algebra can be identified with the 1-shifted affine Yangian of $\mathfrak{u}(1)$ from \cite{Kodera:2016faj,Rapcak:2020ueh}.} In other words, the algebra of operators of the worldline theory must admit an algebra morphism from $A$. 
\item The couplings defining gauge-invariant surface defects involve a collection of chiral currents $W_n(z)$ of conformal dimension $n$ for all positive integers $n$. The OPE of the currents must take the form of the $\cW_\infty$ Vertex Operator Algebra studied extensively by many authors. See e.g. \cite{Hornfeck:1994is,Gaberdiel:2012ku,Prochazka:2014gqa,Linshaw:2017tvv}. In other words, the VOA  of operators of the worldsheet theory must admit a VOA morphism from $\cW_\infty$.\footnote{$\cW_{1+\infty}$ may be a more conventional notation}
\end{itemize}

Classically, gauge transformations take the form of holomorphic Hamiltonian symplectomorphisms which are locally constant in the topological direction. As the bulk is locally trivial, gauge transformations act on defect fields only. The classical action of gauge transformations is modified quantum-mechanically in an intricate way.\footnote{After integrating away most of the bulk fields, all that is left is a single scalar ghost field of ghost number $1$. The original interactions of the bulk theory are captured by a collection of higher operations, computed in the original theory by applying descent to the ghost field and integrating the result over various cycles in the configuration space of points. See \cite{Costello:2020ndc} for a discussion of how higher operations may look like in an holomorphic-topological setup. The important point is that the higher operations determine how the BRST operator is deformed by turning on additional interactions, such as the couplings to line and surface defects. This structure leads to the Koszul duality statements in \cite{Costello:2016nkh,Costello:2017fbo} and presumably would explain the specific gauge-invariance constraints for junctions which we use in this paper.}

In the neighbourhood of a line defect, we would expand the gauge parameter $\lambda$ 
in powers of the transverse directions and obtain modes $\lambda_{n,m}$. The action of gauge transformations 
on the worldline theory would take the form 
\begin{equation}
\delta_\lambda = \sum_{n,m} \lambda_{n,m} t_{n,m} 
\end{equation}
for some operators $t_{n,m}$. This is classically consistent if the $t_{n,m}$ satisfy the same commutation relations as the 
corresponding gauge transformations. Quantum corrections make them into a representation of $A$,\footnote{More precisely, the coupling of the line defect is obtained from a descent procedure from 
\begin{equation}
\delta_c = \sum_{n,m} c_{n,m} t_{n,m}
\end{equation}
where $c_{n,m}$ are modes of the ghost field. BRST invariance requires $\delta_c$ to satisfy a Maurer-Cartan equation 
\begin{equation}
Q \delta_c+ (\delta_c, \delta_c)_{\mathbb{R}}+ (\delta_c, \delta_c,\delta_c)_{\mathbb{R}} + \cdots = 0
\end{equation}
involving the bulk higher operations associated to the topological direction. 
Expanding the MC equation in a basis $c_{a,b} c_{c,d}$ gives the commutation relations defining $A$. }
which is indeed 
defined by non-linear commutation relations which deform the commutation relations of symplectomorphisms:
\begin{equation}
[t_{a,b},t_{c,d}] = (ad - bc) t_{a+c-1,b+d-1} + \cdots
\end{equation}

In the neighbourhood of a surface defect, we would expand the gauge parameters in 
powers of the transverse holomorphic direction and obtain modes $\lambda_n(z)$ on the defect. 
The action of gauge transformations on the worldsheet theory would take the form 
\begin{equation}
\delta_\lambda = \sum_n \oint dz \lambda_n(z) W_{n+1}(z)
\end{equation}
for some currents whose OPE is also determined by the gauge algebra. These OPE are deformed quantum-mechanically 
into the OPE for the $\cW_\infty$ algebra,\footnote{Here there is an extra important subtlety: we consider surface defects which may be disorder defects analogous to 't Hooft defects in gauge theory. As a consequence, the higher operations involving operators on the defect are modified by an amount controlled by the charge of the defect, denoted as $\psi_0$ below. 
We expect the higher operations to be actually ``curved'', in the sense that the MC equation has a source proportional to the charge which has to be cancelled by coupling to extra degrees of freedom unless the charge vanishes. 
Again, the coupling of the surface defect to the worldsheet VOA is obtained from a descent procedure from 
\begin{equation}
\delta_c =\sum_n \oint dz c_n(z) W_{n+1}(z)
\end{equation}
where $c_{n}(z)$ are modes of the ghost field along the defect. BRST invariance requires $\delta_c$ to satisfy a Maurer-Cartan equation 
\begin{equation}
\omega^{\psi_0} + Q^{\psi_0} \delta_c+ (\delta_c, \delta_c)^{\psi_0}_{\mathbb{C}}+ (\delta_c, \delta_c,\delta_c)^{\psi_0}_{\mathbb{C}} + \cdots = 0
\end{equation}
involving the higher operations and source $\omega^{\psi_0}$ associated to the holomorphic direction along the disorder defect. 
Expanding the MC equation in an appropriate basis should give the OPE of $\cW_\infty$.} which is indeed a non-linear deformation of a classical $\cW_\infty$ algebra which is associated to symplectomorphisms of a neighbourhood of the surface. 

\begin{figure}
    \centering
        \includegraphics[width=9cm]{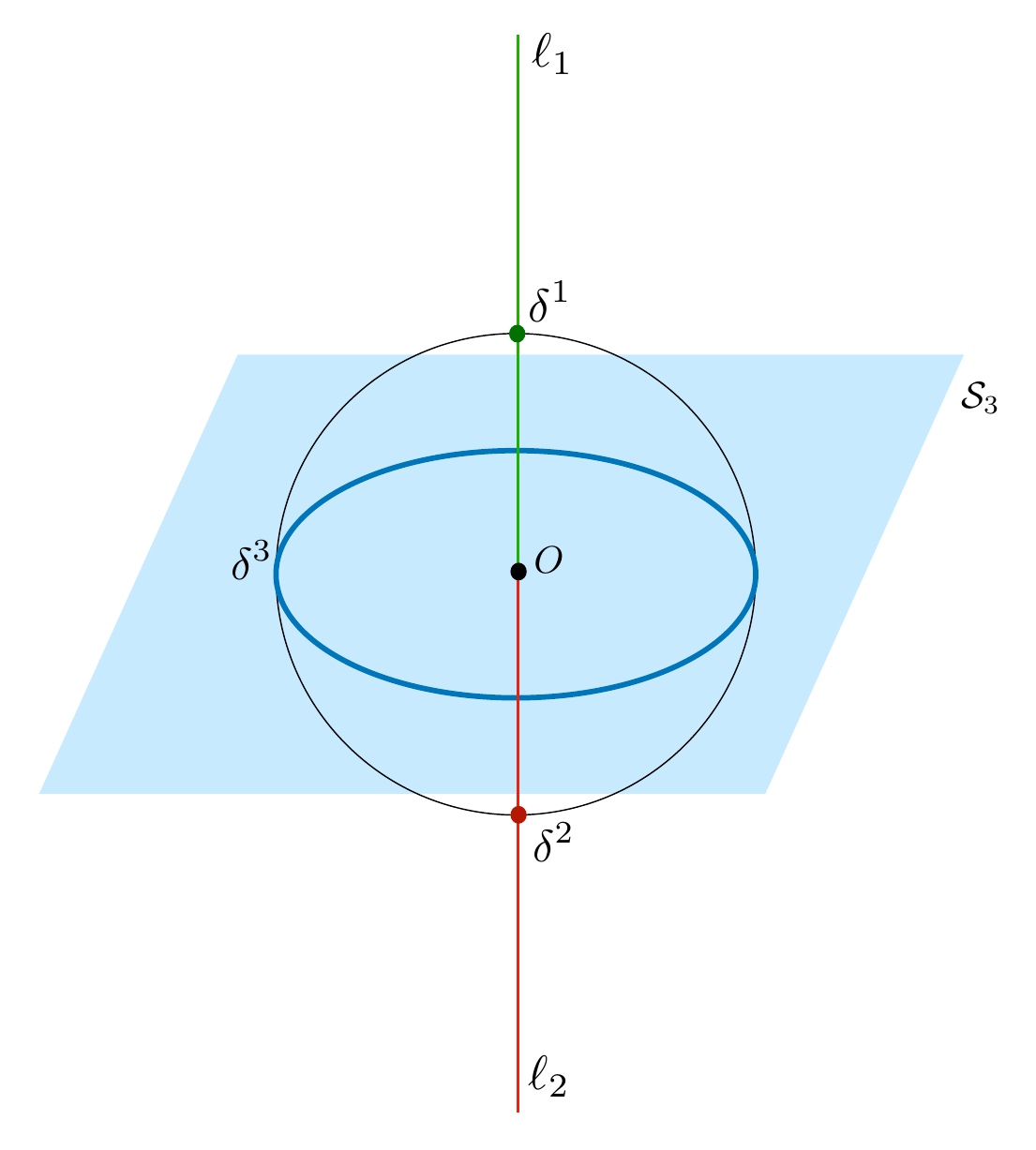}
   \caption{A junction between line defects $L_1$ and $L_2$ and surface defect $\mathcal{S}_3$. Classically, the gauge variation of the junction local operator receives contributions $t^{(1)}_{n,m}$ and $t^{(2)}_{n,m}$ from the north and south pole, $W^{(3)}_{n+1,m-n}$ from the contour integral along the equator. Quantum-mechanically there are extra mixed terms, which arise from bulk and defect interaction vertices integrated within the sphere.}
\label{picdefect}
\end{figure}

Classically, a junction $O$ of several defects will be gauge-invariant if the sum of the gauge variations coming from all defects 
vanishes. One can visualize gauge transformation across a small sphere surrounding the junction, 
picking contributions $\delta^{(i)}_\lambda$ wherever the sphere intersects the $i$-th defect: points for line defects and 
small circles for surface defects:
\begin{equation}
\sum_i \delta^{(i)}_\lambda O = 0
\end{equation}
We now specialize to defects which lie within a specific $\mathbb{R} \times \mathbb{C}$, so that a junction will 
involve generically a ``left'' line defect $L_1$ extended along the positive real numbers, 
a ``right'' line defect $L_2$ extended along the negative real numbers and a surface defect $\mathcal{S}_3$ as in figure \ref{picdefect}.

Then the classical gauge-invariance constraint will take the form 
\begin{equation}
t^{(1)}_{n,m} \cdot O  - O \cdot t^{(2)}_{n,m}  + W^{(3)}_{n+1,m-n} \cdot O =0 \qquad \qquad n,m \geq 0
\end{equation}
where we use a left- or right- module structure for the line defects coming from either direction and we define the modes 
\begin{equation}
W_{n+1,m-n} \equiv \oint z^{m} W_{n+1}(z)
\end{equation}

Quantum mechanically, we expect the gauge-invariance constraint to be deformed to a non-linear constraint, 
possibly involving polynomials in the $t^{(i)}_{n,m}$ generators and all the modes of the $W^{(3)}_{n}$ currents.\footnote{The BRST variation of a local operator receives contributions from higher operations in all directions. As we are looking at operators of ghost number $0$, and we assume that the defect theories live in ghost number $0$, the BRST variation takes the schematic form 
\begin{equation}
\sum_{n_1,n_2,n_3} \left((\delta^{(1)}_c)^{\otimes n_1};(\delta^{(3)}_c)^{\otimes n_3}; O;(\delta^{(2)}_c)^{\otimes n_2}\right)^{\psi_0}_{\mathbb{R}\times \mathbb{C}} =0
\end{equation}
involving higher operations combining the descendants of all defect interactions. Expanding out in a basis of ghost number $1$ modes $c_{n,m}$ will give the gauge-invariance constraints on $O$ which we seek. The detailed form of higher operations 
depend on choices of renormalization scheme and can be reorganized by certain polynomial field redefinitions. 
Below, we assume that one can find such a redefinition which sets to zero all terms with $n_2>1$. 
}
Although the form of the gauge-invariance constraints will be deformed, we do not expect to find new constraints. 
The precise form of the constraints is also subject to re-definition: we could shift a constraint by the image of another constraint 
under any combination of generators and modes. Without loss of generality, we should be able to use such redefinitions to bring the constraints to a canonical form:
\begin{equation}
t^{(1)}_{n,m} \cdot O + W^{(3)}_{n+1,m-n} \cdot O + (\cdots) \cdot O = O \cdot t^{(2)}_{n,m}\qquad \qquad n,m \geq 0
\end{equation}
where the ellipsis denotes polynomials in the $t^{(1)}_{a,b}$ generators and (all) the modes of the $W^{(3)}_{a}$ currents.\footnote{The reader may be disturbed by the apparent asymmetry between the role of line defects on the two sides of the junction. This is only happening because we choose to organize the gauge-invariance conditions by specifying which combination of the left $A$ action and VOA $\cW_\infty$ action cancels 
specific gauge anomalies coming form the right. We can take the gauge invariance constraints and rearrange them to focus on the cancellation of gauge anomalies coming from the right to obtain an equivalent set of conditions. We will discuss this rearrangment in further detail in 
Section \ref{sec:inv}. }

Self-consistency of these constraints require the combination of operators on the left hand side 
\begin{equation}
t_{n,m}  + W_{n+1,m-n} + \cdots
\end{equation}
to have the same commutation relations as the $t_{n,m}$: they must define an algebra morphism
\begin{equation}
\Delta_{A,W_\infty} : A \to A \otimes W_\infty
\end{equation}
where $W_\infty$ is the mode algebra of $\cW_\infty$. We will denote this morphism as the 
{\it mixed coproduct}. We propose a conjectural expression for the mixed coproduct in Section 
\ref{sec:compose}, by giving the image of the generators $t_{2,0}$ and $t_{0,n}$: 
\begin{align}
t_{0,n}&\to t_{0,n}+W_{1,n} \cr
t_{2,0}&\to t_{2,0}+V_{-2}+\sigma_3\sum_{n=1}^{\infty} n W_{1,-n-1}W_{1,n-1}+2 \sigma_3 \sum_{n=1}^{\infty}n W_{1,-n-1}t_{0,n-1}
\end{align}
where we use a mode from the composite quasi-primary 
\begin{equation}
V=W_3+\frac{2}{\psi_0}:W_1W_2:-\frac{2}{3}\frac{1}{\psi_0^2}:W_1W_1W_1: 
\end{equation}
and $\psi_0$ is a central parameter discussed further below.

It is also useful to specialize to a situation where one of the line defects is trivial, i.e. to look at endpoints of a line defect onto a surface defect, by setting either $t^{(1)}_{n,m}$ or $t^{(2)}_{n,m}$ equal to $0$. In particular, the mixed coproduct composed with $t_{n,m} \to 0$ reduces to a morphism  
\begin{equation}
\Delta_{W_\infty} : A \to W_\infty
\end{equation}
mapping the generators $t_{n,m}$ of $A$ to a non-linear deformation of the modes $W_{n+1,m-n}$ 
which annihilate the vacuum vector of $\cW_\infty$.\footnote{Note that the modes $W_{n+1,m-n}$ with $m \geq 0$ do not form an algebra, as their commutators involve normal-ordered sums of modes with generic $m$. Thus the non-linear deformation in $\Delta_{W_\infty}$ is necessary for an algebra morphism
to even exist. The algebra $A$ thus plays the role of a ``positive subalgebra'' of $W_\infty$, annihilating the vacuum vector. Our analysis in Section \ref{sec:boxes} strongly suggests that degenerate modules of $W_\infty$ are induced from modules of $A$ in this sense.}  This embedding will help us give a novel interpretation of Pandharipande-Thomas box counting in Section \ref{sec:boxes}

\subsection{Various coproducts}
As we study defects within an $\mathbb{R} \times \mathbb{C}$ slice of space-time, we encounter two natural fusion operations\footnote{In a derived setting, the fusion of defects is represented by yet another collection of higher operations which controls how the MC element of the fused defect is written as a polynomial in the MC elements of the two defects. See e.g. \cite{Gaiotto:2015aoa} for examples. These operations should lead to the coproducts below.}: 
\begin{enumerate}
\item Two parallel surface defects $\mathcal{S}_1$ and $\mathcal{S}_2$, sitting at different points in the $\mathbb{R}$ direction, may be fused topologically into a single composite surface defect $\mathcal{S}_1 \circ \mathcal{S}_2$. The couplings of the composite defect must be built from the 
couplings of the two original defects. This results in the definition of a VOA coproduct 
\begin{equation}
\Delta_{\cW_\infty,\cW_\infty}: \cW_\infty \to \cW_\infty \otimes \cW_\infty
\end{equation}
which expresses every current as a polynomial in (derivatives) of two independent sets of currents. We aim to identify it with the canonical coproduct built in the very definition of $\cW_\infty$.
\begin{figure}[h]
    \centering
        \includegraphics[width=9.5cm]{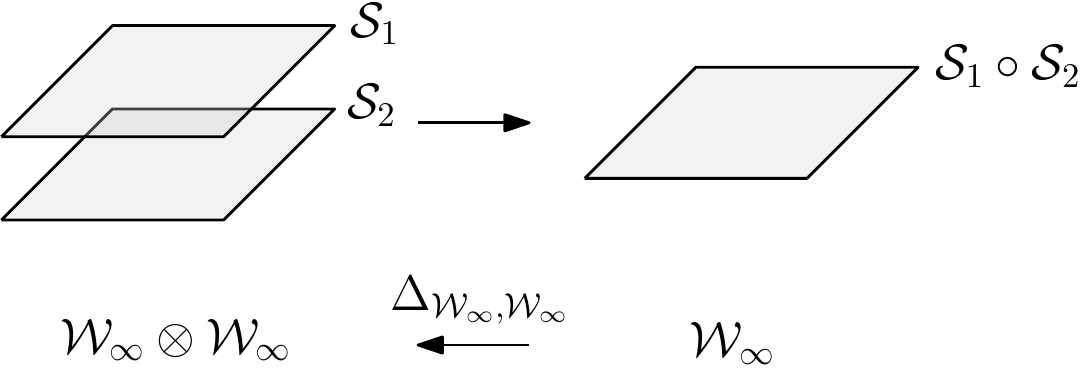}
   \caption{Fusion of surface defects $\mathcal{S}_1$ and $\mathcal{S}_2$ giving rise to the coproduct $\Delta_{\cW_\infty,\cW_\infty}$.}
\label{picdefects}
\end{figure}
\item Two parallel line defects $L_1$ and $L_2$ at different points $0$, $w$ in the $\mathbb{C}$ plane may be fused into a single line defect $L_1 \circ_w L_2$. Because the transverse direction is holomorphic rather than topological, the fusion depends on the separation $w$ and may well diverge as $w \to 0$. 
The couplings of the composite defect must be built from the couplings of the two original defects. This results in the definition of a meromorphic family of coproducts 
\begin{equation}
\Delta_{A,A}(w): A \to A \otimes A
\end{equation}
which expresses every generator as a polynomial in two independent sets of generators, with coefficients which are Laurent polynomials in $w$. We propose a conjectural expression for the mixed coproduct in Section 
\ref{sec:mero}.
\begin{figure}[h]
    \centering
        \includegraphics[width=6.7cm]{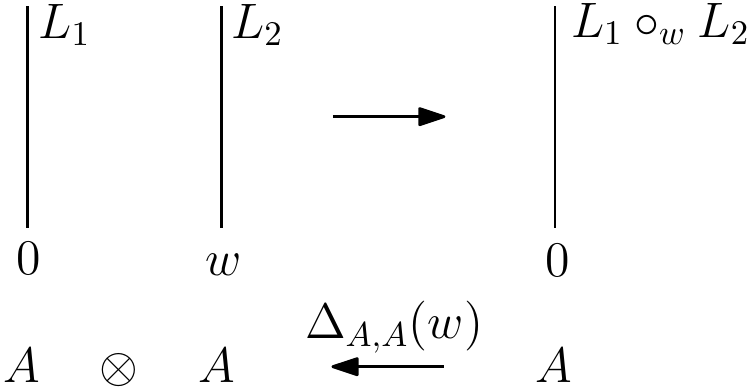}
   \caption{Fusion of line defects $L_1$ and $L_2$ giving rise to the coproduct $\Delta_{A,A}(w)$.}
\label{picdefects}
\end{figure}

\end{enumerate}

If a line defect crosses two surface defects in a gauge-invariant way and we fuse the surface defects, the two junctions should 
also fuse to a new gauge-invariant junction. This requires a compatibility between the mixed coproduct and the VOA coproduct: 
\begin{equation}
[\Delta_{A,W_\infty} \otimes 1_{W_\infty}] \circ \Delta_{A,W_\infty} = [1_A \otimes \Delta_{\cW_\infty,\cW_\infty} ] \circ \Delta_{A,W_\infty} 
\end{equation}
as the action of $t_{n,m}$ from the right on the composition $O_1 O_2$ can be mapped in two ways to a left action in $A \otimes W_\infty \otimes W_\infty$, resulting in the same constraint. 

\begin{figure}[h]
    \centering
        \includegraphics[width=10cm]{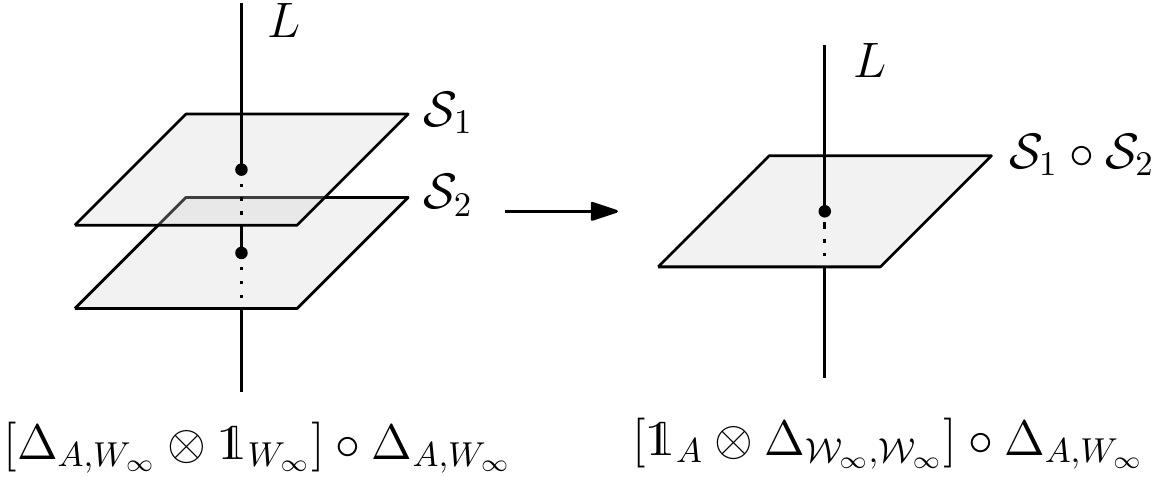}
   \caption{Consistency of the gauge invariance with the fusion of surface defects leads to a condition on compatibility between the mixed coproduct and the VOA coproduct.}
\label{picdefects}
\end{figure}

Similar considerations apply to the OPE of two junctions along the same surface defect, accompanied by the fusion of line defects. Now, we can produce  a $w$-dependent map $A \to A \otimes W_\infty \otimes A \otimes W_\infty$ in two ways:
\begin{equation}
[\Delta_{A,W_\infty} \otimes \Delta_{A,W_\infty} ] \circ \Delta_{A,A}(w) = [\Delta_{A,A}(w) \otimes \Delta_{W_\infty,W_\infty}(w) ] \circ \Delta_{A,W_\infty} 
\end{equation}
where $\Delta_{W_\infty,W_\infty}(w)$ is the linear coproduct on $W_\infty$ defined by the contour integral around two points:
\begin{align}
&W_{n+1,m-n} \circ O_1(0) O_2(w) \equiv \oint z^{m} W_{n+1}(z) O_1(0) O_2(w) = \cr &= \left[\oint z^{m} W_{n+1}(z) O_1(0) \right]O_2(w) 
+  O_1(0) \left[\oint (z+w)^{m} W_{n+1}(z+w)  O_2(w) \right]
\end{align}
though some care may be needed with the formal power series in $w^{-1}$. 

\begin{figure}[h]
    \centering
        \includegraphics[width=11.8cm]{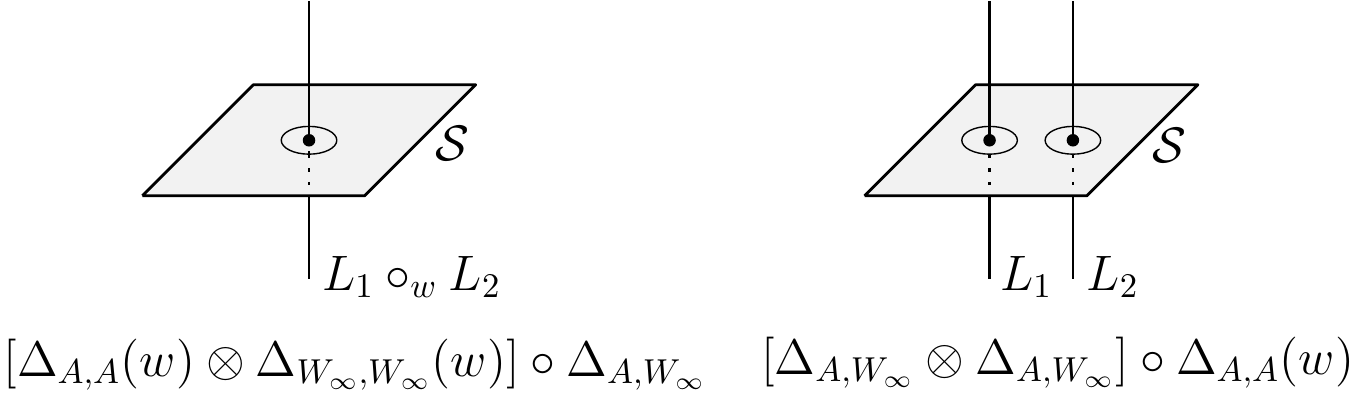}
   \caption{Consistency of the gauge-invariance condition with the fusion of line defects leads to a condition on compatibility between the mixed coproduct and the meromorphic coproduct. One can either implement the gauge-invariance condition by $\Delta_{A,W_{\infty}}$ and then separate the lines or impose the gauge invariance condition on already separated lines. Note that the former case requires the $\Delta_{W_{\infty},W_{\infty}}(w)$ coproduct since the result must contain $W_\infty$ generators acting on the two separated lines.}
\label{picdefects}
\end{figure}

Both associativity relations play an important role in our construction of the coproducts in Section \ref{sec:coproducts}.

\subsection{Defects from branes}
In order to explain our results in greater detail, it is useful to recall a few more details about the specific twisted M-theory setup
we consider in this paper.  
\begin{itemize}
\item The setup involves the compactification of M-theory on an $\Omega$-deformed $\mathbb{C}^3$ geometry, 
which we denote as $\mathbb{C}_{\epsilon_1} \times\mathbb{C}_{\epsilon_2} \times\mathbb{C}_{\epsilon_3}$. Here 
the $\epsilon_i$ are the $\Omega$-deformation parameters. They add up to $0$ to preserve a Calabi-Yau condition:
\begin{equation}
\epsilon_1 + \epsilon_2 + \epsilon_3 = 0
\end{equation}
The volume of the internal $\mathbb{C}_{\epsilon_i}$ directions is effectively $\epsilon_i^{-1}$, so the $\epsilon_i$ control quantum corrections. In particular, non-linearities are controlled by the product $\sigma_3 = \epsilon_1 \epsilon_2 \epsilon_3$. 
\item The remaining 5d geometry can be taken to be $\mathbb{R} \times \mathbb{C}^2$. There is some amount of non-commutativity in the $\mathbb{C}^2$ directions, which motivates our restriction to an $\mathbb{R} \times \mathbb{C}$ subspace. It would be interesting to relax our constraint.
\item M2 branes can wrap submanifolds of the form $\mathbb{C}_{\epsilon_i} \times \mathbb{R}$. The world-line algebras 
for individual M2's will be denoted as $A_{1,0,0}$, $A_{0,1,0}$, $A_{0,0,1}$. These are just Weyl algebras, with the two generators representing the position of the branes in the transverse $\mathbb{C}^2$. The world-line algebras for a generic collection of M2's are denoted as $A_{n_1,n_2,n_3}$. They must be equipped with an algebra morphism 
\begin{equation} 
\ell_{n_1,n_2,n_3}: A \to A_{n_1,n_2,n_3}
\end{equation} 
which maps
\begin{equation}
t_{0,0} \to \frac{n_1}{\epsilon_1} + \frac{n_2}{\epsilon_2} +\frac{n_3}{\epsilon_3}
\end{equation}
aka the ``electric charge'' of the line defect. 
Until this work, only $A_{n,0,0}$ was known. We will derive an explicit description of $A_{n_1,n_2,n_3}$ in Section \ref{sec:mero} by resumming the meromorphic coproducts $\Delta_{A,A}(w)$. 
\item M5 branes can wrap submanifolds of the form $\mathbb{C}_{\epsilon_i} \times \mathbb{C}_{\epsilon_j} \times \mathbb{C}$.
The world-sheet algebras for individual M5's will be denoted as $\cW_{1,0,0}$, $\cW_{0,1,0}$, $\cW_{0,0,1}$. These are just $\widehat{\mathfrak{u}}(1)$ Kac-Moody algebras, with the current generators representing the position of the branes in the transverse $\mathbb{C}$. The world-sheet algebras for a generic collection of M5's are denoted as $\cW_{N_1,N_2,N_3}$ and admit a VOA map 
\begin{equation} 
s_{N_1,N_2,N_3}: \cW_\infty \to \cW_{N_1,N_2,N_3}
\end{equation}
They should coincide with the ``corner VOAs'' of \cite{Gaiotto:2017euk,Prochazka:2017qum}. See also related work of \cite{Bershtein,Litvinov:2016mgi,Prochazka:2014gqa,Creutzig:2017uxh,Prochazka:2017qum,Linshaw:2017tvv,Prochazka:2018tlo,Rapcak:2018nsl,Harada:2018bkb,Creutzig:2020zaj}. They can be built via the standard coproduct \cite{Prochazka:2018tlo} which we identify with $\Delta_{\cW_\infty,\cW_\infty}$. They have a central parameter 
\begin{equation}
\psi_0 \to -\frac{N_1}{\epsilon_2 \epsilon_3} - \frac{N_2}{\epsilon_1 \epsilon_3} -\frac{N_3}{\epsilon_1 \epsilon_2}
\end{equation} 
aka the ``magnetic charge'' of the surface defect. 
\item M2 branes can end supersymmetrically on M5 branes wrapping compatible submanifolds of $\mathbb{C}_{\epsilon_1} \times\mathbb{C}_{\epsilon_2} \times\mathbb{C}_{\epsilon_3}$. The endpoints are expected to be built from ``degenerate modules'' of $\cW_{N_1,N_2,N_3}$ \cite{Gaiotto:2017euk} as well as Verma modules for $A_{N_1,N_2,N_3}$ \cite{Gaiotto:2019wcc}. We will verify and make this expectation precise, as well as identify more general gauge-invariant intersections. Screening charges will make a surprise appearance.
\end{itemize} 

\begin{figure}[h]
    \centering
        \includegraphics[width=7.5cm]{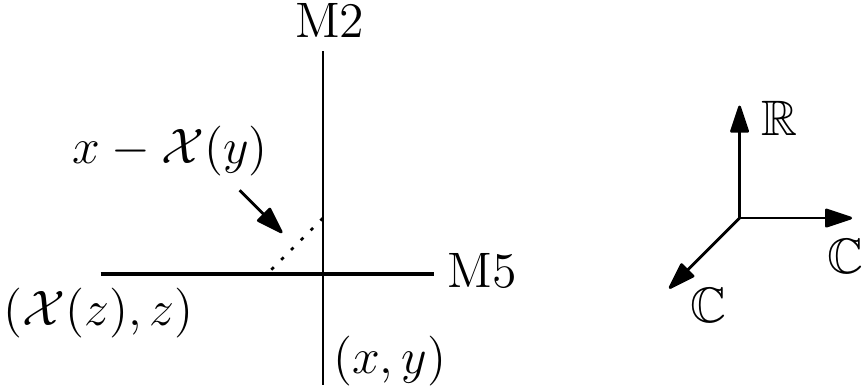}
   \caption{The distance between M2 brane at $(x,y)\in \mathbb{C}^2$ and M5 brane spanning $(\mathcal{X}(\mathcal{z}),z)\in \mathbb{C}^2$ leads to the Miura operator.}
\label{pic35}
\end{figure}

The simplest possible intersection involves fully transverse branes. For example, imagine a single M2 brane wrapping $\mathbb{C}_{\epsilon_3}$ and crossing an M5 brane wrapping $\mathbb{C}_{\epsilon_1}\times \mathbb{C}_{\epsilon_2}$. The intersection is a point-like even in the physical M-theory. Basic string dualities dictate that the intersection point should support some fermion zeromodes, with a mass controlled by the transverse separation between the branes. Assuming we have only a pair of zeromodes whose mass is controlled by the separation, the expectation value of the intersection will be proportional to the distance between the 
branes along the $\mathbb{C}$ direction transverse to both as in the figure \ref{pic35}. We could denote it classically as 
\begin{equation}
x - {\cal X}(y)
\end{equation}
where $(x,y)$ are the coordinates of the M2 brane in $\mathbb{C}^2$ and ${\cal X}(z)$ is the field which controls the transverse position of the M5 brane. 

Quantum mechanically, we propose that the intersection point is represented by the {\it Miura operator}:
\begin{equation}
\cL^{0,0,1}_{0,0,1} \equiv \epsilon_3 \partial_z-\epsilon_1\epsilon_2 J^{(3)}(z)
\label{intromiura}
\end{equation} 
where $(\epsilon_3 \partial_z, z)$ quantize $(x,y)$ into a Weyl algebra and $J^{(3)}(z)$ is a $\widehat{\mathfrak{u}}(1)$ current of level $- (\epsilon_1\epsilon_2)^{-1}$ 
which quantized the transverse field ${\cal X}(z)$. In particular, we will demonstrate in Section \ref{sec:miura} that the Miura operator is ``gauge invariant'' 
in the sense determined by the $\Delta_{A,W_\infty}$ mixed coproduct. 

\begin{figure}[h]
    \centering
        \includegraphics[width=14cm]{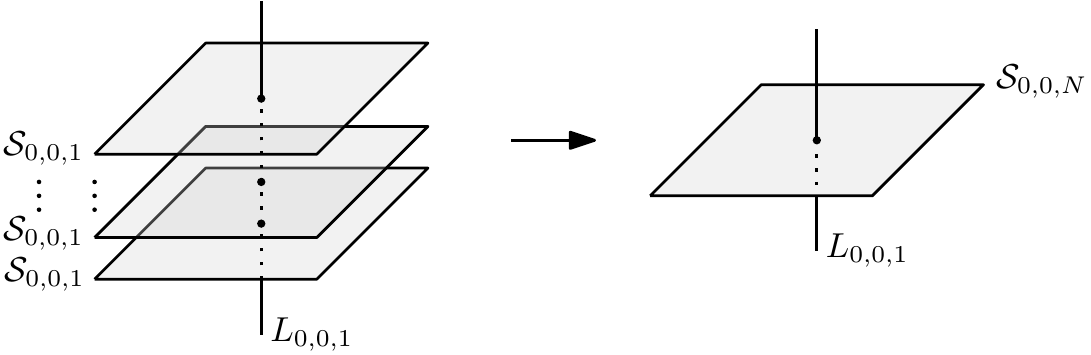}
   \caption{Composition of $N$ Miura operators associated to a single M2 brane and different M5 branes leads to $\mathcal{L}^{0,0,1}_{0,0,N}$.}
\label{picdefects}
\end{figure}

The Miura operator is often employed to {\it define} the $\cW_\infty$ algebra: 
the composition of multiple Miura operators 
\begin{equation}
\cL^{0,0,1}_{0,0,N} \equiv \left(\epsilon_3 \partial_z-\epsilon_1\epsilon_2 J_1^{(3)}(z) \right) \cdots \left(\epsilon_3 \partial_z-\epsilon_1\epsilon_2 J_N^{(3)}(z) \right),
\end{equation} 
gives a differential operator whose coefficients generate $\cW_{0,0,N}$, and $\cW_\infty$ is recovered by continuing $N$ to a generic parameter \cite{Drinfeld:1984qv,Fateev:1987zh,Prochazka:2014gqa}. The construction equips $\cW_\infty$ with a natural coproduct $\Delta_{\cW_\infty,\cW_\infty}$, given by the composition of Miura operators. 
Essentially, that means we can rediscover the $\cW_\infty$ algebra by fusing elementary gauge-invariant M2-M5 intersections to represent a single M2 brane crossing a stack of M5 branes. 

Alternatively, we can take the OPE along the $z$ direction of two or more Miura operators associated to separate M2 branes\footnote{The fact that composition of simple Miura operators leads to Calogero Hamiltonians was first observed in \cite{Prochazka:2019dvu}.}:
\begin{equation}
\cL^{0,0,n}_{0,0,1} \equiv \left(\epsilon_3 \partial_{z_1}-\epsilon_1\epsilon_2 J^{(3)}(z_1) \right) \cdots \left(\epsilon_3 \partial_{z_n}-\epsilon_1\epsilon_2 J^{(3)}(z_n) \right),
\end{equation} 
Expanding this out in a basis for the $\widehat{\mathfrak{u}}(1)$ vacuum module, the coefficients will be Calogero-like differential operators 
in the $n$ variables $z_n$. These differential operators essentially generate $A_{0,0,n}$ and can be used to reconstruct $A$.
Again, this construction would also determine the meromorphic coproducts $\Delta_{A,A}(w)$. 

\begin{figure}[h]
    \centering
        \includegraphics[width=12cm]{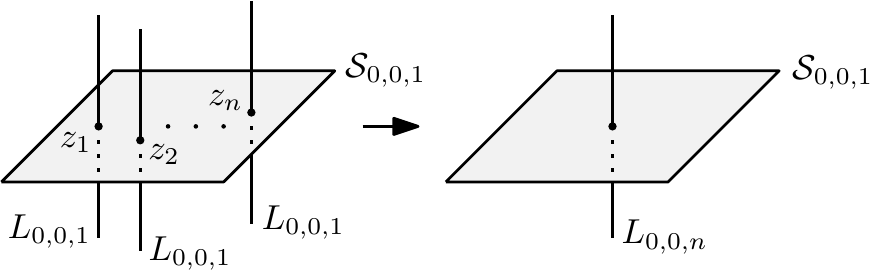}
   \caption{Composition of $n$ Miura operators associated to a single M5 brane and different M2 branes at positions $z_i$ leads to $\mathcal{L}^{0,0,n}_{0,0,1}$.}
\label{picdefects}
\end{figure}

In other word, the protected sector for the strongly-coupled CFTs of multiple M2 branes or multiple M5 branes can 
be fully reconstructed from the knowledge of the gauge-invariant intersection of a single M2 brane and a single M5 brane. 

\subsection{Structure of the paper}
Although we could in principle derive all the coproducts from a careful analysis of the Miura operator 
and its behaviour under composition and OPE, we find it simpler to first derive the coproducts using an alternative route and 
then verify that the Miura operator is gauge-invariant. Our starting point will be the relation between $A$ and the Calogero model: in Section \ref{sec:coproducts} we will employ it to derive coproducts which we identify with $\Delta_{A,A}(w)$ and $\Delta_{A,W_\infty}$, as well as some simple gauge-invariant intersections. We then introduce the Miura operator as a gauge-invariant intersection in Section \ref{sec:miura} and discuss general gauge-invariant endpoints and a relation to PT invariants in 
Section \ref{sec:boxes}. We conclude with some open questions in Section \ref{sec:future}, an Appendix \ref{app:winfty} devoted to $\cW_\infty$ and an Appendix \ref{app:box} devoted to examples of box-counting modules. 

\section{The Calogero representations of $A$ and various coproducts}\label{sec:coproducts}
The algebra $A$ has many different useful presentations. It can be challenging to 
relate them explicitly. 

In this paper we will find useful to generate the whole algebra from the commuting collection $t_{0,d}$ together with $t_{2,0}$,
which is the raising operator of an $U(\mathfrak{sl}_2)$ subalgebra \cite{Gaiotto:2020vqj}.
All other generators are obtained recursively from 
\begin{equation}
[ t_{2,0}, t_{c,d} ] = 2 d\, t_{c+1,d-1}
\end{equation}
Furthermore, all commutators can be computed recursively from the basic set
\begin{equation}
[ t_{3,0}, t_{0,d} ] = 3 d\, t_{2,d-1} + \sigma_2 \frac{d (d - 1) (d - 2)}{4} t_{0,d-3}+ \frac32 \sigma_3 \sum_{m=0}^{d-3} (m + 1)(d - m - 2) t_{0,m} t_{0,d-3-m}
\end{equation}
which introduces the explicit dependence on the deformation parameters $\sigma_2$ and $\sigma_3$, 
\begin{equation}
\sigma_2 = \epsilon_1^2 + \epsilon_1 \epsilon_2 + \epsilon_2^2 \qquad \qquad \sigma_3 = \epsilon_1  \epsilon_2  \epsilon_3
\end{equation}
with the property that $\epsilon_i^3 = \sigma_2 \epsilon_i + \sigma_3$. 

\subsection{Weyl representations}
We can give a simple representation of $A$ in terms of polynomial differential operators in one variable, starting from
\begin{align}
t_{0,d} &= \epsilon_1^{-1} z^d \cr
t_{2,0} &= \epsilon_1 \partial_z^2
\end{align}
We compute easily 
\begin{align}
t_{1,d-1} &= z^{d-1} \partial_z + \frac{d-1}{2} z^{d-2} \cr
t_{2,d-1} &= \epsilon_1 z^{d-1} \partial_z^2 + (d-1) \epsilon_1 z^{d-2} \partial_z + \frac{(d-1)(d-2)}{4} \epsilon_1 z^{d-3} \cr
t_{3,0} \quad&= \epsilon^2_1  \partial_z^3 
\end{align}
The seed commutators
\begin{equation}
[ t_{3,0}, t_{0,d} ] =  3 d\, t_{2,d-1} + \epsilon_1 \frac{d(d-1)(d-2)}{4} z^{d-3} 
\end{equation}
hold as long as $\epsilon_1^3 = \sigma_2 \epsilon_1 + \sigma_3$.

These formulae give an algebra morphism $\ell_{1,0,0}: A \to A_{1,0,0}$ from $A$ to a Weyl algebra, so that $t_{ab}$ is the symmetrized version of $\epsilon_1^{-1} z^b (\epsilon_1 \partial_z)^a$. It governs the coupling of the line defect $L_{1,0,0}$ built from a single M2 brane 
with orientation $\mathbb{R} \times \mathbb{C}_{\epsilon_1}$. Permutations of the $\epsilon_i$ give two other representations $\ell_{0,1,0}: A \to A_{0,1,0}$ and $\ell_{0,0,1}: A \to A_{0,0,1}$ which are associated to the two line defects $L_{0,1,0}$ and $L_{0,0,1}$ built from a single M2 brane with orientations $\mathbb{R} \times \mathbb{C}_{\epsilon_2}$ 
or $\mathbb{R} \times \mathbb{C}_{\epsilon_3}$ respectively. 

We should think about the variable $z$ as the position of the M2 brane in the complex plane along which we will place surface defects and along which we will fuse line defects. This identification suggests that an endpoint of $\ell_{1,0,0}$ on some surface defect should be represented by some vertex operator $O(z)$ for the surface defect VOA, placed at position $z$. 

This interpretation is actually consistent with the idea that an endpoint should be an element in a space of local operators 
which is a module for $A_{1,0,0}$ as well as a VOA module for the surface defect world-sheet chiral algebra. Indeed, we can Taylor-expand
\begin{equation}
O(z) = O(0) + z \, \partial O(0) + \frac{z^2}{2} \, \partial^2 O(0) + \cdots
\end{equation}
to identify explicitly $O(z)$ as an element of the VOA module generated by $O(0)$ and its descendants tensored with the space of polynomials in $z$, which is indeed a (left- or right-) highest weight module for $A_{1,0,0}$.\footnote{According to a proposal in \cite{Gaiotto:2019wcc}, this is a special case of a general principle: endpoints of line defects on surface defects  should always be built from highest weight modules of $A_{n_1,n_2,n_3}$. This will indeed be the case in all endpoints we build in this paper. }

Consider, for example, an endpoint for a line defect stretched along the negative real axis. Gauge-invariance will follow from relations of the form 
\begin{align}
\Delta_{W_\infty}[t_{0,n}] O(z) &= \epsilon_1^{-1} z^n O(z) \cr
\Delta_{W_\infty}[t_{2,0}] O(z) &=\epsilon_1 \partial^2 O(z)
\end{align}
likely imposing some shortening conditions on the VOA module. Indeed, when the surface defect is built from M5 branes, $O(z)$ is expected to be one of the simplest degenerate modules for $\cW_{N_1,N_2,N_3}$ \cite{Gaiotto:2017euk}. At this point, we could just take the known structure of such degenerate modules and work out the explicit form of $\Delta_{W_\infty}$. 
This is somewhat tedious and we will follow a different route. 

\subsection{Calogero representations}
The world-line algebra $A_{n,0,0}$ for the line defect $L_{n,0,0}$ built from $n$ M2 branes with the same orientation 
was derived in \cite{Costello:2017fbo} and described further in \cite{Gaiotto:2019wcc}. It is the quantization of the ADHM moduli space of $U(1)$ instantons in $\mathbb{C}^2$, aka the spherical DAHA, and inherits from it various integrable structures. In particular, it inherits a relation to the Calogero model \cite{Etingof,Opdam,Kodera:2016faj} which we will be very useful to us.

As a warm-up, consider the following tentative ``Calogero'' representation of $A$ as meromorphic differential operators in two variables:
\begin{align}
t_{0,d} &= \epsilon_1^{-1} z_1^d +\epsilon_1^{-1} z_2^d \cr
t_{2,0} &= \epsilon_1 \partial_1^2+ \epsilon_1 \partial_2^2 + \frac{\epsilon_2 \epsilon_3}{\epsilon_1} \frac{2}{(z_1-z_2)^2}
\end{align}
We can compute 
\begin{equation}
t_{3,0} = \epsilon^2_1 \partial_1^3+ \epsilon^2_1 \partial_2^3 + \frac{\epsilon_2 \epsilon_3}{\epsilon_1} \frac{3}{(z_1-z_2)^2} (\partial_1 + \partial_2)
\end{equation}
and verify that the necessary relations hold. 

More generally, we have a Calogero representation on $n$ variables:
\begin{align}
t_{0,d} &= \epsilon_1^{-1} \sum_{i=1}^n z_i^d \cr
t_{2,0} &= \epsilon_1 \sum_{i=1}^n \partial_i^2+ \frac{\epsilon_2 \epsilon_3}{\epsilon_1} \sum_{i<j} \frac{2}{(z_i-z_j)^2}
\end{align}
The differential operators on the right hand side actually generate $A_{n,0,0}$. This defines the algebra morphism $\ell_{n,0,0}: A \to A_{n,0,0}$. In particular, the $t_{d,0}$ generators are the commuting Hamiltonians of the Calogero model. 

Intuitively, we expect the $z_n$ to represent the individual positions of the $n$ M2 branes which build the line defect $L_{n,0,0}$.
Based on this idea, we expect two facts to hold:
\begin{itemize}
\item The endpoints for $L_{n,0,0}$ should take the form $\prod_a O(z_a)$ where $O(z)$ is an endpoint for $L_{1,0,0}$.
\item The fusion of $L_{n,0,0}$ and $L_{n',0,0}$ should give the $L_{n+n',0,0}$ line defect. Correspondingly, 
composing the coproduct $\Delta_{A,A}(w)$ with the Calogero representations for $A_{n,0,0}$ and $A_{n',0,0}$
should result in the Calogero representation for $A_{n+n',0,0}$, with variables $z_a$ and $z'_a+w$. 
\end{itemize} 
We will now use the latter proposal to determine the meromorphic coproducts, a Calogero-like representation for $A_{n_1,n_2,n_3}$ as well as the image of $\Delta_{W_\infty}$ in $\cW_{0,0,1}$. 

\subsection{The meromorphic coproducts and general M2 brane algebras}\label{sec:mero}
We will now derive the form of the meromorphic coproducts from the Calogero representation. Split the $z_i$ into two groups, $z'_a$ and $w+ \tilde z_b$ and expand 
\begin{equation}
\frac{1}{(w + \tilde z_b-  z'_a)^2} = \sum_{n\geq 0}\sum_{m\geq 0} \frac{(m+n+1)!}{m! n!} (-1)^n (z'_a)^m (\tilde z_b)^n w^{-m-n-2} 
\end{equation}
to write 
\begin{align}
t_{0,n}[w] &= t'_{0,n}+ \sum_{m=0}^n {n \choose m} w^{n-m} \tilde t_{0,m}\cr
t_{2,0}[w] &=  t'_{2,0}+ \tilde t_{2,0} + 2 \sigma_3 \sum_{n\geq 0}\sum_{m\geq 0} \frac{(m+n+1)!}{m! n!} (-1)^n  t'_{0,m} \tilde t_{0,n} w^{-m-n-2}
\end{align}
It is straightforward to verify that these new generators satisfy the relations for $A$ as formal power series in $w^{-1}$ and thus define 
the family of coproducts $\Delta_{A,A}(w): A \to A \otimes A$. By construction, they are associative much in the same sense an an OPE is associative. 

There is another useful construction. Although the coproducts $\Delta_{A,A}(w)$ are written as formal power series, 
if we compose them with certain truncations of the algebra, the formal power series can be resummed and evaluated back at $w=0$.
As an intermediate step, we can go back to the Calogero representation for $A_{n+1,0,0}$ and expand the differential operators in inverse powers of one of the variables, say $z$. This gives us partially resummed expressions for the composition $\left[\ell_{1,0,0} \otimes 1 \right] \circ \Delta_{A,A}(w)$ evaluated back at $w=0$. 
\begin{align}
t_{0,d} &= t'_{0,d}+  \epsilon_1^{-1} z^d \cr
t_{2,0} &= t'_{2,0} + \epsilon_1 \partial_z^2+ 2 \epsilon_2 \epsilon_3\sum_{k=0}^\infty (k+1) z^{-k-2} t'_{0,k}
\end{align}
These expressions map $A$ to the algebra of differential operators on $\mathbb{C}^*$ valued in $A$.

We can apply a permutation of the $\epsilon_i$ to get resummed expressions for the meromorphic coproducts with $A_{0,1,0}$ and 
$A_{0,0,1}$. 

Applying these repeatedly and resumming at every step, we finally arrive at one of our main results for this section: a Calogero-like representation of $A$ depending on three non-negative integers $n_1$, $n_2$, $n_3$:
\begin{align}
t_{0,d} &=  \epsilon_1^{-1} \sum_{i=1}^{n_1} z_i^d + \epsilon_2^{-1} \sum_{i=1}^{n_2} (z'_i)^d + \epsilon_3^{-1} \sum_{i=1}^{n_3} (z''_i)^d  \cr
t_{2,0} &=  \epsilon_1 \sum_{i=1}^{n_1} \partial_{z_i}^2+ \frac{\epsilon_2 \epsilon_3}{\epsilon_1} \sum_{i<j} \frac{2}{(z_i-z_j)^2}+ \epsilon_1 \sum_{i,j} \frac{2}{(z'_i-z''_j)^2} +\cr
&+ \epsilon_2 \sum_{i=1}^{n_2} \partial_{z'_i}^2+ \frac{\epsilon_1 \epsilon_3}{\epsilon_2} \sum_{i<j} \frac{2}{(z'_i-z'_j)^2}+ \epsilon_2 \sum_{i,j} \frac{2}{(z_i-z''_j)^2} +\cr
&+ \epsilon_3 \sum_{i=1}^{n_3} \partial_{z''_i}^2+ \frac{\epsilon_1 \epsilon_2}{\epsilon_3} \sum_{i<j} \frac{2}{(z''_i-z''_j)^2}+ \epsilon_3 \sum_{i,j} \frac{2}{(z_i-z'_j)^2} 
\end{align}
We will denote as $A_{n_1,n_2,n_3}$ the algebra generated by the differential operators on the right hand side. These expressions define an algebra morphism $\ell_{n_1,n_2,n_3}: A \to A_{n_1,n_2,n_3}$ with ``charge'' 
\begin{equation}
t_{0,0} = \frac{n_1}{\epsilon_1}+ \frac{n_2}{\epsilon_2}+ \frac{n_3}{\epsilon_3}
\end{equation}
It is a natural candidate to represent the couplings for a line defect $L_{n_1,n_2,n_3}$ associated to the superposition of three stacks of M2 branes, wrapping the three possible complex planes in the $\mathbb{C}_{\epsilon_1} \times\mathbb{C}_{\epsilon_2} \times\mathbb{C}_{\epsilon_3}$ internal geometry. 

\subsection{The oscillator representation and degenerate fields} \label{sec:deg}
As we expect $\Delta_{A,A}(w)$ to represent the fusion of line defects along the complex direction, we also expect the Calogero representation 
of $A_{n_1,n_2,n_3}$ to be an ingredient in the gauge variation of local operators of the schematic form 
\begin{equation}
\prod_a O_1(z_a) \prod_b O_2(z'_a) \prod_c O_3(z''_c)
\end{equation}
representing the endpoint of $L_{n_1,n_2,n_3}$ on some surface defect. 

The simplest type of endpoint predicted by the corner constructions of \cite{Gaiotto:2017euk} is an endpoint of $A_{n_1,n_2,0}$ onto a surface defect ${\cal S}_{0,0,1}$ associated to a single M5 brane wrapping $\mathbb{C}\times \mathbb{C}_{\epsilon_1} \times \mathbb{C}_{\epsilon_2}$. The world-sheet theory of such a defect is 
a $\widehat{\mathfrak{u}}(1)_{\kappa_3}$ Kac-Moody algebra at level $\kappa_3 = - (\epsilon_1 \epsilon_2)^{-1}$, aka
$\cW_{0,0,1}$, and is equipped with a VOA map $s_{0,0,1}$ from $\cW_\infty$. 

The endpoints of individual M2's are expected to be $\widehat{\mathfrak{u}}(1)$ primary vertex operators with specific charge. We will write them as 
\begin{equation}
O_1(z_a) = e^{\pm \epsilon_2  \phi(z_a)} \qquad \qquad O_2(z_a) = e^{\pm \epsilon_1  \phi(z_a)}
\end{equation}
where the sign depends on the orientation of the line defect and we bosonize the Kac-Moody current as $J = \partial \phi$.

In order to derive the image of $\Delta_{W_\infty}$ in $\cW_{0,0,1}$, which enters the gauge-invariance condition for these endpoints, it will be sufficient to consider at first the case of $A_{n,0,0}$. 

For $n=1$, we identified the Taylor expansion of a vertex operator $O(z)$ as an element in the tensor product of 
a VOA module and of the $A_{1,0,0}$ module consisting of polynomials of $z$. As we try to generalize this statement, we encounter an apparent obstacle: the Calogero operators for 
$A_{n,0,0}$ do not act nicely on the space of polynomials in the $z_a$, because of the meromorphic dependence on the $z_i$. 

This is actually OK: a product of vertex operators $\prod_a O_1(z_a)$ for a $\widehat{\mathfrak{u}}(1)$ Kac-Moody algebra has 
a dependence in the $z_a$ which precisely solves the problem: 
\begin{equation}
\prod_a O_1(z_a) = \prod_{b<c} (z_b-z_c)^{-\frac{\epsilon_2}{\epsilon_1}} :\prod_a O_1(z_a):\,.
\end{equation}
As a consequence, an expression such as $\prod_a O_1(z_a)$ can be identified as an element of the $\widehat{\mathfrak{u}}(1)_{\kappa_3}$ module generated from the primary field $e^{\pm n \epsilon_2  \phi(0)}$, tensored with 
a space of functions of the form 
\begin{equation}
p(z_1 ,\cdots, z_n) \prod_{b<c} (z_b-z_c)^{-\frac{\epsilon_2}{\epsilon_1}}
\end{equation}
with $p$ being a symmetric polynomial in the $z_a$. 

We will see momentarily that the Calogero differential operators act naturally (from the left or from the right) 
on that space of functions, which is thus equipped with the structure of a (highest weight left- or right-) module for $A_{n,0,0}$. 
It thus makes sense to write down a gauge-invariance condition for $\prod_a O_1(z_a)$.

We are now ready for a concrete calculation. We will start from endpoints of line defects along the negative real axis. Recall that differential operators act on functions from the right as 
\begin{equation}
f(z) \circ \partial_z =- \partial_z f(z) 
\end{equation}
It is useful to introduce a generating function 
\begin{equation}
e^{\epsilon_2 \sum_{i=1}^n \sum_{k=1}^\infty \frac{\tau_k}{k} z_i^k}  \prod_{c<d} (z_c-z_d)^{-\frac{\epsilon_2}{\epsilon_1}}
\end{equation}
which will later be identified with a normal-ordered product of vertex operators. 

It satisfies 
\begin{align}
\left[ e^{\epsilon_2 \sum_{i=1}^n \sum_{k=1}^\infty \frac{\tau_k}{k} z_i^k}\right] \circ \partial_{z_a} &=  - \epsilon_2 \left[  \sum_{k=1}^\infty \tau_k z_a^{k-1} \right] \left[ e^{\epsilon_2\sum_{i=1}^n \sum_{k=1}^\infty  \frac{\tau_k}{k} z_i^k}\right] \cr
t \partial_{\tau_t} \circ \left[ e^{\epsilon_2 \sum_{i=1}^n \sum_{k=1}^\infty  \frac{\tau_k}{k} z_i^k}\right] &=  \epsilon_2 \left[ \sum_{i=1}^n z_i^t \right] \left[ e^{\epsilon_2 \sum_{i=1}^n \sum_{k=1}^\infty  \frac{\tau_k}{k} z_i^k}\right] 
\end{align}
We have
\begin{align}
& \left[\epsilon_1 \sum_{a=1}^n \partial^2_{z_a} + 2 \epsilon_2 \sum_{a<b}   (\partial_{z_a}-\partial_{z_b})\frac{1}{z_a-z_b}\right]\prod_{c<d} (z_c-z_d)^{-\frac{\epsilon_2}{\epsilon_1}} = \cr
&= \prod_{c<d} (z_c-z_d)^{-\frac{\epsilon_2}{\epsilon_1}} \left[\epsilon_1 \sum_{i=1}^n \partial_i^2+ \frac{\epsilon_2 \epsilon_3}{\epsilon_1} \sum_{i<j} \frac{2}{(z_i-z_j)^2}\right]
\end{align}
so that we can trade the right Calogero action on the generating function for a left action of $\tau_i$ derivatives
\begin{align}
&\left[e^{\epsilon_2 \sum_{i=1}^n \sum_{k=1}^\infty \frac{\tau_k}{k} z_i^k}  \prod_{c<d} (z_c-z_d)^{-\frac{\epsilon_2}{\epsilon_1}} \right]\circ t_{2,0} = \cr
 &=  \Big[\epsilon_1 \epsilon_2  \sum_{k=1}^\infty \sum_{t=1}^\infty \tau_k \tau_t (k+t-2)\partial_{\tau_{k+t-2}}-
 \sum_{k=1}^\infty \sum_{t=0}^{k-2}\tau_k  t \partial_{\tau_{t}} (k-2-t) \partial_{\tau_{k-2-t}} +\cr
 &- \epsilon_3 \sum_{k=2}^\infty (k-1) \tau_k (k-2) \partial_{\tau_{k-2}} 
 \Big]\circ \left[e^{\epsilon_2 \sum_{i=1}^n \sum_{k=1}^\infty \frac{\tau_k}{k} z_i^k} \prod_{c<d} (z_c-z_d)^{-\frac{\epsilon_2}{\epsilon_1}} \right]
 \end{align}
 where any ``$0 \partial_{\tau_0}$'' in the equations above should be interpreted as the constant $\epsilon_2 n$.
 
Introducing explicit $\widehat{\mathfrak{u}}(1)_{\kappa_3}$ generators
\begin{equation}
J_{-k} := -\tau_k \qquad \qquad J_k = \frac{1}{\epsilon_1 \epsilon_2} k \partial_{\tau_k} \qquad \qquad J_0 = \frac{1}{\epsilon_1} n
\end{equation}
we find a new ``oscillator'' representation of the $A$ generators, i.e. an algebra map $A \to \widehat{\mathfrak{u}}(1)_{\kappa_3}$ 
\begin{align}
t_{0,k} &= J_{k} \cr
t_{2,0} &= \frac{1}{3 \kappa_3^2}  \sum_{k=-\infty}^\infty \sum_{m=-\infty}^\infty :J_{-m-k-2} J_m J_k:+  \frac{\sigma_3}{2}  \sum_{k=-\infty}^\infty |k| :J_{1-k} J_{k+1}:
\end{align}
with $\kappa_3^{-1} = -\epsilon_1 \epsilon_2$. This representation has a nice symmetry under the exchange $\epsilon_1 \leftrightarrow \epsilon_2$, even though the original Calogero representation did not. 

We also find that the element 
\begin{equation}
\prod_{a=1}^n e^{-\epsilon_2  \phi(z_a)} |0 \rangle =  e^{-\epsilon_2 \sum_{a=1}^n \sum_{k=1}^\infty k^{-1} z_a^k J_{-k} } |\epsilon_1^{-1} n \rangle \prod_{b<c} (z_b-z_c)^{-\frac{\epsilon_2}{\epsilon_1}}
\end{equation}
is an intertwiner between the oscillator representation and the Calogero representation of $A$:
\begin{equation}
t_{ab}\circ \left[ \prod_{k=1}^n e^{-\epsilon_2  \phi(z_k)} |0 \rangle \right]= \left[\prod_{k=1}^n e^{-\epsilon_2  \phi(z_k)} |0 \rangle  \right] \circ t_{ab}\end{equation}
where we use the $\widehat{\mathfrak{u}}(1)$ image of $t_{ab}$ on the left side and the $A_{n,0,0}$ image 
on the right hand side. For clarity, we used the state-operator map to discuss the VOA module associated to a vertex operator
at the origin.

We thus identify the map $A \to \widehat{\mathfrak{u}}(1)_{\kappa_3}$ as the composition $s_{0,0,1} \circ \Delta_{W_\infty}$ 
and the intertwining condition as the condition for gauge invariance for an endpoint of $L_{n,0,0}$ onto  ${\cal S}_{0,0,1}$ from the right. 

As a test of this identification, we verify that $\Delta_{W_\infty}$ is a non-linear deformation of $t_{n,m} \to W_{n+1,m-n}$. 
Indeed, the map $s_{0,0,1}$ maps the $W_n(z)$ generators to certain local polynomials in $J(z)$ and its derivatives.  
The oscillator expression for $t_{0,k}$ coincides with the image of $W_{1,k}$ in $\widehat{\mathfrak{u}}(1)_{\kappa_3}$. The
oscillator expression for $t_{2,0}$, instead, differs from the image of $W_{3,-2}$ by a bilinear term in the currents proportional to the deformation parameter $\sigma_3$, as expected.   

It is not hard to generalize the intertwining relation to a collection of vertex operators with two types of charges:
\begin{equation}
t_{ab}\circ \left[ \prod_{k=1}^n e^{-\epsilon_2  \phi(z_k)} \prod_{k'=1}^{n'} e^{-\epsilon_1  \phi(z'_{k'})} |0 \rangle \right]= \left[\prod_{k=1}^n e^{-\epsilon_2  \phi(z_k)} \prod_{k'=1}^{n'} e^{-\epsilon_1  \phi(z'_{k'})} |0 \rangle \right]  \circ t_{ab}\end{equation}
where we use the $\widehat{\mathfrak{u}}(1)_{\kappa_3}$ image of $t_{a,b}$ on the left side and the $A_{n,n',0}$ image 
on the right hand side.

This relation can be verified directly, or derived recursively with the help of the resummed meromorphic coproduct:
if $|v \rangle$ intertwines the oscillator representation and some other of representation $A \to R$, 
then $e^{-\epsilon_2  \phi(z)} |v \rangle$ intertwines the oscillator representation and the representation of $A$ into the tensor product of $R$ and the algebra differential operators acting on $z \in \mathbb{C}^*$.  

Before moving on, we should mention a perspective which will be useful in Section \ref{sec:boxes}.
Consider the right Verma module for $\widehat{\mathfrak{u}}(1)$ with momentum $n \epsilon_2$. We have a map 
\begin{equation}
\pi_n: \langle v| \to \langle v| \prod_{k=1}^n e^{-\epsilon_2  \phi(z_k)} |0 \rangle
\end{equation}
projecting a vector in the Verma module to a function of the $z_k$. The intertwining relations show that this is a morphism of right $A$-modules. For example, if we use the generator of the Verma module, the correlation function is
\begin{equation}
 \prod_{a<b} (z_a-z_b)^{-\frac{\epsilon_2}{\epsilon_1}} 
 \end{equation}
which is indeed annihilated by the $t_{2,0}$ Calogero Hamiltonian generator in $A_{n,0,0}$.

Similarly, the projection 
\begin{equation}
\pi_{n,n'}: \langle v| \to \langle v| \prod_{k=1}^n e^{-\epsilon_2  \phi(z_k)}  \prod_{k'=1}^{n'} e^{-\epsilon_1  \phi(z'_{k'})}|0 \rangle
\end{equation}
from the right Verma module with momentum $n \epsilon_2+ n' \epsilon_1$ to functions of the $z_k$ and the $z'_k$ is a morphism of right $A$-modules. For example, if we use the generator of the Verma module, the correlation function is
\begin{equation}
 \prod_{a<b} (z_a-z_b)^{-\frac{\epsilon_2}{\epsilon_1}}  \prod_{c<d} (z'_c-z'_d)^{-\frac{\epsilon_1}{\epsilon_2}}\prod_{a,c} (z_a-z'_c)^{-1}
\end{equation}
which is indeed annihilated by the $t_{2,0}$ Calogero-like Hamiltonian generator in $A_{n,n',0}$. Similar relations 
were known before, see e.g. \cite{Cardy:2003td,Doyon:2006ph,Estienne:2009mp,Estienne:2010as,Estienne:2011qk}. 

Notice that we do not expect to have endpoints of $L_{0,0,1}$ onto ${\cal S}_{0,0,1}$, so we cannot produce in the same manner a ground-state wavefunction for the $t_{2,0}$ Calogero-like Hamiltonian generator in $A_{n,n',n''}$, unless we employ the degenerate operators for a more complicated VOA such as $\cW_{0,1,1}$. Note also that generators $t_{m,0}$ of $A_{n,n',n''}$ form an infinite set of mutually commuting differential operators of order $m$ leading to a family of "tripled" Calogero-like integrable systems generalizing the known "doubled" models from \cite{Cardy:2003td,Doyon:2006ph,Estienne:2009mp,Estienne:2010as,Estienne:2011qk}.
 
\subsection{The $A \to A \otimes \widehat{\mathfrak{u}}(1)$ coproduct}
We can use the map $A \to \widehat{\mathfrak{u}}(1)_{\kappa_3}$ we just found as inspiration for a mixed coproduct $A \to A \otimes \widehat{u}(1)_{\kappa_3}$:
\begin{align}
t_{0,n} &= t'_{0,n} + J_{n} \cr
t_{2,0} &=t'_{2,0} +2 \sigma_3 \sum_{n=1}^\infty n t'_{0,n-1} J_{-n-1} +  \frac{1}{3 \kappa_3^2}  \sum_{n=-\infty}^\infty \sum_{m=-\infty}^\infty :J_{m+n-2} J_{-m} J_{-n}:\cr &+  \frac{ \sigma_3}{2 }  \sum_{n=-\infty}^\infty |n| :J_{-1-n} J_{n-1}:
\label{coproduct1}
\end{align}
which we expect to coincide with the composition $\left[1_A \otimes s_{0,0,1}\right] \circ \Delta_{A,W_\infty}$ and thus enter the condition for gauge-invariance of local operators on ${\cal S}_{0,0,1}$ attached to generic line defects from the left and from the right.

As a first test, we can look at left endpoints for $L_{n,0,0}$. Compose $A \to A \otimes \widehat{u}(1)_{\kappa_3}$ 
with the Calogero representation $A_{n,0,0}$ for $t'_{ab}$, to get an action of $t_{ab}$ on 
functions of $z_a$ valued in a module for $\widehat{u}(1)_{\kappa_3}$.
The condition for an element $O$ to be annihilated by all $t_{ab}$ is that it is annihilated by the combinations
\begin{align}
t'_{0,n} &+J_n \cr
t'_{2,0} &+  \frac{1}{3 \kappa_3^2}  \sum_{n=-\infty}^\infty \sum_{m=-\infty}^\infty :J_{m+n-2} J_{-m} J_{-n}:- \frac{\sigma_3}{2}  \sum_{n=-\infty}^\infty |n| :J_{-1-n} J_{n-1}: 
\end{align}

Now, realize $t'_{2,0}$ as a Calogero Hamiltonian, and act on the conjugate vector 
\begin{equation}
\prod_{a=1}^n e^{\epsilon_2  \phi(z_a)} |0 \rangle =  e^{\epsilon_2 \sum_{a=1}^n \sum_{k=1}^\infty k^{-1} z_a^k J_{-k} } |-\epsilon_1^{-1} n \rangle \prod_{b<c} (z_b-z_c)^{- \frac{\epsilon_2}{\epsilon_1}}
\end{equation}
By the same analysis as before, we recognize this vector is indeed annihilated by all $t_{ab}$. We thus recognize that the 
local operator $\prod_{a=1}^n e^{\epsilon_2  \phi(z_a)}$ gives a gauge-invariant endpoint, as expected from \cite{Gaiotto:2017euk}:
the combination of the gauge anomalies arising from the line and the surface defect cancels the vanishing gauge anomaly coming from the trivial defect.  

Similarly, 
\begin{equation}
\prod_{a=1}^n e^{\epsilon_2  \phi(z_a)} \prod_{a=1}^{n'} e^{\epsilon_1 \phi(z'_a)} |0 \rangle 
\end{equation}
should be annihilated by the action of $A$ obtained by composing $A \to A \otimes \widehat{u(1)}_{\kappa_3}$ 
with the Calogero representation $A_{n,n',0}$ for $t'_{a,b}$, representing the endpoint of a line defect $L_{n,n',0}$.

\begin{figure}[h]
    \centering
        \includegraphics[width=12.5cm]{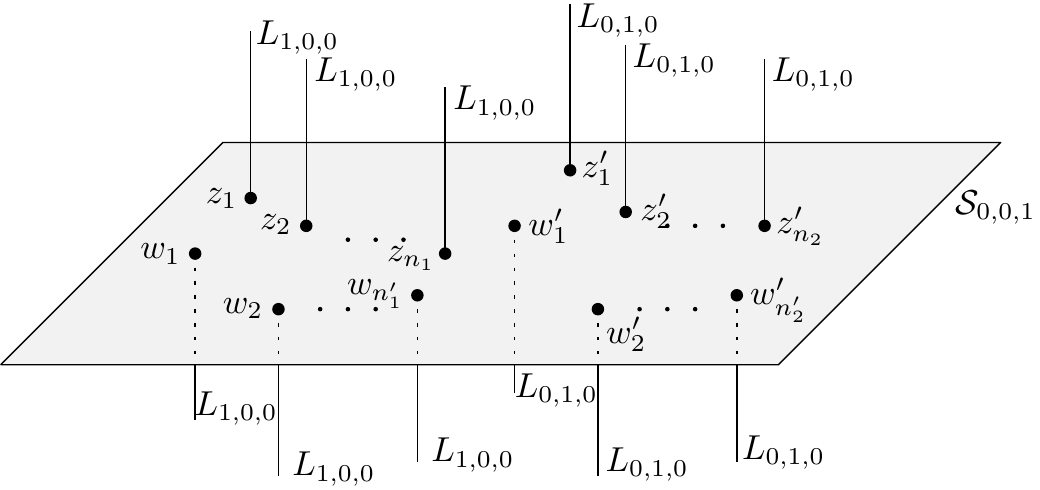}
   \caption{General configuration of lines $L_{1,0,0}$ and $L_{0,1,0}$ ending from both sides on the surface defect $\mathcal{S}_{0,0,1}$ associated to operator (\ref{generalendpoinds}).}
\label{picdefects}
\end{figure}

More generally,  the combination of vertex operators 
\begin{equation}
\prod_{a=1}^{n_1} e^{\epsilon_2  \phi(z_a)} \prod_{a=1}^{n_2} e^{\epsilon_1 \phi(z'_a)}\prod_{a=1}^{n'_1} e^{-\epsilon_2  \phi(w_a)} \prod_{a=1}^{n'_2} e^{-\epsilon_1 \phi(w'_a)} |0 \rangle 
\label{generalendpoinds}
\end{equation}
intertwines the right Calogero representation $\ell_{n'_1,n'_2,0} : A \to A_{n'_1,n'_2,0}$ with the combination 
\begin{equation}
\left(\ell_{n_1,n_2,0} \otimes s_{0,0,1}\right) \circ \Delta_{A,W_\infty} 
\end{equation}
of the oscillator representation and the left Calogero representation $A \to A_{n_1,n_2,0}$
and thus gives a gauge-invariant junction between $L_{n_1,n_2,0}$, $L_{n'_1,n'_2,0}$ and ${\cal S}_{0,0,1}$.

This is somewhat tedious to show directly, but follows from an associativity relation between the mixed coproduct 
$A \to A \otimes \widehat{u}(1)_{\kappa_3}$ and the meromorphic coproduct, which implies that the OPE of gauge-invariant 
junctions with ${\cal S}_{0,0,1}$ produces a gauge-invariant junction. We leave the details to an enthusiastic reader.

\subsection{The mixed coproduct} \label{sec:compose}
We will now determine the full mixed coproduct $\Delta_{A,W_\infty}$ by fusing elementary defects along the topological direction. 

As an appetizer, consider the composition of the $A \to A \otimes \widehat{\mathfrak{u}}(1)_{\kappa_3}$ coproduct with itself
to get a morphism $A \to A \otimes \widehat{\mathfrak{u}}(1)_{\kappa_3}\otimes \widehat{\mathfrak{u}}(1)_{\kappa_3}$:
\begin{align}
t_{0,n} &= t'_{0,n} + \left[J_{n}+ J'_{n}\right] \cr
t_{2,0} &=t'_{2,0} +2 \sigma_3 \sum_{n=1}^\infty n t'_{0,n-1} \left[J_{-n-1}+J'_{-n-1} \right]+  \frac{1}{3 \kappa_3^2}  \sum_{n=-\infty}^\infty \sum_{m=-\infty}^\infty :J_{m+n-2} J_{-m} J_{-n}:+\cr 
&- \sigma_3 \sum_{n=-\infty}^\infty n J_{n-1} J'_{-n-1}+  \frac{1}{3 \kappa_3^2}  \sum_{n=-\infty}^\infty \sum_{m=-\infty}^\infty :J'_{m+n-2} J'_{-m} J'_{-n}: \cr
&+  \frac{\sigma_3}{2}  \sum_{n=-\infty}^\infty |n| :\left[J_{-1-n}+J'_{-1-n}\right]\left[ J_{n-1}+J'_{n-1}\right]:
\end{align}
We have grouped the generators of $\widehat{\mathfrak{u}}(1)_{\kappa_3}\otimes \widehat{\mathfrak{u}}(1)_{\kappa_3}$
in combinations which can be recognized as generators of a standard free-field realization of 
$W_{0,0,2}$ in terms of two copies of $W_{0,0,1} \otimes W_{0,0,1}$. We can thus read off a mixed coproduct 
$A \to A \otimes W_{0,0,2}$ which we identify with the composition $\left[1_A \otimes s_{0,0,2}\right] \circ \Delta_{A,W_\infty}$ 

The same magic happens even if we compose coproducts associated to different types of surface defects. For example, we can get a  morphism $A \to A \otimes \widehat{\mathfrak{u}}(1)_{\kappa_3}\otimes \widehat{\mathfrak{u}}(1)_{\kappa_2}$:
\begin{align}
t_{0,n} &= t'_{0,n} + \left[J_{n}+ J'_{n}\right] \cr
t_{2,0} &=t'_{2,0} +2 \sigma_3 \sum_{n=1}^\infty n t'_{0,n-1} \left[J_{-n-1}+J'_{-n-1} \right]+  \frac{1}{3 \kappa_2^2}  \sum_{n=-\infty}^\infty \sum_{m=-\infty}^\infty :J_{m+n-2} J_{-m} J_{-n}:+\cr 
&- \sigma_3 \sum_{n=-\infty}^\infty n J_{n-1} J'_{-n-1}+  \frac{1}{3 \kappa_3^2}  \sum_{n=-\infty}^\infty \sum_{m=-\infty}^\infty :J'_{m+n-2} J'_{-m} J'_{-n}: \cr
&+  \frac{\sigma_3}{2}  \sum_{n=-\infty}^\infty |n| :\left[J_{-1-n}+J'_{-1-n}\right]\left[ J_{n-1}+J'_{n-1}\right]:
\end{align}
This has the correct form to factor through the coproduct $A \to A \otimes W_{0,1,1}$ and the known free field realizations $W_{0,1,1} \to \widehat{\mathfrak{u}}(1)_{\kappa_3}\otimes \widehat{\mathfrak{u}}(1)_{\kappa_2}$, say from \cite{Prochazka:2018tlo}.

We can easily reconstruct along these lines the full form of the $\Delta_{A,W_\infty}: A \to A \otimes W_\infty$ coproduct: 
\begin{align}
t_{0,n} &= t'_{0,n} + W_{1,n} \cr
t_{2,0} &=t'_{2,0} +2 \sigma_3 \sum_{n=1}^\infty n t'_{0,n-1} W_{1,-n-1}+ V_{-2}
+ \frac{\sigma_3}{2}  \sum_{n=-\infty}^\infty |n| :W_{1,-1-n}W_{1,n-1}:
\end{align}
where $V_{-2}$ is a mode of a quasi-primary field $V$ and $W_{1}$ is the spin-one current. An explicit formula for the field $V(z)$ in terms of the primary generators $W_{n}$ of  $\mathcal{W}_{\infty}$ can be easily found analogously to the above two examples.  Using the coproduct $A\rightarrow A\otimes \widehat{\mathfrak{u}}(1)_{\kappa_3}$ successively $N$ times, one finds an explicit expression
\begin{align}
V =& \frac{1}{3\kappa_3^2}\sum_{i=1}^{N}\sum_{k=-\infty}^{\infty} \sum_{l=-\infty}^{\infty}:J^{(i)}_{-k-l-2}J^{(i)}_{k}J^{(i)}_{l}:\cr
&-\sigma_3\sum_{i=1}^{N}\sum_{j=i+1}^{N}\sum_{k=-\infty}^{\infty}k:J^{(i)}_{-k-1}J^{(j)}_{k-1}:
\end{align}
where $J^{(i)}$ is the $\widehat{\mathfrak{u}}(1)_{\kappa_3}$ generator from the $i$'th use of the coproduct. This expression can be easily rewritten in terms of fields $W_1,W_2,W_3$ from the standard free-field realization of $\mathcal{W}_{N}$, leading to the formula 
\begin{equation}
V=W_3+\frac{2}{\psi_0}:W_1W_2:-\frac{2}{3}\frac{1}{\psi_0^2}:W_1W_1W_1: 
\end{equation}
where we identified $\sigma_3\psi_0=-\epsilon_3N$. Working uniformly in $N$, this gives an explicit identification of $V$ in terms of standard generators of $\mathcal{W}_{\infty}$. See appendix \ref{utop} for further details.

With a bit of patience, one can also directly verify that this is an algebra morphism. This is the main result of this section.

\subsection{An important involution} \label{sec:inv}
Consider the gauge-invariance condition $\Delta_{A,W_\infty}[t_{ab}] O =O t_{ab}$, i.e. 
\begin{align}
O t_{0,n} &= (t'_{0,n} + J_{n})O \cr
O t_{2,0} &=\left[t'_{2,0} -2 \sigma_3 \sum_{n=1}^\infty n t'_{0,n-1} J_{-n-1}- V_{-2}
-  \frac{\sigma_3}{2}  \sum_{n=-\infty}^\infty |n| :J_{-1-n}J_{n-1}:\right] O \cr
\cdots &= \cdots
\end{align}
The $\cW_\infty$ algebra has an involution, which flips the sign of the currents of odd dimension. It flips the signs of both $V_{-2}$ and $J_n$.
There is also an $A \to A^{\mathrm{op}}$ algebra morphism, which sends $t_{a,b} \to (-1)^a t_{ab}$. 

If we twist the actions of $A$ and $W_\infty$ by these morphisms, we get constraints of the form
\begin{align}
t_{0,n} \hat O  &= \hat O t'_{0,n} - J_{n} \hat O \cr
t_{2,0} \hat O &= \hat O t'_{2,0} +2 \sigma_3 \sum_{n=1}^\infty n J_{-n-1}\hat O  t'_{0,n-1}+V_{-2} \hat O
-  \frac{\sigma_3}{2}  \sum_{n=-\infty}^\infty |n| :J_{-1-n}J_{n-1} \hat O \cr
\cdots &= \cdots
\end{align}
where we denote the image of $O$ in the twisted module  as $\hat O$ and identify an $A^{\mathrm{op}}$ left action by the right $A$ action and vice versa.

These constraints are actually equivalent to the constraints 
\begin{align}
t_{0,n} \hat O + J_{n} \hat O &= \hat O t'_{0,n}  \cr
t_{2,0} \hat O - V_{-2} \hat O +  \frac{\sigma_3}{2}  \sum_{n=-\infty}^\infty |n| :J_{-1-n}J_{n-1} \hat O- 2 \sigma_3 \sum_{n=1}^\infty n J_{-n-1}(t_{0,n-1}+ J_{n-1})\hat O  &= \hat O t'_{2,0}   \cr
\cdots &= \cdots
\end{align}
which are just the usual gauge-invariance constraints, but with the roles of left and right line defects exchanged. 

\section{The Miura operator} \label{sec:miura}
In the previous Section \ref{sec:coproducts}, we have derived an expected expression for the mixed coproduct
\begin{equation}
\Delta_{A,W_\infty}: A \to A \otimes W_\infty
\end{equation}
encoding the gauge invariance condition 
\begin{equation}
\left(\left(\ell_{L_1} \otimes s_{\cal S} \right) \circ \Delta_{A,W_\infty} \right)[t_{ab}] \cdot O = O \cdot \ell_{L_2}[t_{ab}]
\end{equation}
for local operators on a surface defect ${\cal S}$ attached to line defects $L_1$ and $L_2$. Here $\ell_{L_i}$ are the algebra maps 
$A\to A_{L_i}$ encoding the action of gauge symmetries in the worldline theories and $s_{\cal S}$ is the 
VOA map $\cW_\infty \to \cW_{\cal S}$ encoding the action of gauge symmetries in the worldsheet theory. The gauge-invariant local operator $O$ 
is an element of some space of local operators which carry a left action of $\ell_{L_1}$, a right action of $\ell_{L_2}$ and a VOA action of $s_{\cal S}$.

We have also provided gauge-invariant endpoints for line defects $L_{n_1,n_2,0}$ and $L_{n'_1,n'_2,0}$
onto the ${\cal S}_{0,0,1}$ surface defect, in the form of a collection of vertex operators for the associated $\widehat{\mathfrak{u}}(1)_{\kappa_3}$ vertex algebra. 

In this section, we will add one further ingredient which will allow for the construction of the most generic junctions involving the ${\cal S}_{0,0,1}$ surface defect: a canonical gauge-invariant intersection between $L_{0,0,1}$ and ${\cal S}_{0,0,1}$. Notice that in M-theory, the corresponding M2 branes simply cannot end on the M5 brane and have to continue on the other side. Hence we do not expect to have ${\it endpoints}$ of $L_{0,0,1}$ on ${\cal S}_{0,0,1}$.

Afterwards, we will rediscover the relation between the Miura operator and the free-field construction of W-algebras, as well as the Coulomb gas construction of conformal blocks.

\subsection{Elementary Miura operator}

Let us introduce the elementary Miura operator
\begin{equation}
\cL^{0,0,1}_{0,0,1} \equiv \epsilon_3 \partial_z -\epsilon_1\epsilon_2 \sum_{n=-\infty}^{\infty}\frac{J_n^{(3)}}{z^{n+1}}=\epsilon_3 \partial_z-\epsilon_1\epsilon_2 J^{(3)}(z)
\label{simplemiura}
\end{equation} 
with the $\widehat{\mathfrak{u}}(1)_{\kappa_3}$ generators normalized as above
\begin{equation}
[J^{(3)}_k, J^{(3)}_l] = - \frac{1}{\epsilon_1 \epsilon_2}k\delta_{k,-l}
\end{equation}
The superscripts in $\cL^{0,0,1}_{0,0,1}$ denotes the support of the M2 brane associated to the line operator to be $\mathbb{C}_{\epsilon_3} \subset \mathbb{C}_{\epsilon_1}\times\mathbb{C}_{\epsilon_2}\times\mathbb{C}_{\epsilon_3}$ whereas the subscripts labels an orientation of the M5 brane associated to the surface defect to be $\mathbb{C}_{\epsilon_1}\times\mathbb{C}_{\epsilon_2}$.

The Miura operator is naturally an element of the tensor product of the vacuum module for $\widehat{\mathfrak{u}}(1)_{\kappa_3}$ and the Weyl algebra $A_{0,0,1}$. This space carries a bi-module action for $A_{0,0,1}$
given by the left- and right- multiplication. 

We can thus ask if $\cL^{0,0,1}_{0,0,1}$ can represent a gauge-invariant 
intersection of $L_{0,0,1}$ and ${\cal S}_{0,0,1}$, i.e. if it intertwines the right action of $A$ associated to the map $\ell_{0,0,1}: A \to A_{0,0,1}$ 
and the left action of $A$ arising from the composition 
\begin{equation}
(\ell_{0,0,1} \otimes s_{0,0,1}) \circ \Delta_{A,W_\infty}: A \to A_{0,0,1} \otimes \widehat{\mathfrak{u}}(1)_{\kappa_3}
\end{equation}

It is straightforward to check that this is indeed the case. We will use the state-operator map to 
describe the module, i.e. describe the intersection operator as $\cL^{0,0,1}_{0,0,1} |0\rangle$, where $|0\rangle$ is the vacuum state, annihilated by all non-negative modes of the current. 

We can first verify that
\begin{align}
\cL^{0,0,1}_{0,0,1}|0\rangle \frac{z^d}{\epsilon_3} & =\left  (\epsilon_3\partial -\epsilon_1\epsilon_2\sum_{m=1}^{\infty}z^{m-1}J^{(3)}_{-m}\right )|0\rangle \frac{z^d}{\epsilon_3}\cr
&=\frac{z^d}{\epsilon_2}\cL^{0,0,1}_{0,0,1}|0\rangle +dz^{d-1}|0\rangle =\left ( \frac{z^d}{\epsilon_3}  +J^{(3)}_d \right ) \cL^{0,0,1}_{0,0,1}|0\rangle 
\end{align}
and then observe that the right action of  $t_{2,0}= \epsilon_3\partial^2$ gives
\begin{align}
\cL^{0,0,1}_{0,0,1}|0\rangle \epsilon_3\partial^2= \epsilon_3\partial^2\cL^{0,0,1}_{0,0,1}|0\rangle +2\sigma_3\partial J^{(3)}(z)|0\rangle \partial+\sigma_3\partial^2 J^{(3)}(z)|0\rangle
\end{align}
 the left action of $t_{2,0}$ under the map $A\to  \widehat{\mathfrak{u}}(1)_{\kappa_3}$ produces
\begin{align}
t_{2,0}\cL^{0,0,1}_{0,0,1}|0\rangle&=\left (\epsilon_1^2\epsilon_2^2 \sum_{n=1}^{\infty}\sum_{m=1}^{\infty}J^{(3)}_{-n}J^{(3)}_{-m}J^{(3)}_{m+n-2}+\sigma_3\sum_{n=1}^{\infty}(n-1)J^{(3)}_{-n}J^{(3)}_{n-2}\right )\cL^{0,0,1}_{0,0,1}|0\rangle\cr
&= \epsilon_1^2\epsilon_2^2 \sum_{n,m=1}^{\infty}(m+n-2)J^{(3)}_{-n}J^{(3)}_{-m}z^{m+n-3}|0\rangle+\sigma_3\partial^2 J^{(3)}(z)|0\rangle
\end{align}
and finally the mixed term is
\begin{align}
2\sigma_3\sum_{m=1}^{\infty}mJ^{(3)}_{-m-1} \frac{1}{\epsilon_3}z^{m-1}\cL^{0,0,1}_{0,0,1}|0\rangle=& -\epsilon_1^2\epsilon_2^2\sum_{n=1}^{\infty}\sum_{m=1}^{\infty}J^{(3)}_{-m}J^{(3)}_{-n}(m+n-2)z^{m+n-3}|0\rangle\cr
&+ 2\sigma_3  \partial J(z)|0\rangle \partial
\end{align}
Combining the above formulas leads to
\begin{align}
\cL^{0,0,1}_{0,0,1} |0\rangle\frac{1}{\epsilon_3}z^n&=\left (\frac{1}{\epsilon_3}z^n+s_{0,0,1}[t_{0,n}]\right) \cL^{0,0,1}_{0,0,1} |0\rangle \cr
\cL^{0,0,1}_{0,0,1} |0\rangle \epsilon_3\partial^2 &= \left ( \epsilon_3\partial^2+s_{0,0,1}[t_{2,0}] + 2\sigma_3\sum_{n=1}^{\infty}n s_{0,0,1}[t_{0,-n-1}]\frac{1}{\epsilon_3}z^{n-1} \right )\cL^{0,0,1}_{0,0,1} |0\rangle
\end{align}
consistent with formulae (\ref{coproduct1}).

\subsection{Some OPE checks and predictions}
As a further check, we can take the OPE between two gauge-invariant junctions
\begin{align}
\cL^{0,0,1}_{0,0,2} &\equiv \left(\epsilon_3 \partial_{z_1}-\epsilon_1\epsilon_2 J^{(3)}(z_1) \right)\left(\epsilon_3 \partial_{z_2}-\epsilon_1\epsilon_2 J^{(3)}(z_2) \right) =\epsilon^2_3 \partial_{z_1} \partial_{z_2}- \frac{\epsilon_1 \epsilon_2}{(z_1-z_2)^2}+\cr &-\left(\sigma_3 J^{(3)}(z_1)\partial_{z_2}+\sigma_3 J^{(3)}(z_2)\partial_{z_1} \right)
+ \epsilon_1^2 \epsilon^2_2 :J^{(3)}(z_1)J^{(3)}(z_2):
\end{align} 
This is an element in the tensor product of $A_{0,0,2}$ and the vacuum module of $\widehat{\mathfrak{u}}(1)_{\kappa_3}$!
The singular term in the OPE of the currents contributes crucially to produce the coefficient of the identity 
\begin{equation}
\epsilon^2_3 \partial_{z_1} \partial_{z_2}- \frac{\epsilon_1 \epsilon_2}{(z_1-z_2)^2} =\frac{1}{2} \left(\epsilon^2_3 t_{1,0}^2 - \epsilon_3t_{2,0}\right) 
\end{equation}
This is the image under Weyl reflection of 
\begin{equation}
\frac{1}{2} \left(\epsilon^2_3 t_{0,1}^2 -\epsilon_3 t_{0,2}\right) = z_1 z_2
\end{equation}

More generally, we expect the OPE of $n$ Miura operators to give a gauge-invariant junction $\cL^{0,0,1}_{0,0,n}$ in the tensor product 
of $A_{0,0,n}$ and the vacuum module of $\widehat{\mathfrak{u}}(1)_{\kappa_3}$, starting with the Weyl image of $\prod_i z_i$ \cite{Gaiotto:2019wcc}. 

We should also be able to take the OPE between the Miura junction and the endpoints discussed in previous sections. For example, the OPE
\begin{equation}
\left(\epsilon_3 \partial_{z_1}-\epsilon_1\epsilon_2 J^{(3)}(z_1) \right)e^{\epsilon_2  \phi^{(3)}(z_2)} = \left(\epsilon_3 \partial_{z_1}+\frac{\epsilon_2}{z_1 - z_2} \right) e^{\epsilon_2  \phi^{(3)}(z_2)} - \epsilon_1\epsilon_2 :J^{(3)}(z_1)e^{\epsilon_2  \phi^{(3)}(z_2)}:
\end{equation} 
should be a gauge-invariant junction between $L_{1,0,1}$, $L_{0,0,1}$ and ${\cal S}_{0,0,1}$, involving a
$A_{1,0,1}$-$A_{0,0,1}$-bimodule described by certain differential operators in $z_1$, which depend meromorphically on $z_2$. 
We leave the check to an enthusiastic reader.

\subsection{Miura for $\mathcal{S}_{0,0,N}$}

The main role of the Miura operator in the literature is to produce a free field realization of the $\mathcal{W}_N$ VOA via Miura transformation
\begin{align}
\cL^{0,0,N}_{0,0,1} &\equiv \left[\epsilon_3 \partial -\epsilon_1\epsilon_2 J^{(3)}_1(z) \right] \left[\epsilon_3 \partial -\epsilon_1\epsilon_2 J^{(3)}_2(z) \right]\cdots \left[\epsilon_1 \partial - \epsilon_1\epsilon_2 J^{(3)}_N(z) \right]\cr
 &= (\epsilon_1 \partial )^N + U^{(3)}_1(z) (\epsilon_1 \partial)^{N-1}+ U^{(3)}_2(z) (\epsilon_1 \partial)^{N-2} + \cdots + U^{(3)}_N(z) 
\end{align}
The $\mathcal{W}_N$ VOA generated by fields $U^{(3)}_k$ depends polynomially on the parameters $\epsilon_i$ and integer $N$. For the convenience of the reader, we list OPEs of first few $U^{(3)}_k$ in our normalization in appendix \ref{uope}. We will call the differential operator on the right hand side the Miura operator associated to the intersection of $L_{0,0,1}$ with $\mathcal{S}_{0,0,N}$. By definition, composition of the Miura differential operators gives a coassociative coproduct 
\begin{equation}
 \mathcal{W}_{N+N'} \to \mathcal{W}_N \otimes \mathcal{W}_{N'}
\end{equation}
Essentially by construction, $\cL^{0,0,N}_{0,0,1} |0\rangle$ gives a gauge-invariant intersection point of $L_{0,0,1}$ with $\mathcal{S}_{0,0,N}$. 

\subsection{Pseudo-differential Miura operators}
The $\mathcal{W}_\infty$ VOA is an uniform-in-$N$ limit of $\mathcal{W}_N$. Concretely, it is a VOA with an infinite tower of generators 
$U^{(3)}_i(z)$ and an extra central element $\nu= - \sigma_3 \psi_0$ (besides the $\epsilon_i$), equipped with canonical truncation maps $\mathcal{W}_\infty \to \mathcal{W}_N$ which send $\nu$ to $N \epsilon_3$. Note that there exist three different $U$-bases depending on the choice of $\cL^{0,0,1}_{0,0,1}$, $\cL^{0,1,0}_{0,1,0}$ or $\cL^{1,0,0}_{1,0,0}$ as an elementary building block.

The Miura operator can be generalized to a pseudo-differential operator
\begin{equation}
\cL^{0,0,1} \equiv (\epsilon_3 \partial)^{\frac{\nu}{\epsilon_3}}+ U^{(3)}_1(z) (\epsilon_3 \partial)^{\frac{\nu}{\epsilon_3}-1}+ U^{(3)}_2(z) (\epsilon_3 \partial)^{\frac{\nu}{\epsilon_3}-2} + \cdots
\end{equation}
Such pseudo-differential operators can be composed
\begin{align}
&\left[(\epsilon_3 \partial)^{\frac{\nu}{\epsilon_3}}+ U_1(z) (\epsilon_3 \partial)^{\frac{\nu}{\epsilon_3}-1} + \cdots \right]
\left[(\epsilon_3 \partial)^{\frac{\nu'}{\epsilon_3}}+ U'_1(z) (\epsilon_3 \partial)^{\frac{\nu'}{\epsilon_3}-1} + \cdots \right]
\cr &= \left[(\epsilon_3 \partial)^{\frac{\nu+\nu'}{\epsilon_3}}+ U''_1(z) (\epsilon_3 \partial)^{\frac{\nu+\nu'}{\epsilon_3}-1}+ U''_2(z) (\epsilon_3 \partial)^{\frac{\nu+\nu'}{\epsilon_3}-2} + \cdots \right]
\end{align}
to give a coassociative coproduct 
\begin{equation}
\Delta_{\cW_\infty, \cW_\infty}: \mathcal{W}_\infty \to \mathcal{W}_\infty \otimes \mathcal{W}_\infty
\end{equation}
For clarity, we omitted the $(3)$ superscripts on the generators. It is far from obvious, but true, that algebra $\cW_\infty$ together with this coproduct are invariant under permutation of the $\epsilon_i$. See the manifestly triality-invariant primary basis from appendix \ref{utop}.

The space of pseudo-differential operators is a bimodule for $A_{0,0,1}$, with the obvious action of the 
Weyl algebra $(z, \epsilon_3 \partial)$ from the left or from the right of the pseudo-differential operator. It thus makes sense to ask if $\cL^{0,0,1}|0 \rangle$ 
could represent a gauge-invariant intersection of $L_{0,0,1}$ with a generic surface defect, i.e. if 
it could intertwine the action of $t_{ab}$ from the right in  $A_{0,0,1}$ and the action of $t_{ab}$ from the left in  $A_{0,0,1} \otimes W_\infty$. This is very likely automatically true, as the intertwining relations for $\cL^{0,0,1}_{0,0,N}|0 \rangle$ for generic $N$ can likely be cast in a form polynomial in $N$ and continued to non-integral $N$. 

An important test would be to look at the  $\cL^{0,0,1}_{0,1,0}$ and $\cL^{0,0,1}_{1,0,0}$ specializations, which involve again $\widehat{\mathfrak{u}}(1)$ current algebras and are explicitly known \cite{Prochazka:2018tlo,Prochazka:2019dvu}:
\begin{align}
\cL^{0,0,1}_{0,0,1}&\equiv :\exp [\frac{\epsilon_1\epsilon_2}{\epsilon_3}\phi^{(3)}(z)](\epsilon_3\partial)\exp [-\frac{\epsilon_1\epsilon_2}{\epsilon_3}\phi^{(3)}(z)]:\cr
\cL^{0,0,1}_{0,1,0}&\equiv :\exp [\epsilon_1\phi^{(2)}(z)](\epsilon_3\partial)^{\frac{\epsilon_2}{\epsilon_3}}\exp [-\epsilon_1\phi^{(2)}(z)]:\cr
\cL^{0,0,1}_{1,0,0}&\equiv :\exp [\epsilon_2\phi^{(1)}(z)](\epsilon_3\partial)^{\frac{\epsilon_1}{\epsilon_3}}\exp [-\epsilon_2\phi^{(1)}(z)]:
\label{miurasofanyorientation}
\end{align}
with the modes of $J^{(i)}=\partial \phi^{(i)}$ normalized as
\begin{align}
[J^{(i)}_k, J^{(i)}_l] = - \frac{\epsilon_i}{\sigma_3}k\delta_{k,-l}
\end{align}
The expression for $\cL^{0,0,1}_{0,0,1}$ can be  obviously simplified to (\ref{simplemiura}). On the other hand, the other two produce more complicated pseudo-differential operators with coefficients being various combinations of the field $J^{(i)}$ and its derivatives.

\begin{figure}[h]
    \centering
        \includegraphics[width=11cm]{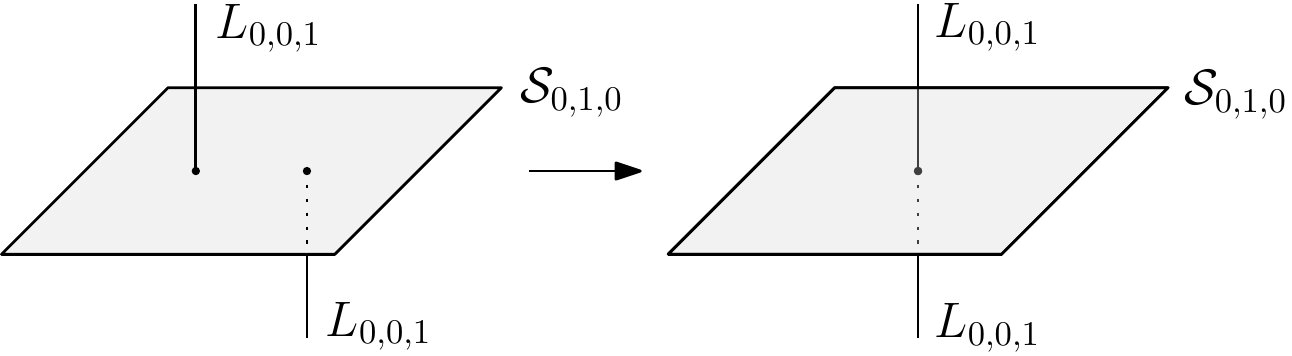}
   \caption{The Miura operator $\mathcal{L}_{0,1,0}^{0,0,1}$ associated to line $L_{0,0,1}$ intersecting surface $\mathcal{S}_{0,1,0}$ is expected to come from the fusion of two lines $L_{0,0,1}$ ending from opposite side.}
\label{picdefects}
\end{figure}

We can re-derive these two pseudo-differential Miura operators in a surprising way, which automatically verifies their gauge-invariance. 
The trick is to use an alternative way to represent the Weyl bimodule of pseudo-differential operators: use a Fourier transform 
to map them to integral kernels, i.e. functions of $z$, $z'$ with the action of the ``$z$'' generator from the left and from the right mapped to multiplications by $z$ and by $z'$. Then the element $\partial^\alpha$ is mapped to an integral kernel proportional to $(z-z')^{-\alpha-1}$ and we can write
\begin{align}
\cL^{0,0,1}_{0,1,0}&\sim :\exp [\epsilon_1\phi^{(2)}(z)](z-z')^{\frac{\epsilon_1}{\epsilon_3}}\exp [-\epsilon_1\phi^{(2)}(z')]:=\cr
&=e^{\epsilon_1\phi^{(2)}(z)} e^{-\epsilon_1\phi^{(2)}(z')} 
\end{align}
This is just the OPE fusion of the gauge-invariant endpoints of $L_{0,0,1}$ on ${\cal S}_{0,1,0}$ from either sides. 
It is thus obviously gauge invariant. 

\subsection{Topological composition and screening charges}
The composition of elementary $\cL^{0,0,1}_{0,0,1}$ Miura operators along the topological direction to 
produce $\cL^{0,0,1}_{0,0,N}$ is only the simplest example of a general process of composition of gauge-invariant 
junctions. 

In general, given an $O^{L_1,L_2}_{\cal S}$ gauge-invariant junction between some $L_1$, $L_2$ and ${\cal S}$ and an 
$O^{L_2,L_3}_{\cal S'}$ gauge-invariant junction between some $L_2$, $L_3$ and ${\cal S'}$ defects, 
we can obtain a gauge-invariant junction
\begin{equation}
O^{L_1,L_2}_{\cal S} \otimes_{A_{L_2}} O^{L_2,L_3}_{\cal S'}
\end{equation} 
between $L_1$, $L_3$ and the surface defect obtained from the fusion of ${\cal S}$ and ${\cal S'}$ along the topological $\mathbb{R}$ direction,
such that the gauge action in the surface defect is the composition $\left(s_{\cal S} \otimes s_{\cal S'}\right) \circ \Delta_{\cW_\infty, \cW_\infty}$.

If our objective is to build such junctions, we will be given $L_1$ and $L_3$ but we have a choice of $L_2$. We will now demonstrate by examples that the different choices of $L_2$ are analogous to the different choices of screening charges one may employ in a free-field realization of a VOA.

Notice that the VOA $\cW_{\cal SS'}$ associated to the composition of surface defects is embedded inside $\cW_{\cal S} \otimes \cW_{\cal S'}$ as a sub-algebra generated by the image of the coproduct map of $\cW_\infty$ generators. Consider now the local operator obtained by stretching an $L_{1,0,0}$ segment between ${\cal S}$ and ${\cal S'}$. We have some right endpoint $V(z)$ and some left endpoint $V'(z)$,
with the property that the Weyl action of $A$ on the appropriate side is intertwined with the images of $A$ in $W_{\cal S}$ and  $W_{\cal S'}$.

The tensor product between the highest weight left- and right- modules for the Weyl algebra vanishes, so one may naively think that $V'(z)\otimes_{A_{1,0,0}}V(z)$ will necessarily vanish as well. But if we think about these endpoints as part of a more complicated collection of vertex operators, and about the $L_{1,0,0}$ segment as a part of a more complicated collection of line defects, the relevant modules actually have interesting tensor products. 

Indeed, consider the formal contour integral 
\begin{equation}
(VV') \equiv \oint V(z) V'(z) dz
\end{equation}
which clearly has the desired tensor product property that 
\begin{align}
\oint V(z) \left(z V'(z) \right) dz &= \oint \left( z V(z) \right) V'(z) dz \cr
\oint V(z) \left(\partial_z V'(z) \right) dz &= \oint \left( - \partial_z V(z) \right) V'(z) dz 
\end{align}
In the absence of other vertex operators, the integration contour could be shrunk to a point, leading to a vanishing tensor product. 
In the presence of other vertex operators, though, one will have some collection of non-trivial integration contours, 
which can potentially pick up the non-trivial possible tensor products. 

If we combine the gauge-invariance properties of $V(z)$ and $V'(z)$, we learn a simple, striking fact: $(VV')$ is annihilated by the 
image of $A$ in the coproduct $\cW_{\cal SS'}$, which implies it has non-singular OPE with the $\cW_{\cal SS'}$ generators. 
This is precisely the expected property of a screening charge!
We thus get up to three distinct screening charges, associated to $L_{1,0,0}$, $L_{0,1,0}$ or $L_{0,0,1}$ segments.

We will now illustrate the appearance of screening charges in three examples involving various configurations of a single line $L_{0,0,1}$ and two elementary surface operators $\mathcal{S}$ and $\mathcal{S}'$.

\subsubsection{Example: Degenerate modules for $\mathcal{S}_{2,0,0}$}

\begin{figure}
    \centering
        \includegraphics[width=14.5cm]{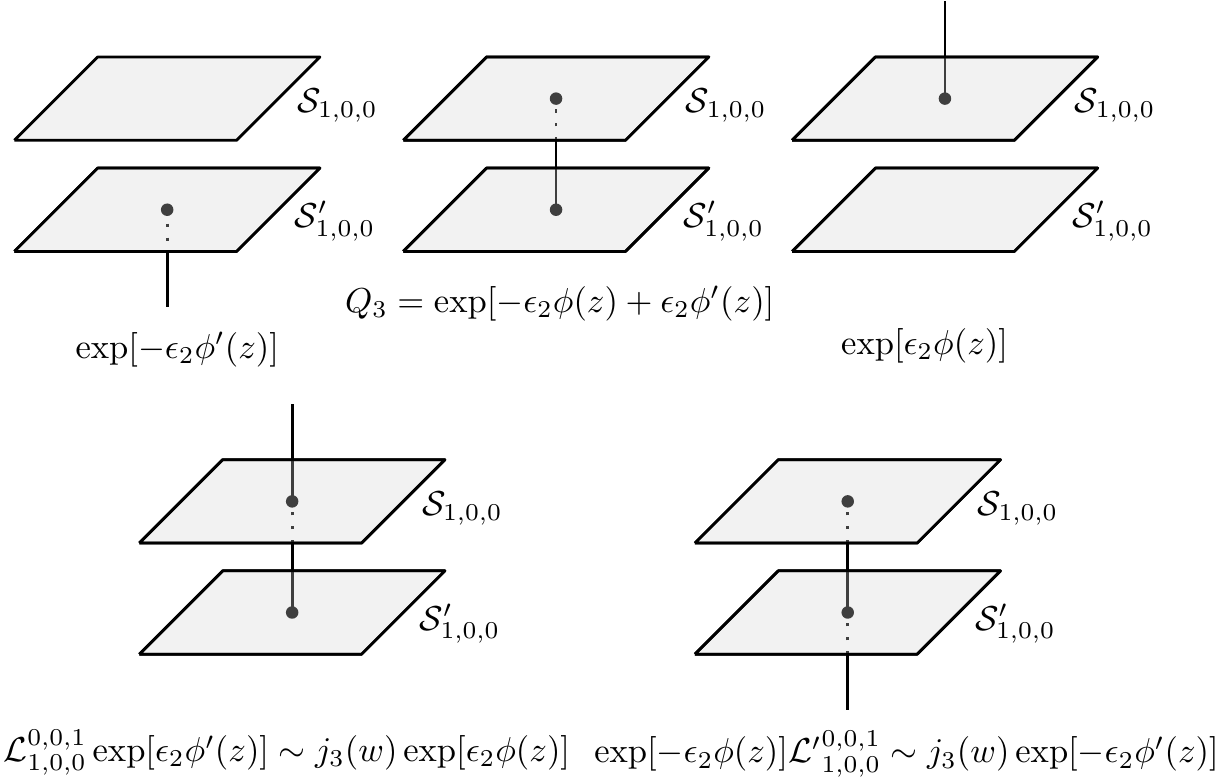}
 \caption{Allowed configurations of a single line $L_{0,0,1}$ and a pair of surfaces $\mathcal{S}_{1,0,0}$ and $\mathcal{S}'_{1,0,0}$ with an explicit realization of the operator describing the corresponding intersection.}
\label{pic22}
\end{figure}

Let us start with a pair of surface operators of the same orientation $\mathcal{S}_{1,0,0},\tilde{\mathcal{S}}_{1,0,0}$. Five possible ways how can a single line $L_{0,0,1}$ interact with $\mathcal{S}_{1,0,0},\tilde{\mathcal{S}}_{1,0,0}$ are depicted in figure \ref{pic22}. 

If the line $L_{0,0,1}$ intersect both surfaces,  each of the two intersections are described in therms of the Miura operator
\begin{align}
\mathcal{L}_{1,0,0}^{0,0,1}&=&\left (1-\epsilon_1\epsilon_2J^{(1)}(\epsilon_3\partial)^{-1}+\frac{\epsilon_1\epsilon_2(\epsilon_1-\epsilon_3)}{2}(\epsilon_2 (J^{(1)})^2-\partial J^{(1)})(\epsilon_3\partial)^{-2}+\dots\right )(\epsilon_3\partial)^{\frac{\epsilon_1}{\epsilon_3}}
\end{align}
and analogously for $\mathcal{L}'{}_{1,0,0}^{0,0,1}$. Their composition produces the Miura operator associated to the junction of $\mathcal{S}_{2,0,0}$ with $L_{0,0,1}$:
\begin{eqnarray}\nonumber
\mathcal{L}_{1,0,0}^{0,0,1}\mathcal{L}'{}_{1,0,0}^{0,0,1}&=&\Big [1-\epsilon_1\epsilon_2(J+J')(\epsilon_3\partial)^{-1}+\frac{\epsilon_1\epsilon_2}{2}\Big (\epsilon_2(\epsilon_1-\epsilon_3) (J^2+J'{}^2)+ 2\epsilon_1\epsilon_2 JJ'\\
&&-(\epsilon_1-\epsilon_3)\partial J-(3\epsilon_1-\epsilon_3)\partial J')\Big )(\epsilon_3\partial)^{-2}+\dots\Big ](\epsilon_3\partial)^{\frac{2\epsilon_1}{\epsilon_3}}
\end{eqnarray}
where we omitted indices $(1)$ for simplicity. The fields $U_i$ appearing in front of the derivatives $(\epsilon_3\partial)^{\frac{2\epsilon_1}{\epsilon_3}-i}$ are generators of algebra $Y_{2,0,0}$ in the $U$-basis associated to the composition of $\mathcal{L}_{0,0,1}^{0,0,1}$ Miura operators. It is easy to check that fields $U_i$ lie in the kernel of the zero modes of the screening currents
\begin{eqnarray}\nonumber
j_3&=& :\exp [-\epsilon_2\phi (z)+\epsilon_2\phi' (z)]:\\
j_2&=& :\exp [-\epsilon_3\phi (z)+\epsilon_3\phi' (z)]:
\end{eqnarray}
with $ \exp [-\epsilon_2\phi (z)]$ and $ \exp [-\epsilon_3\phi (z)]$ describing the right endpoint of $L_{0,0,1}$ and $L_{0,1,0}$ on the surface $\mathcal{S}_{1,0,0}$ together with   $\exp [\epsilon_2\phi' (z)]$ and $\exp [\epsilon_2\phi' (z)]$ describing the left endpoint of  $L_{0,0,1}$ and $L_{0,1,0}$ on the surface $\mathcal{S}'_{1,0,0}$ as expected from the general discussion above. The subscript of $j_i$ indicates the orientation of the line stretched between $\mathcal{S}_{1,0,0}$ and $\mathcal{S}'_{1,0,0}$.

The vertex operator $\exp [\epsilon_2 \phi (z)]$ associated to the right endpoint of $L_{0,0,1}$ on $\mathcal{S}_{1,0,0}$ and the vertex operator $\exp [-\epsilon_2\phi'(z)]$ associated to the left endpoint of $L_{0,0,1}$ on $\mathcal{S}'_{1,0,0}$ both describe degenerate modules on the fused defect $\mathcal{S}_{2,0,0}$. Analogously, we get degenerate modules for $\mathcal{S}_{2,0,0}$ describing endpoints of the line $L_{0,1,0}$ by switching  $\epsilon_2\leftrightarrow \epsilon_3$.

Let us finish the discussion by describing operators associated to the degenerate modules but with an addition of a finite-line segment stretched between the two surfaces. This configuration can be built in two different ways. Fusing along the topological direction, we can multiply the Miura operator associated to the intersection of a line with a surface defect with a degenerate module describing the endpoint of the line. Fusing along the holomorphic direction, we can insert the degenerate field describing the endpoint of a line on the surface defect with the screening current associated to the finite-line segment. Equality of results of these two fusions is a highly non-trivial test of our proposal. Indeed, using the Fourier transform to rewrite the derivative in the definition of $\mathcal{L}_{1,0,0}^{0,0,1}$ in terms of an integral kernel, we can write
\begin{eqnarray}\nonumber
\mathcal{L}_{1,0,0}^{0,0,1} :\exp[\epsilon_2\phi' (z)]:&\sim& \frac{1}{(z-w)^{\frac{\epsilon_1}{\epsilon_3}+1}} :\exp[\epsilon_2\phi(z)-\epsilon_2\phi (w)+\epsilon_2\phi' (w)]:\\
&\sim& j_3(w) :\exp[\epsilon_2\phi (z)]:
\end{eqnarray}
giving us the screened vertex operator $ \exp[\epsilon_2\phi (z)]$. Analogously, we have for the right action of $\tilde{\mathcal{L}}_{0,0,1}^{1,0,0}$:
\begin{eqnarray}
:\exp[-\epsilon_2\phi (z)]:\mathcal{L}'{}_{0,0,1}^{1,0,0} \sim j_3(w) :\exp[-\epsilon_2\phi' (z)]:
\end{eqnarray}

\subsubsection{Example: Degenerate modules for $\mathcal{S}_{1,1,0}$ along third direction}

Let us perform the same analysis involving the line $L_{0,0,1}$ and two M5 branes of different orientations $\mathcal{S}'_{1,0,0}$ and $\mathcal{S}_{0,1,0}$. The left intersection is again described by the Miura operator $\mathcal{L}'{}_{1,0,0}^{0,0,1}$ as above whereas the other one is now given by
 \begin{align}
\mathcal{L}_{0,1,0}^{0,0,1}=\left (1-\epsilon_1\epsilon_2J^{(2)}(\epsilon_3\partial)^{-1}+\frac{\epsilon_1\epsilon_2(\epsilon_2-\epsilon_3)}{2}(\epsilon_1 (J^{(2)})^2-\partial J^{(2)})(\epsilon_3\partial)^{-2}+\dots\right )(\epsilon_3\partial)^{\frac{\epsilon_2}{\epsilon_3}}
\end{align}
Composing the two Miura operators produces 
\begin{eqnarray}\nonumber
\mathcal{L}_{0,1,0}^{0,0,1}\mathcal{L}'{}_{1,0,0}^{0,0,1}&=&\Big [1-\epsilon_1\epsilon_2(J+J')(\epsilon_3\partial)^{-1}+\frac{\epsilon_1\epsilon_2}{2}\Big (\epsilon_1(\epsilon_2-\epsilon_3)J^2+\epsilon_2(\epsilon_1-\epsilon_3)J'{}^2\\
&&2\epsilon_1\epsilon_2JJ'+(\epsilon_3-\epsilon_2)\partial J+(2\epsilon_3-\epsilon_2)\partial J'\Big )(\epsilon_3\partial)^{-2}+\dots\Big ](\epsilon_3\partial)^{-1}
\end{eqnarray}
where we have again omitted superscripts $(1)$ and $(2)$.

Fields $U_i$ appearing in front of the derivatives $(\epsilon_3\partial)^{-1-i}$ are now generators of algebra $Y_{1,1,0}$ in the $U$-basis associated to the composition of $\mathcal{L}_{0,0,1}^{0,0,1}$. Fields $U_i$ now lie in the kernel of the zero mode of a single screening current
\begin{eqnarray}
j_3&=&:\exp [-\epsilon_1\phi (z)+\epsilon_2\tilde{\phi} (z)]:
\end{eqnarray}
since only the line $L_{0,0,1}$ can be stretched stretched between the pair of surfaces $\mathcal{S}'_{1,0,0}$ and $\mathcal{S}_{0,1,0}$! The screening current is again given in terms of the product of vertex operators associated to $L_{0,0,1}$ ending from the left on $\mathcal{S}_{0,1,0}$ and from the right on $\mathcal{S}'_{1,0,0}$.

Similarly to the previous example, the degenerate module of $\mathcal{S}_{0,1,0}$ given in terms of $\exp [\epsilon_1 \phi (z)]$ and the degenerate module of $\mathcal{S}'_{1,0,0}$ given by $\exp [-\epsilon_2\tilde{\phi}(z)]$ are again degenerate modules of the fused defect $\mathcal{S}_{1,1,0}$. Composition of the Miura operator $\mathcal{L}_{0,1,0}^{0,0,1}$ describing the intersection of $\mathcal{S}_{0,1,0}$ with $L_{0,0,1}$ and the degenerate field $\exp [\epsilon_2\phi'(z)]$  of $\mathcal{S}'_{1,0,0}$ associated to the left endpoint gives
\begin{eqnarray}\nonumber
:\mathcal{L}_{0,1,0}^{0,0,1} \exp[\epsilon_2\phi' (z)]:&\sim& \frac{1}{(z-w)^{\frac{\epsilon_2}{\epsilon_3}+1}} :\exp[\epsilon_1\phi(z)-\epsilon_1\phi (w)+\epsilon_2\phi' (w)]:\\
 &\sim& j_3(w) :\exp[\epsilon_1\phi (z)]:
\end{eqnarray}
producing a screened vertex operator $ \exp[\epsilon_1\tilde{\phi} (z)]$ nicely agreeing with our expectations and similarly for the right action of $\mathcal{L}'{}_{1,0,0}^{0,0,1}$ on the degenerate field  $\exp[-\epsilon_1\phi (z)]$.

\subsubsection{Example: Degenerate modules for $\mathcal{S}_{1,0,1}$ along third direction}

Let us finally investigate junctions involving a single line $L_{0,0,1}$ and two surfaces $\mathcal{S}'_{1,0,0}$ and $\mathcal{S}_{0,0,1}$. Together with $\mathcal{L}'{}_{1,0,0}^{0,0,1}$ from above, we need the simple Miura operator
\begin{eqnarray}
\mathcal{L}_{0,0,1}^{0,0,1}=\epsilon_3\partial-\epsilon_1\epsilon_2J^{(3)}(z) 
\end{eqnarray}
Composing the two Miura operators produces
\begin{eqnarray}\nonumber
\mathcal{L}_{0,0,1}^{0,0,1}\mathcal{L}'{}_{1,0,0}^{0,0,1}&=&\Big [1-\epsilon_1\epsilon_2(J+J')(\epsilon_3\partial)^{-1}\\
&&+\frac{\epsilon_1\epsilon_2^2}{2} \Big (2\epsilon_1JJ'+(\epsilon_1-\epsilon_3)J'^2+\partial J'\Big )(\epsilon_3\partial)^{-2}+\dots\Big ](\epsilon_3\partial)^{1+\frac{\epsilon_1}{\epsilon_3}}.
\end{eqnarray}
Fields appearing in front of the derivatives $(\epsilon_3\partial)^{1+\frac{\epsilon_1}{3}-i}$ are generators of the algebra $Y_{1,1,0}$ in the $U$-basis coming from the composition of $\mathcal{L}_{0,0,1}^{0,0,1}$. These fields lie in the kernel of the zero mode of the screening current
\begin{eqnarray}
j_3&=& :\exp [-\epsilon_1\phi (z)+\epsilon_3\tilde{\phi} (z)]:
\end{eqnarray}
analogously to the previous example.

In the present case, the line $L_{0,0,1}$ cannot end on $\mathcal{S}_{0,0,1}$ and we do not get the corresponding degenerate module. On the other hand the degenerate module $\exp [-\epsilon_2\phi'(z)]$ associated to the right endpoint on $\mathcal{S}'_{1,0,0}$ is again a degenerate field for the fused surface $\mathcal{S}_{1,0,1}$. Furthermore, the composition of the Miura operator $\mathcal{L}_{0,0,1}^{0,0,1}$ describing the intersection of $\mathcal{S}_{0,0,1}$ with $L_{0,0,1}$ and the degenerate field $\exp [\epsilon_2\phi'(z)]$ associated to the right endpoint on $S'_{1,0,0}$ gives
\begin{eqnarray}
:\mathcal{L}_{0,0,1}^{0,0,1}\exp [\epsilon_2\tilde{\phi}(z)]:=:\epsilon_2(\epsilon_3\tilde{J}-\epsilon_1 J)\exp [\epsilon_2\tilde{\phi}(z)]: 
\end{eqnarray}
One can check that the right-hand side produces a realization of a degenerate field for the fused surface $\mathcal{S}_{1,0,1}$! Note that compared to the previous two examples, it is not given in terms of a composition with a screening current since the line $L_{0,0,1}$ cannot stretch between surfaces $\mathcal{S}'_{1,0,0}$ and $\mathcal{S}_{0,0,1}$.

\section{The restricted coproduct and the affine Yangian} \label{sec:boxes}
In this section, we will switch gear and explore an alternative perspective on the 
map $A \to W_\infty$ which controls gauge-invariant endpoints of line defects onto surface defects. 

An important property of the $W_\infty$ mode algebra is that it can be identified with the $\mathfrak{u}(1)$ {\it affine Yangian} algebra $Y$. The algebra $Y$ is generated by three semi-infinite collections
of generators $e_i$, $\psi_i$, $f_i$, for $i=0,1,\dots$, subject to a large collection of relations \cite{Tsymbaliuk:2014fvq,Schiffmann:2012gf} that we spell out in appendix \ref{yrelations}.
The Yangian algebra presentation is very convenient in the study of certain modules for $W_\infty$. 
It is also explicitly invariant under permutations of the $\epsilon_i$, as the relations depend on $\sigma_2$ and $\sigma_3$ only. This is obviously not the case for the Miura presentation of $\cW_\infty$, though the $\cW_\infty$ can be presented in a permutation-invariant form \cite{Gaberdiel:2012ku,Prochazka:2014gqa}.

The algebra $A$  has a close relationship to the affine Yangian Y.  It can be essentially obtained from the affine Yangian by dropping the $e_0$ generator \cite{Kodera:2016faj,Rapcak:2020ueh}.\footnote{The ``Coulomb branch'' presentation of $A$ from \cite{Kodera:2016faj,Gaiotto:2019wcc} can be found by dropping $e_0$ and performing a triangular linear redefinition of the $f_i$. The algerba A is usually referred to as the 1-shifted affine Yangian of $\mathfrak{u}(1)$.} As a consequence, there is a rather natural algebra morphism $A \to Y$. It is easy to show that it precisely coincides with the algebra morphism $\Delta_{W_\infty}: A \to W_\infty$ which describes gauge-invariant endpoints of line defects on surface defects. 

Indeed, the map $A \to W_{\infty}$ sends $t_{0,n}$ to modes of the primary generator of spin one $W_{1,n}$ that can be identified using formulas from appendix \ref{yrelations} with the $Y$  generators $t_{0,1}=-f_0$, $t_{0,2}=[f_1,f_0]$ and repeated commutators of $f_1$ with $f_0$,
\begin{align}
t_{0,n} &\to W_{1,n} \to -\frac{1}{(n-1)!}\mbox{ad}_{f_1}^{n-1} f_0
\label{ref1}
\end{align}
for general $n>1$. The central element $t_{0,0}$ maps to the $\widehat{\mathfrak{u}}(1)$ charge 
\begin{align}
t_{0,0} &\to W_{1,0} \to\psi_1
\end{align}
Furthermore, the map $A\to W_{\infty}$ sends $t_{2,0}$ to
\begin{align}
t_{2,0} & \to  V_{-2}+ \frac{\sigma_3}{2}  \sum_{n=-\infty}^\infty |n| :W_{1,-1-n}W_{1,n-1}:
\label{simpleemb}
\end{align}
with $V$ a quasi-primary field that can be characterized by its OPEs with $W_1,W_2$ listed in appendix \ref{utop}. Looking at formulas from \cite{Gaberdiel:2017dbk}, we can now identify our expression (\ref{simpleemb}) with the commutator
\begin{align}
t_{2,0} & \to -[e_2,e_1]
\end{align}
This uniquely characterizes a map $A\to Y$. Furthermore, it is straightforward to argue that commuting $W_{1,n}$ for $n\geq 0$ together with $[e_2,e_1]$ generates all the generators $e_i,f_i,\psi_i$ except of $e_0$ and the resulting map $A \to W_{\infty}$ is identical to the standard realization of the 1-shifted affine Yangian $A$ in terms of a subalgebra of $Y$.

Analogously to $t_{0,n}$, generators $t_{n,0}$ can be realized in terms of repeated commutators of $e_2$ with $e_1$,
\begin{align}
t_{n,0} \to -\frac{1}{(m-1)!}\mbox{ad}_{e_2}^{m-1} e_1
\end{align}
As a check that formulas given in this section give a well-defined embedding $A\to W_{\infty}$, one can start directly from (\ref{ref1}) and (\ref{simpleemb}) and verify the defining relations of $A$. We collect relevant formulae in appendix \ref{ainy}. 

\subsection{Fundamental and anti-fundamental endpoints}
Armed with this identification, we may seek VOA modules $\cM$ for $\cW_\infty$ and right modules $M$ for $A$ such that the 
gauge-invariance condition can be satisfied by an element of $\cM \otimes M$, representing endpoints of M2 line defects 
on M5 surface defects. 

There is a computationally slightly more transparent perspective: 
an element in $\cM \otimes M$ is the same as a map $M^\vee \to \cM$. Gauge-invariance is equivalent to the condition that 
the map should be a morphism of left $A$ modules, or that the dual $\cM^\vee \to M$ should be a morphism of 
right $A$-modules. We have seen examples in Section \ref{sec:deg} of the latter statement for degenerate fields of $\cW_{0,0,1}$.

Corner VOA considerations \cite{Gaiotto:2017euk} or the free field realizations \cite{Prochazka:2018tlo} we discussed in the previous section suggest that the endpoints we seek should involve certain ``degenerate'' modules  of $\cW_\infty$, labelled by three Young diagrams. The identification between $W_\infty$ and $Y$ has been used in the past to build 
explicitly such a collection of degenerate modules $\cM_{\lambda, \mu,\nu}$, which we will dub ``fundamental'' degenerate modules. These have a very neat combinatorial description in the Yangian description \cite{Prochazka:2015deb,Li:2020rij,Galakhov:2020vyb}. 

The algebra $\cW_\infty$ admits a well-known involution $W_n \to (-1)^n W_n$, where $W_n$ is a primary generator of spin $n$ for $n=1,3,4,\dots$ and $W_2$ is the stress-energy tensor. The transformation between the $U$-basis from the Miura representation of the algebra and the primary basis $W_n$ is sketched in appendix \ref{utop}. A remarkable fact is that this involution does {\it not} have a simple form in the affine Yangian description, but rather corresponds to a very non-trivial non-linear redefinition of the Yangian generators. 

Indeed, from the explicit relations between Yangian generators and $W_{\infty}$ spelled out in the appendix, we can easily read off the charge conjugation of some of the Yangian generators
\begin{eqnarray}\nonumber
&\tilde{e}_0=-e_0\qquad \tilde{e}_1=e_1\\ \nonumber
&\tilde{f}_0=-f_0\qquad \tilde{f}_1=f_1\\
&\tilde{\psi}_0=\psi_0\qquad \tilde{\psi}_1=-\psi_1\qquad \tilde{\psi}_2=\psi_2
\end{eqnarray}
To fully specify the conjugation automorphism in the Yangian basis, we also need to determine the conjugation of $\psi_3$. Performing the charge conjugation in the explicit expression (\ref{psi}) produces
\begin{equation}
\tilde{\psi}_3=-\psi_3-\sigma_3(\psi_0\psi_2-\psi_1^2)+6\sigma_3\sum_{m>0}mW_{1,-m}W_{1,m}
\label{psi3conj}
\end{equation}
Note the non-trivial tail present in charge-conjugation formulas in Yangian basis compared to the simple expressions in the primary basis of $\mathcal{W}_{\infty}$. All the other conjugated generators can be identified by commuting the above generators. We list some of them in appendix \ref{ainy}. 

Acting on fundamental degenerate modules with the involution of $\cW_\infty$, we get a second collection of degenerate modules $\cM_{\bar \lambda, \bar \mu,\bar \nu}$, which we will dub ``anti-fundamental'' degenerate modules. Much of this section will be devoted to the formulation of a combinatorial description of the Yangian action on anti-fundamental degenerate modules. 

One of the simplest consequences of this combinatorial description will emerge as we look at the $A$ action associated to the natural $A \to Y$ embedding: the anti-fundamental degenerate modules $\cM_{\bar \lambda, \bar \mu,\bar \nu}$ have a relatively small sub-modules 
$M_{\lambda, \mu, \nu}$ for the subalgebra $A$, which are dual to the $A$ modules which are expected to describe endpoints of M2 brane line defects. 

As a consequence, we have a collection of candidate gauge-invariant endpoints in $\cM_{\bar \lambda, \bar \mu,\bar \nu} \otimes M^\vee_{\lambda, \mu,\nu}$, satisfying the constraint that $t_{ab}$ acting from the left according to $\Delta_{W_\infty}: A \to W_\infty$ agrees with $t_{ab}$ acting from the right.

Recall that the $A \to A \otimes W_\infty$ coproduct had a nice hidden symmetry: the $\Delta_{A,W_\infty}[t'_{ab}] O = O t_{ab}$ 
relations can be rearranged to $\Delta_{A,W_\infty}[t_{ab}] \hat O = \hat O t'_{ab}$ by a combination of the involution of $\cW_\infty$ and the $A \to A^{\mathrm{op}}$ map $t_{ab} \to (-1)^a t_{ab}$. 

A particularly simple consequence is that given a gauge-invariant element in $\cM_{\bar \lambda, \bar \mu,\bar \nu} \otimes M^\vee_{\lambda, \mu,\nu}$ we can apply the involution to obtain a gauge-invariant element of 
$\cM_{\lambda, \mu,\nu} \otimes \hat M_{\lambda, \mu,\nu}$, where $\hat M_{\lambda, \mu,\nu}$ is the $A$-module obtained from $M^\vee_{\lambda, \mu,\nu}$ by the $t_{ab} \to (-1)^a t_{ab}$ twist. These are natural candidate gauge-invariant endpoints of line defects on surface defects from the left, built from fundamental degenerate modules. 

To be precise, in order to give endpoints of concrete line defects and surface defects we need to employ modules for the appropriate truncations $\cW_{N_1,N_2,N_3}$ and $A_{n_1,n_2,n_3}$. The degenerate modules $\cM_{\lambda, \mu,\nu}$ are known to admit truncations to modules 
$\cM^{N_1,N_2,N_3}_{\lambda, \mu,\nu}$ for $\cW_{N_1,N_2,N_3}$, as long as Young diagrams fit within certain shapes. Analogously, based on the VOA perspective, we expect (but lack their precise combinatorial description) that degenerate modules $\cM_{\bar \lambda, \bar \mu,\bar \nu}$ admit truncations to modules $\cM^{N_1,N_2,N_3}_{\bar \lambda, \bar \mu,\bar \nu}$  for $\cW_{N_1,N_2,N_3}$.

We expect, but have not proven, that $M_{\lambda, \mu,\nu}$ remain $A$ sub-modules of 
$\cM^{N_1,N_2,N_3}_{\bar \lambda, \bar \mu,\bar \nu}$. Furthermore, we expect the $A$ action on $M_{\lambda, \mu,\nu}$ and $M_{\bar \lambda, \bar \mu,\bar \nu}$ to factor through 
$A_{|\lambda|,|\mu|,|\nu|}$, so that we have the expected gauge-invariant endpoints for actual M2 and M5 brane defects. We have verified this in several examples.


\subsection{The combinatorial description of fundamental modules of $\mathcal{W}_{\infty}$}

 \Yboxdim{5pt}

Let us start with a review of fundamental $\mathcal{W}_{\infty}$ modules  $\mathcal{M}_{\lambda,\mu,\nu}$ parametrized by a triple of Young diagrams $\lambda,\mu,\nu$. To define these modules, it is useful to introduce generating functions
\begin{eqnarray}
e(z)=\sum_{i=0}^{\infty}\frac{e_i}{z^{i+1}}\qquad f(z)=\sum_{i=0}^{\infty}\frac{f_i}{z^{i+1}}\qquad \psi (z)=1+\sigma_3\sum_{i=0}^{\infty}\frac{\psi_i}{z^{i+1}}
\end{eqnarray}
Relations of the affine Yangian can be conveniently written in terms of these functions as reviewed in appendix \ref{relationsfunctions}.

The basis vectors of the highest-weight module $\mathcal{M}_{\lambda,\mu,\nu}$ are labelled by particular configurations of 3d partitions. 
In a precise sense, the configurations of boxes encode the eigenvalues of $\psi_i$ acting on the state. The action of the raising operators $e_i$  
and the lowering operators $f_i$ changes the configurations of boxes in a simple combinatorial manner by adding or removing a box. 

\begin{figure}
    \centering
    \begin{subfigure}[b]{0.55\textwidth}
        \includegraphics[width=\textwidth]{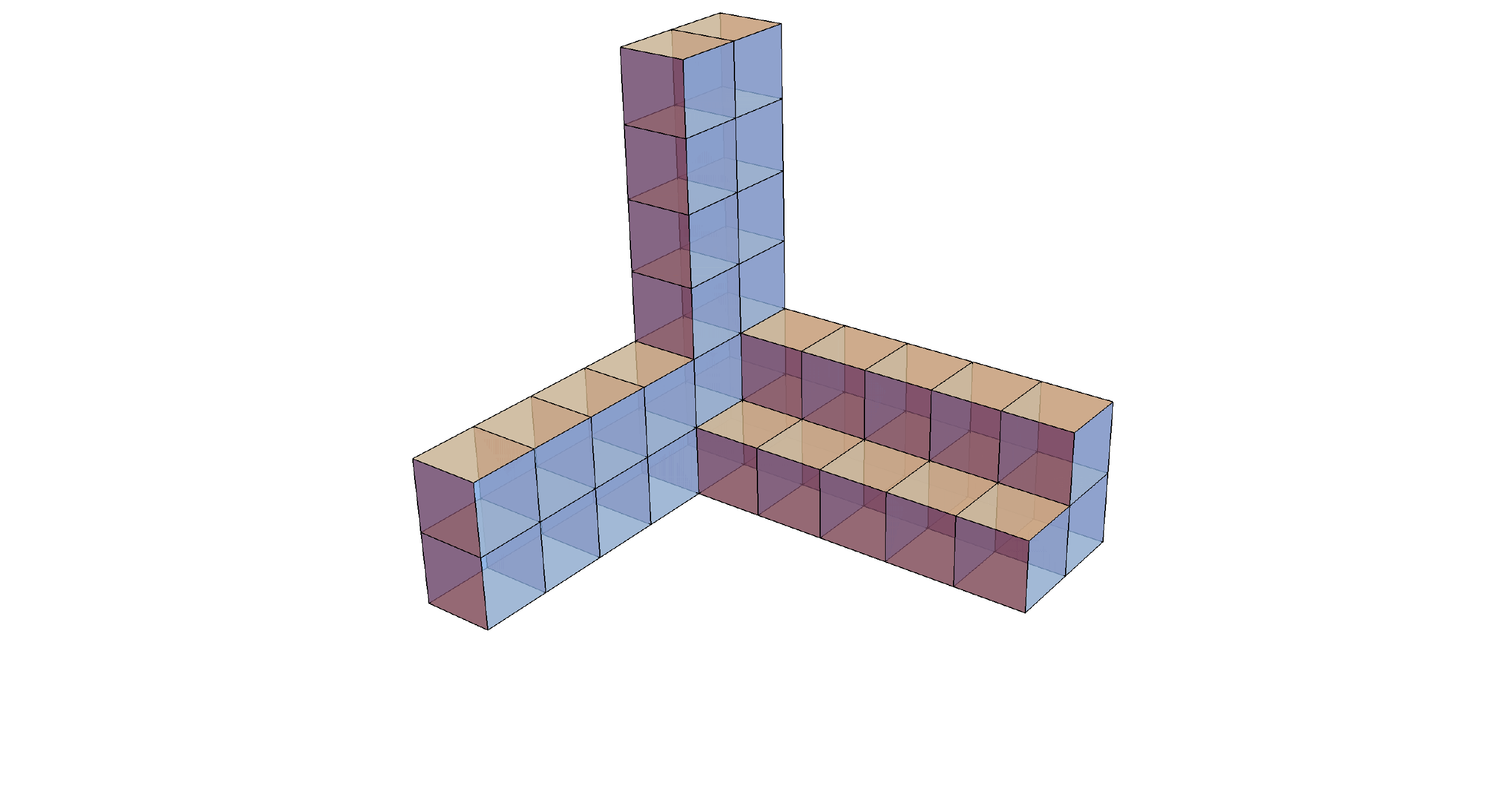}
    \end{subfigure}
    \begin{subfigure}[b]{0.15\textwidth}
        \includegraphics[width=\textwidth]{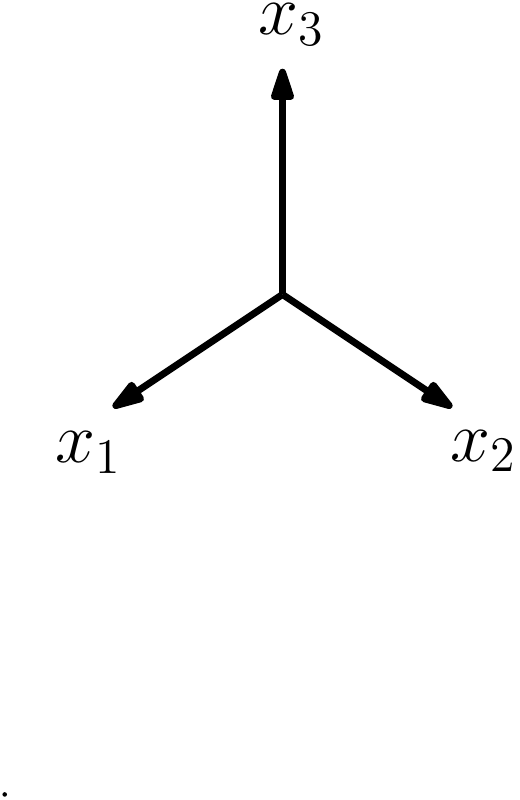}
    \end{subfigure}
\vspace{-0.5cm}
 \caption{Configuration of boxes associated to the vacuum state of the module $\mathcal{M}_{\protect \yng(1,1),\protect \yng(1,2), \protect \yng(2)}$. Note the convention for the axes labels and orientation of Young diagrams used throughout this note.}
\label{pic13}
\end{figure}

To the highest-weight state $|vac\rangle$ in the module $\mathcal{M}_{\lambda,\mu,\nu}$, we associate a configuration of boxes in a 3d corner with a triple collection of semi-infinite rows of boxes in the shape of $\lambda,\mu$ and $\nu$ along the three coordinate axes. To specify the configuration uniquely, one needs to specify the orientation of the Young diagram. We use the convention that $\lambda$ labels the shape along the $x_1$ axis with $x_2$ pointing to the right and $x_3$ pointing up, $\mu$ labels the Young diagram along the $x_2$ axis with $x_3$ pointing to the right and $x_1$ pointing up and $\nu$ labels the Young diagram along the $x_3$ axis with $x_1$ pointing to the right and $x_2$ pointing up. An example of such a configuration for $(\lambda,\mu,\nu)=(\yng(1,1),\yng(1,2),\yng(2))$ is depicted in figure \ref{pic13}. 

The other basis vectors for the module are labelled by all the possible configurations of boxes that can be obtained from the vacuum configuration by adding a finite number of boxes to the basic configuration according to the rules of 3d partitions, i.e. stable stacks in the 3d corner with the gravity acting in direction $(-1,-1,-1)$. For example, there are four configurations one can construct by adding four boxes to our vacuum configuration from figure \ref{pic13}. 

We will associate to a box at position $(n_1,n_2,n_3)$ the ``projected position'' $n_1 \epsilon_1 + n_2 \epsilon_2 + n_3 \epsilon_3$. In the following, we will refer to that simply as the position of a box.
In the example above, the position of the four possible boxes that can be added on top of the vacuum configuration are thus given as $2\epsilon_1+\epsilon_2$, $\epsilon_1+\epsilon_2+\epsilon_3$, $\epsilon_2+2\epsilon_3$ and $2\epsilon_1+2\epsilon_3$. The state corresponding to a given configuration of $n$ boxes is going to be labelled as $|a_1,\dots,a_n\rangle$ modulo ordering of $a_i$'s, where the $a_i$'s label the positions of added boxes. For example the state associated to two boxes located at $2\epsilon_1+\epsilon_2$ and $\epsilon_1+\epsilon_2+\epsilon_3$ in our example is denoted by $|2\epsilon_1+\epsilon_2,\epsilon_1+\epsilon_2+\epsilon_3\rangle$. The total number of boxes on top of the vacuum configuration will be referred to as a level of the configuration.

\Yboxdim{4pt} 

Intuitively, the $\psi(z)$ eigenvalue of a basis vector is simply the product $\prod_a \phi(z-x_a)$ over all the locations $x_a$ of the boxes, with  
\begin{eqnarray}
\phi(z)=\frac{(z+\epsilon_1)(z+\epsilon_2)(z+\epsilon_3)}{(z-\epsilon_1)(z-\epsilon_2)(z-\epsilon_3)}
\end{eqnarray}
multiplied by an overall factor of  
\begin{eqnarray}
\psi_0(z)=\frac{z+\sigma_3\psi_0}{z}
\end{eqnarray}

Because of the infinite arrays of boxes, a bit of care is needed to regularize the product. It turns out that the infinite product admits  many cancellations and each semi-infinite row in direction $1$ at position $x=j \epsilon_2+k\epsilon_3$ in the $2-3$ plane contributes by the factor of $\psi_1(z-x)$, where
\begin{eqnarray}
\psi_1(z)=\frac{(z+\epsilon_1)z}{(z-\epsilon_2)(z-\epsilon_3)}
\end{eqnarray}
Similarly, the semi-infinite rows along the other two coordinate axes with positions $x=i \epsilon_1+k\epsilon_3$ and $x=i \epsilon_1+j\epsilon_2$ respectively contribute by the factor of $\psi_2(z-x)$ and $\psi_3(z-x)$, where
\begin{eqnarray}\nonumber
\psi_2(z)=\frac{(z+\epsilon_2)z}{(z-\epsilon_1)(z-\epsilon_3)}\qquad \psi_3(z)=\frac{(z+\epsilon_3)z}{(z-\epsilon_1)(z-\epsilon_2)}
\end{eqnarray}

As the three semi-infinite rows overlap on a finite collection of boxes, we need to remove the contribution of the overlaps to get our final $\psi(z)$. 
We can distinguish a few cases:
\begin{itemize}
\item The generating function of $\psi(z)$ acting on the highest-weigh state with a single asymptotics is given by the product of such factors associated to each box of the young diagram.  \Yboxdim{8pt} For example, the generating function for the $(\yng(1),0,0)$ module is simply \Yboxdim{4pt}
\begin{eqnarray}
\psi(z)|vac\rangle=\psi_0(z)\psi_1(z)|vac\rangle=\frac{(z+\sigma_3\psi_0)(z+\epsilon_1)}{(z-\epsilon_3)(z-\epsilon_2)}|vac\rangle
\end{eqnarray}
 \Yboxdim{8pt} For $(\yng(2),0,0)$, it would be
\begin{eqnarray}
\psi(z)|vac\rangle=\psi_0(z)\psi_1(z)\psi_1(z-\epsilon_2)|vac\rangle=\frac{(z+\sigma_3\psi_0)(z-2\epsilon_2-\epsilon_3)}{(z-\epsilon_3)(z-2\epsilon_2)}|vac\rangle
\end{eqnarray}
\item To get the generating function of $\psi(z)$ acting on the highest-weigh state with two non-trivial asymptotics, one needs to take the product of the contributions of each asymptotic and subtract the contribution of boxes that lie in their intersection by dividing by the factor $\phi (z-x)$ for each box at position $x=i \epsilon_1+j\epsilon_2+k\epsilon_3$ that appear in the intersection of bot half-cylinders of boxes.
For example, the generating function for $(\yng(1),\yng(1),0)$ is given by
\begin{eqnarray}
\psi (z)|vac\rangle=\frac{\psi_0(z)\psi_1(z)\psi_2(z)}{\phi(z)}|vac\rangle=\frac{(z+\sigma_3\psi_0)z}{(z-\epsilon_3)(z+\epsilon_3)}|vac\rangle
\end{eqnarray}
since there is only a single box located at the origin that lies inside the intersection of the two asymptotic cylinders of shape $\yng(1)$.
\item Finally, with all three asymptotics non-trivial, there are three kind of boxes one needs to take into account. Let us label by $M^1$ the set of boxes  associated to the vacuum configuration of $\mathcal{M}_{\lambda,0,0}$ (cylinder of shape $\lambda$) and analogously $M^2$ and $M^3$ for sets associated to $\mathcal{M}_{0,\mu,0}$ and $\mathcal{M}_{0,0,\nu}$. There are three kinds of boxes one needs to take into consideration. The generating function $\psi(z)$ is then a product of $\psi_i(x)$ factors associated to each asymptotic $M^i$ divided by the factor of $\phi (z-x)$ for each box lying in the intersection of exactly two asymptotics
\begin{eqnarray}
S_2=\left ((M^1\cap M^2)\cup (M^1\cap M^3) \cup (M^2\cap M^3)\right )\backslash (M^1\cap M^2\cap M^3)
\end{eqnarray}
and divide by two factors of $\phi (z-x)$ associated to each box inside the intersection of all three cylinders
\begin{eqnarray}
S_3=M^1\cap M^2\cap M^3
\end{eqnarray}
For example, the generating function for three one-box representations $(\yng(1),\yng(1),\yng(1))$ is given by
\begin{eqnarray}
\psi (z)|vac\rangle =\frac{\psi_0(z) \psi_1(z)\psi_2(z)\psi_3(z)}{\phi(z)^2}|vac\rangle=\frac{(z+\sigma_3\psi_0)z^2}{(z+\epsilon_1)(z+\epsilon_2)(z+\epsilon_3)}|vac\rangle
\label{tribox}
\end{eqnarray}
since there are no boxes inside $S_2$ and there is a single box in the intersection of all three cylinders.
\end{itemize}

The action of $\psi (z)$ on the state associated to a given configuration is given by the vacuum contribution multiplied by the factor of $\phi (z-x_a)$ for each box added at positions $x_a$. For example 
\begin{eqnarray}
\psi (z)|\epsilon_1+\epsilon_2\rangle =\psi_0(z) \psi_1(z)\psi_2(z)\psi_3(z)\phi(z)^{-2}\phi(z-\epsilon_1-\epsilon_2)|\epsilon_1+\epsilon_2\rangle
\end{eqnarray}
for the state with a box at position $\epsilon_1+\epsilon_2$ of the $(\Box,\Box,\Box)$ module.

After describing the action of $\psi (z)$, let us move to the definition of the action of $e(z)$ and $f(z)$. Generators in $e(z)$ act as rising operators that add a single box at each allowed position 
\begin{eqnarray}
e(z)|\Lambda\rangle =\sum_{a\in \Lambda^+}\frac{E(\Lambda \rightarrow \Lambda+a)}{z-x_a}|\Lambda+a\rangle
\label{eaction}
\end{eqnarray}
where $\Lambda^+$ denotes the set of configurations that can be obtained from $\Lambda$ by adding a box and $E(\Lambda \rightarrow \Lambda+a)$ is an amplitude associated to adding a box at a given position $x_a$. Similarly, $f(z)$ removes boxes by
\begin{eqnarray}
f(z)|\Lambda\rangle =\sum_{a\in \Lambda^+}\frac{F(\Lambda-a \rightarrow \Lambda)}{z-x_a}|\Lambda-a\rangle
\label{faction}
\end{eqnarray}
where $\Lambda^-$ again labels configurations that can be obtained from $\Lambda$ by removing a box and $F(\Lambda+a \rightarrow \Lambda)$ is an amplitude for removing the given box. The amplitudes $E(\Lambda \rightarrow \Lambda+a)$ and $F(\Lambda+a \rightarrow \Lambda)$ are constrained by the $fe$ relation
\begin{eqnarray}
E(\Lambda \rightarrow \Lambda+a)F(\Lambda+a \rightarrow \Lambda)=-\frac{1}{\sigma_3}\mbox{res}_{z=x_a}\psi_{\Lambda}(z)
\label{condition1}
\end{eqnarray}
where $\psi(z)|\Lambda\rangle=\psi_\Lambda(z)|\Lambda\rangle$.  They can be fixed by requiring $E(\Lambda \rightarrow \Lambda+a) = F(\Lambda+a\rightarrow \Lambda)$, leading to
\begin{eqnarray}\nonumber
e(z)|\Lambda\rangle &=&\sum_{a\in \Lambda^+}\sqrt{-\frac{\mbox{Res}_{z=x_a}\ \psi_{\Lambda}(z)}{\sigma_3}}\frac{1}{z-x_a}|\Lambda+a\rangle\\
f(z)|\Lambda\rangle &=&\sum_{a\in \Lambda^-}\sqrt{-\frac{\mbox{Res}_{z=x_a}\ \psi_{\Lambda-a}(z)}{\sigma_3}}\frac{1}{z-x_a}|\Lambda-a\rangle
\label{action1}
\end{eqnarray}
One can check that the proposed action of rising generators (\ref{action1}) with the action of the Cartan elements $\psi(z)$ satisfy relations of the affine Yangian \cite{feigin,Prochazka:2015deb}. We are going to see explicit examples of the modules $\mathcal{M}_{\lambda,\mu,\nu}$ later on.

\subsection{Anti-fundamental modules of $\mathcal{W}_{\infty}$}
Next, we would like to find a combinatorial description of the anti-fundamental modules $\mathcal{M}_{\bar{\lambda},\bar{\mu},\bar{\nu}}$ obtained from $\mathcal{M}_{\lambda,\mu,\nu}$ by applying the involution of $\cW_\infty$.

Such a combinatorial description for the single non-trivial asymptotics $\mathcal{M}_{\bar{\lambda},0,0}$ was given in \cite{Gaberdiel:2018nbs}. The case with two non-trivial asymptotics was hinted in \cite{Harada:2018bkb}. The combinatorial descriptions of \cite{Gaberdiel:2018nbs,Harada:2018bkb} employ two independent collections of boxes: one collection follows the standard rules of finite 3d partitions (with trivial asymptotics), while the second collection of boxes at a different location follows a different set of rules. The construction in these special cases relies on the fact that modules $\mathcal{M}_{\bar{\lambda},\bar{\mu},0}$ are consistent with the $\mathcal{W}_N$ truncation of $\mathcal{W}_{\infty}$ for large enough $N$. This is obviously not true for the general anti-fundamental module $\mathcal{M}_{\bar{\lambda},\bar{\mu},\bar{\nu}}$.

We claim that this pattern persists for the general modules $\mathcal{M}_{\bar{\lambda},\bar{\mu},\bar{\nu}}$ and
they also have a combinatorial description involving a standard 3d partition (with trivial asymptotics) together with some other data. The resulting modules thus have a natural sub-space, consisting of vectors labeled by an empty collection of standard boxes 
and an arbitrary collection of boxes of the second type.  As we argue bellow, a simple consequence of this combinatorial rule is that the generators of $A$ in $Y$ preserve this sub-space. Such a subspace is thus equipped with the structure of an $A$-module. This gives a derivation of a combinatorial description of $A$-modules $M_{\lambda,\mu,\nu}$. 

It is worth explaining this point in a greater detail by first giving a general argument why should one expect such a combinatorics to appear and then argue that the action of $A$ stabilizes the subspace $M_{\lambda,\mu,\nu}$. The module $\mathcal{M}_{\bar{\lambda},\bar{\mu},\bar{\nu}}$ is again a highest-weight module. The action of the generating function of conjugated charges
\begin{eqnarray}
\tilde{\psi}(z)=1+\sigma_3\sum_{i=0}^{\infty}\frac{\tilde{\psi}_i}{z^{1+i}}
\end{eqnarray}
on the highest-weight state can be determined (at least up to first couple of coefficients) by recalling the charge conjugation of $\psi_i$ generators, leading to
\begin{eqnarray}
\tilde{\psi}(z)|0\rangle =\left (1+\sigma_3 \left (\frac{\psi_0}{z}-\frac{\psi_1}{z^2}+\frac{\psi_2}{z^3}-\frac{\psi_3+\sigma_3 (\psi_0\psi_2-\psi_1^2)}{z^4}+\dots \right )\right )|0\rangle
\end{eqnarray}
where $\psi_i$ are the charges of the vacuum state of $\mathcal{M}_{\lambda,\mu,\nu}$. This is consistent with the general proposal from \cite{Gaberdiel:2017hcn} of the form
\begin{eqnarray}
\tilde{\psi}(z)|vac\rangle = \frac{1}{\psi (-z-\sigma_3\psi_0)}|vac\rangle
\label{conjfunc}
\end{eqnarray}
Note that this formula is valid only when acting on the highest-weight state and it receives corrections at higher levels due to the tails present in the charge conjugation.

Let us look at the structure of the conjugated $\psi(z)$ acting on the vacuum state. $\psi (z)$ acting on the vacuum state of the fundamental module has generally the following form
\begin{eqnarray}
\psi (z)|vac\rangle =\frac{z+\sigma_3\psi_0}{z} \frac{P(z)}{Q(z)}|vac\rangle
\end{eqnarray}
where $P(z)$ and $Q(z)$ are polynomials with zeros at locations $x_1\epsilon_1+x_2\epsilon_2+x_3\epsilon_3$ for integral $x_i$. The charge conjugation gives
\begin{eqnarray}
\psi (z)|vac\rangle =\frac{z+\sigma_3\psi_0}{z}\frac{Q(-z- \sigma_3\psi_0)}{P(-z- \sigma_3\psi_0)} |vac\rangle
\end{eqnarray}
and polynomial $P(-z- \sigma_3\psi_0)$ only produces poles of $\psi (z)$ at shifted locations $z=x_1\epsilon_1+x_2\epsilon_2+x_3\epsilon_3-\sigma_3\psi_0$ for integers $x_i$.
 
For now, let us assume\footnote{As we are going to see later, these rules require a minor modification in the general case of $\mathcal{M}_{\bar \lambda,\bar \mu,\bar \nu}$ but in a way that the argument still holds.} an existence of a combinatorial description of $\mathcal{M}_{\bar \lambda,\bar \mu,\bar \nu}$, where addition and removal of boxes is governed by poles of $\psi (z)$ and the action of $\psi(z)$ is governed by a multiplication by factors $\phi (z-x_a)$ for each box added at position $x_a$ on top of the vacuum configuration. This is indeed the case for both fundamental modules $\mathcal{M}_{\lambda,\mu,\nu}$ and anti-fundamental modules $\mathcal{M}_{\bar \lambda,\bar \mu,0}$. As a consequence, the rules for the addition of boxes at standard locations and unshifted locations are independent of each other since addition of a box at a standard location does not change poles of $\psi(z)$ at the shifted location and vice versa. Moreover, the factor responsible for the addition of the first box at standard locations is simply $1/z$, i.e. the same one as for the simplest MacMahon module $\mathcal{M}_{0,0,0}$. 
This strongly suggests that the combinatorial description of the anti-fundamental modules will generally involve two collection of boxes:
a collection following the standard rules for 3d partitions and a collection of boxes at shifted locations, following a different set of rules.

We can also give a general argument why the action of $A$ preserves the subspace with no boxes at the standard location.  Assuming that $e(z)$ has poles at positions, where a box can be added, we find that the raising generator $e(z)$ can either have poles at standard locations $x_1\epsilon_1+x_2\epsilon_2+x_3\epsilon_3$ or at shifted locations $x_1\epsilon_1+x_2\epsilon_2+x_3\epsilon_3-\sigma_3\psi_0$. Residues of poles at standard locations are proportional to states with an added box at a standard location. Residues of poles at shifted locations are proportional to states with a box added at a shifted location. As mentioned above, $M_{\lambda,\mu,\nu}$ is defined as the subspace generated by states for which the 3d partition is empty. The action of $e(z)$ on a vector in $M_{\lambda,\mu,\nu}$ involves a term proportional to $z^{-1}$ 
which adds a box to the empty 3d partition, as well as terms where the 3d partition remains empty. The former term does not contribute to $e_1$, $e_2$, etc. The action of $A$ thus stabilizes the $M_{\lambda,\mu,\nu}$ subspace, acting only on the extra combinatorial data attached to each state in $M_{\lambda,\mu,\nu}$.

By construction, $M_{\lambda,\mu,\nu}$ will be an highest weight module, generated from a vacuum vector annihilated by 
all the $f_i$ and more generally all the $t_{ab}$ with $b>a$. Ultimately, we want to identify it with one of the highest weight modules for the truncation $A_{|\lambda|,|\mu|,|\nu|}$. We will not attempt to prove this identification, but we can give a conjectural formula for that analogous to the one in Section \ref{sec:deg}: the map
\begin{equation}
\langle v| \to \langle v| \left[\prod_a \Phi_{1,0,0}(z_a) \right]\left[\prod_a \Phi_{0,1,0}(z'_a) \right]\left[\prod_a \Phi_{0,0,1}(z''_a) \right]|0\rangle
\end{equation}
sending a vector $\langle v|$ in $\mathcal{M}_{\bar{\lambda},\bar{\mu},\bar{\nu}}^{\vee}$ to the space of conformal blocks
on a Riemann sphere, with $\langle v|$ placed at infinity and elementary fundamental degenerate fields placed
at various points, should intertwine the right $A$ action on $M^\vee_{\mu,\nu,\lambda}$ and the right Calogero-like action for 
$A_{|\lambda|,|\mu|,|\nu|}$ acting on the positions of the degenerate fields.

To build and explicit combinatorial description of $\mathcal{M}_{\bar{\lambda},\bar{\mu},\bar{\nu}}$, we will encounter two related subtleties:
\begin{enumerate}
\item The action of $\psi(z)$ is not diagonalizable  and we need to introduce Jordan blocks.
\item  It is possible for $\psi(z),e(z),f(z)$ to have double poles. 
\end{enumerate}
These subtleties only occur for the case with three non-trivial asymptotics. In simpler situations, the combinatorial description involves a finite collection of extra boxes stacked in an auxiliary region with a shape constrained by the 2d partitions. These subtleties force us to 
go beyond such a simple picture. We receive an important hint from the theory of Pandharipande-Thomas invariants.

\subsection{Pandharipande-Thomas box counting and $A$ modules}

\Yboxdim{5pt} 

As justified above, rules for adding boxes at the shifted location should be independent of the actual configuration of boxes at the standard location.\footnote{This is not going to be true for truncations $\mathcal{W}_{N_1,N_2,N_3}$ since boxes at standard and shifted locations are expected to interact. This is a phenomenon that deserves further investigation.} Let us denote the character of $\mathcal{M}_{\lambda,\mu,\nu}$ by $\chi_{\lambda,\mu,\nu}$.  This is also the character of $\mathcal{M}_{\bar{\lambda},\bar{\mu},\bar{\nu}}$ since the generator $\psi_2$ that counts the number of boxes remains unchanged by the charge conjugation. We can factor out the contribution from the standard 3d partitions by writing 
\begin{eqnarray}
\chi_{\lambda,\mu,\nu}=\chi_{\bar{\lambda},\bar{\mu},\bar{\nu}}=\prod_{n=1}^{\infty}\frac{1}{(1-q^n)^n}C_{\lambda,\mu,\nu}
\end{eqnarray}
with $C_{\lambda,\mu,\nu}$ being the character of  $M_{\lambda,\mu,\nu}$. Here, we used the standard counting of 3d partitions by the MacMahon generating function
\begin{eqnarray}
\prod_{n=1}^{\infty}\frac{1}{(1-q^n)^n}
\end{eqnarray}

The topological-vertex partition function $C_{\lambda,\mu,\nu}$ admits a known box-counting construction due to Pandharipande and Thomas. The above argument suggests that the vector space with a basis labelled by Pandharipande-Thomas box configurations associated to $C_{\lambda,\mu,\nu}$ actually carries a structure of an $A$-module. Similarly, the vector space with a basis labelled by a pair of a Pandharipande-Thomas box configuration and a standard 3d partition carries a structure of an anti-fundamental module of $Y$. Before turning to the actual construction of $M_{\lambda,\mu,\nu}$ and $\mathcal{M}_{\bar{\lambda},\bar{\mu},\bar{\nu}}$, let us first rephrase the Pandharipande-Thomas (PT) box-counting prescription \cite{Pandharipande:2007sq}.

\subsection{PT box-counting rules}

Let us proceed slowly by first analysing the box counting rules for $M_{\lambda,0,0}$ followed by $M_{\lambda,\mu,0}$ and the general $M_{\lambda,\mu,\nu}$. 

\begin{figure}
    \centering
    \begin{subfigure}[b]{0.17\textwidth}
        \includegraphics[width=\textwidth]{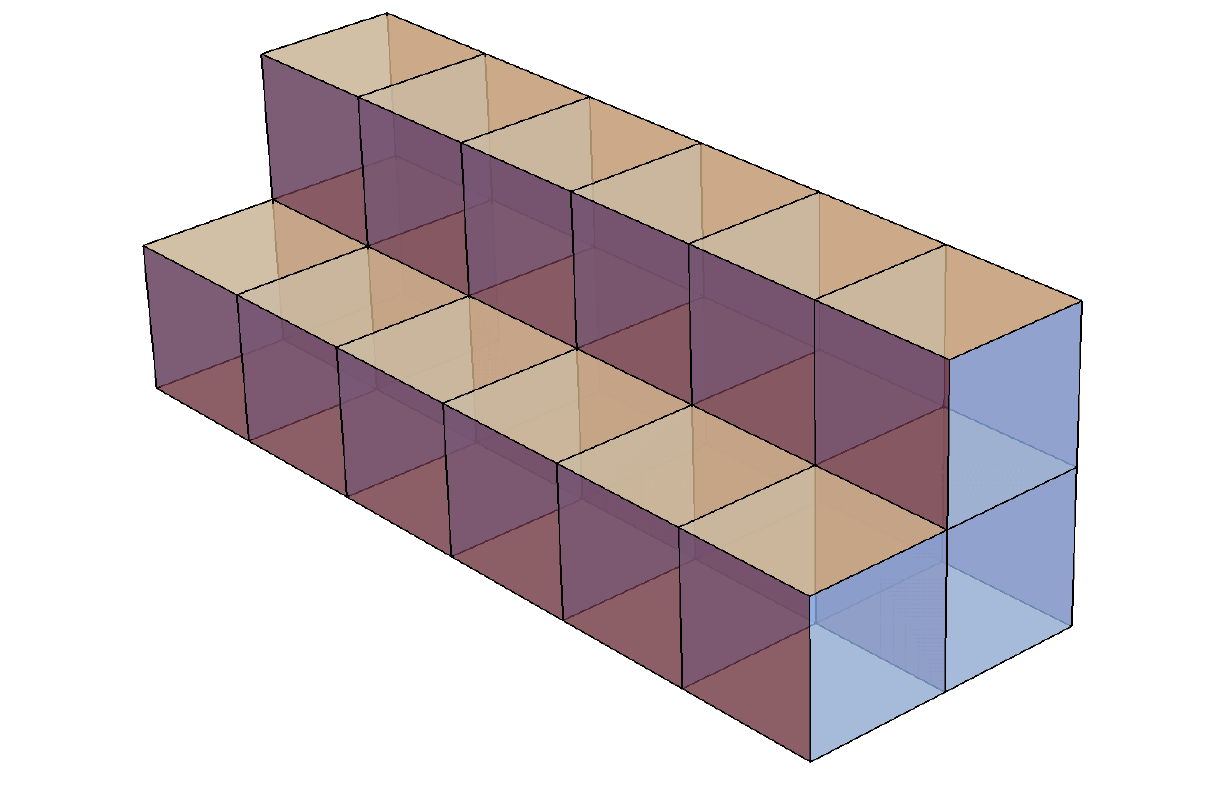}
    \end{subfigure}
 \caption{The vacuum configuration for $(0,\protect \yng(1,2),0)$.}
\label{pic23}
\end{figure}

\begin{figure}
    \centering
    \begin{subfigure}[b]{0.17\textwidth}
        \includegraphics[width=\textwidth]{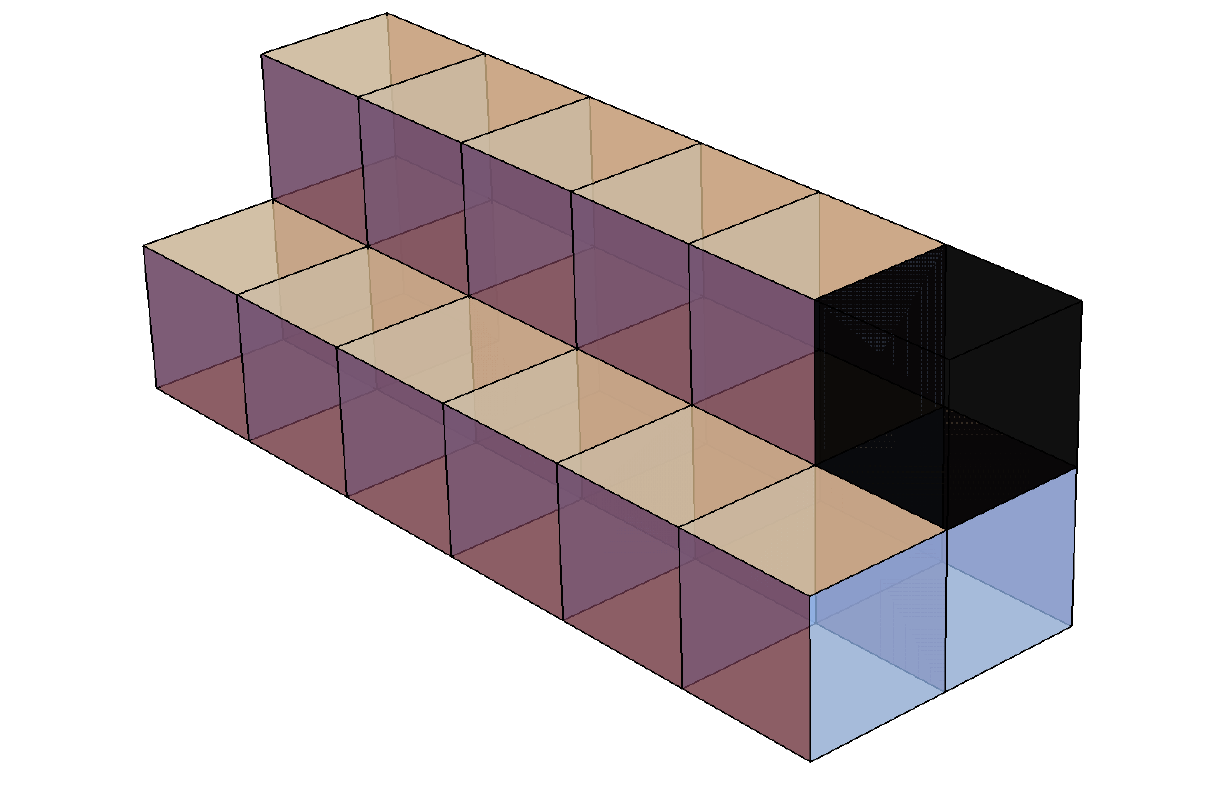}
    \end{subfigure}\quad
    \begin{subfigure}[b]{0.17\textwidth}
        \includegraphics[width=\textwidth]{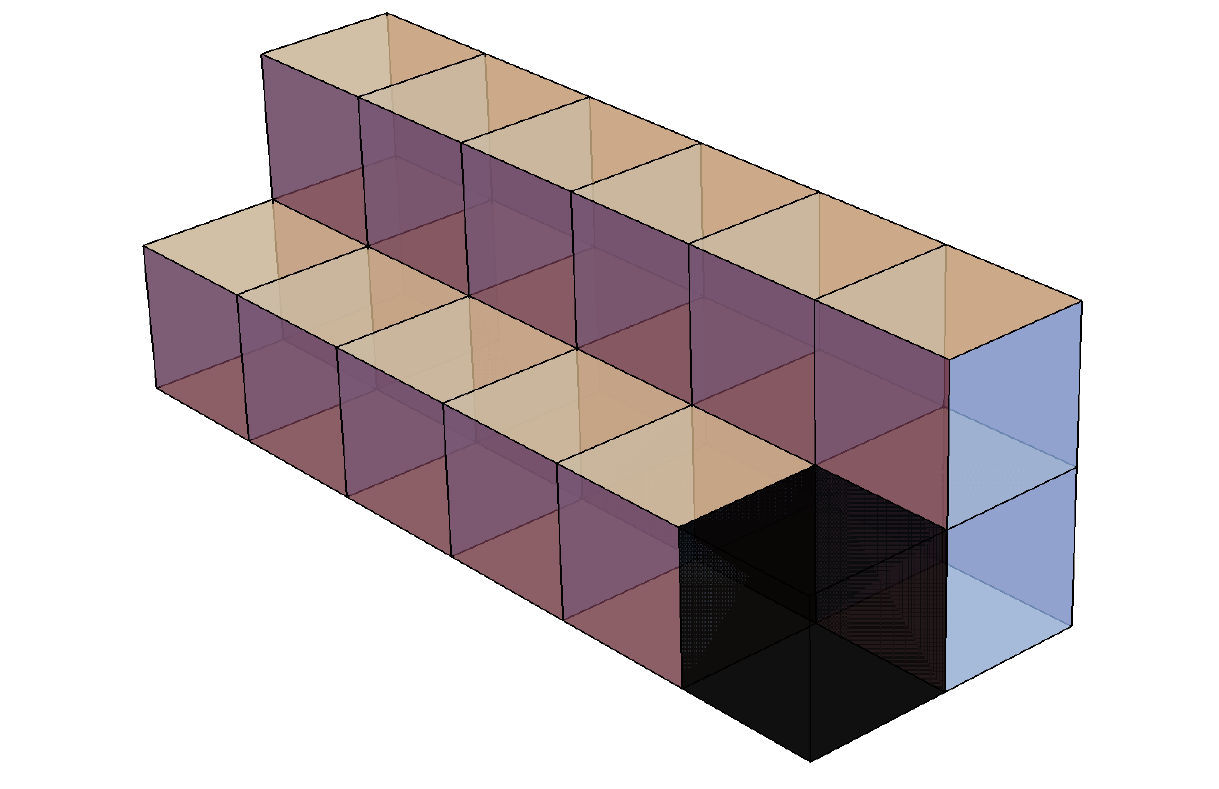}
    \end{subfigure}
 \caption{Two possible configurations containing a single boxe for $(0,\protect \yng(1,2),0)$.}
\label{pic24}
\end{figure}

\begin{figure}
    \centering
    \begin{subfigure}[b]{0.17\textwidth}
        \includegraphics[width=\textwidth]{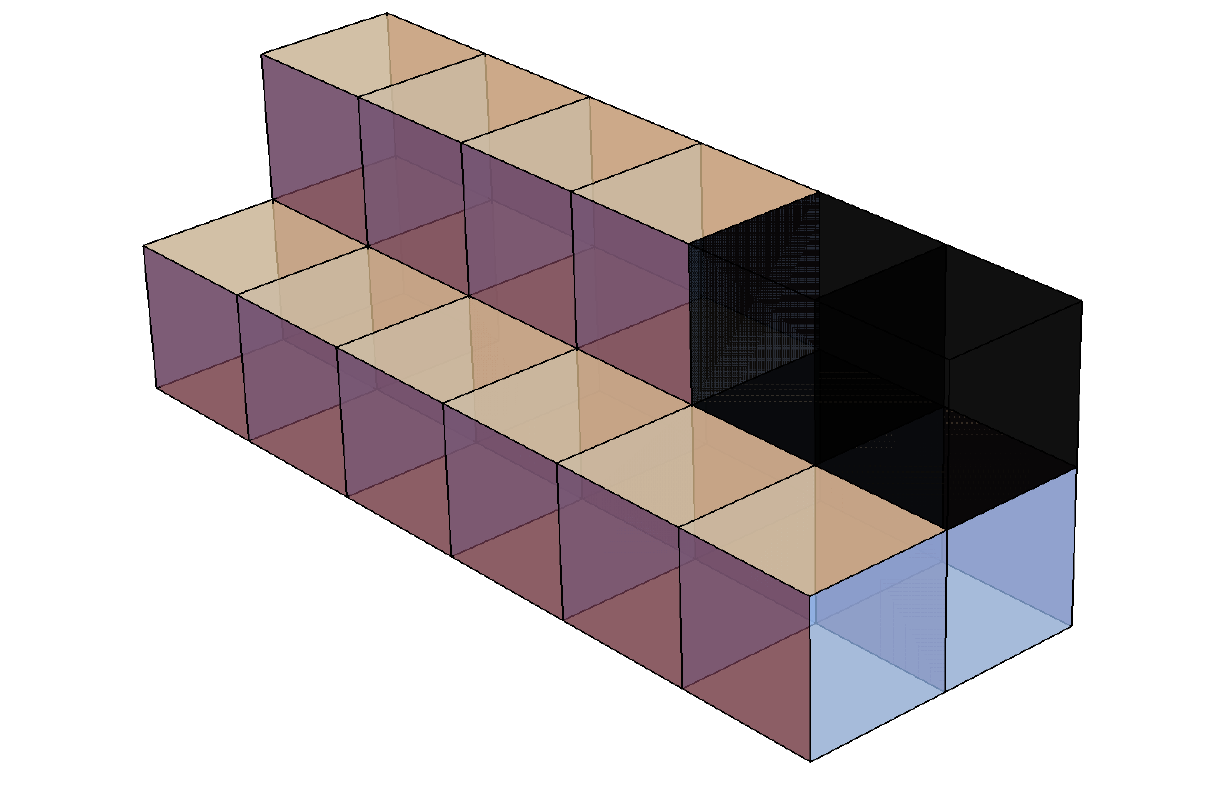}
    \end{subfigure}\quad
    \begin{subfigure}[b]{0.17\textwidth}
        \includegraphics[width=\textwidth]{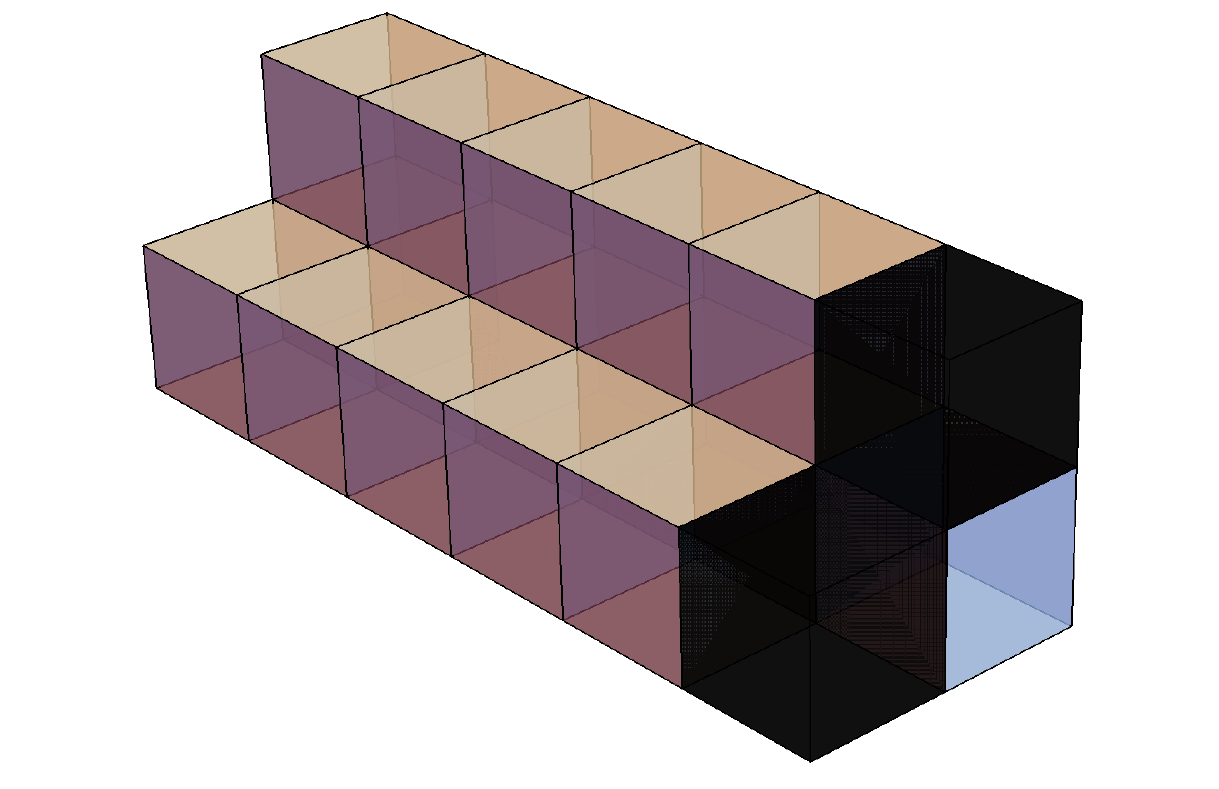}
    \end{subfigure}\quad
    \begin{subfigure}[b]{0.17\textwidth}
        \includegraphics[width=\textwidth]{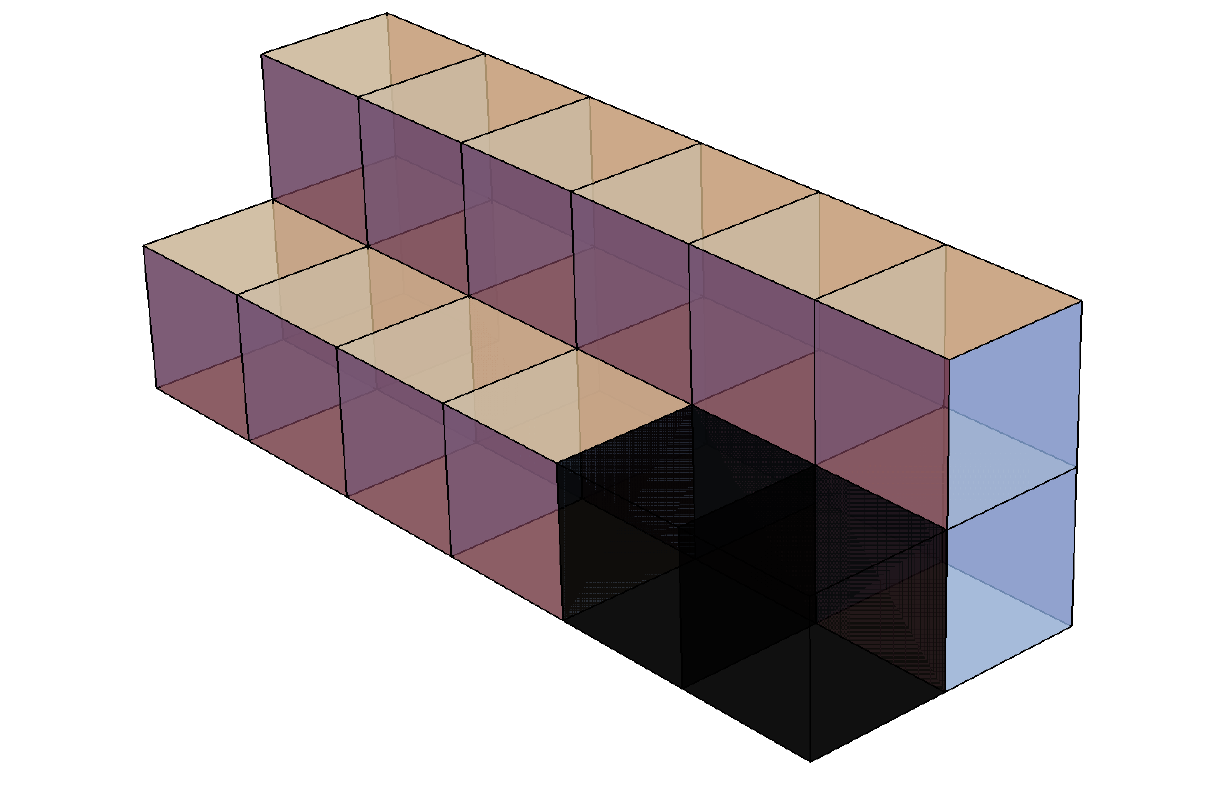}
    \end{subfigure}
 \caption{Three possible configurations containing two boxes for $(0,\protect \yng(1,2),0)$.}
\label{pic26}
\end{figure}

The starting point is the vacuum configuration for the standard 3d partition counting for $\mathcal{M}_{\lambda,0,0}$. An example of such a configuration for $(0,\protect \yng(1,2),0)$ is shown in figure \ref{pic23}. In the first step towards the PT counting, let us extend the the half cylinder of shape $\lambda$ placed along the first coordinate line in the negative direction. In the second step, we remove boxes in the positive quadrant, i.e. all the boxes originally present in the standard box counting. We end up with the same picture \ref{pic23} but now representing a half cylinder starting at the origin an going further in the negative direction. According to PT rules, one can now think about the resulting half cylinder as forming a hollow structure inside which we stack boxes with one crucial difference that the gravity is now pointing towards us. Possible configurations containing one and two boxes in our example of $(0,\protect \yng(1,2),0)$ are shown in figures \ref{pic24} and \ref{pic26}.

\begin{figure}
    \centering
    \begin{subfigure}[b]{0.28\textwidth}
        \includegraphics[width=\textwidth]{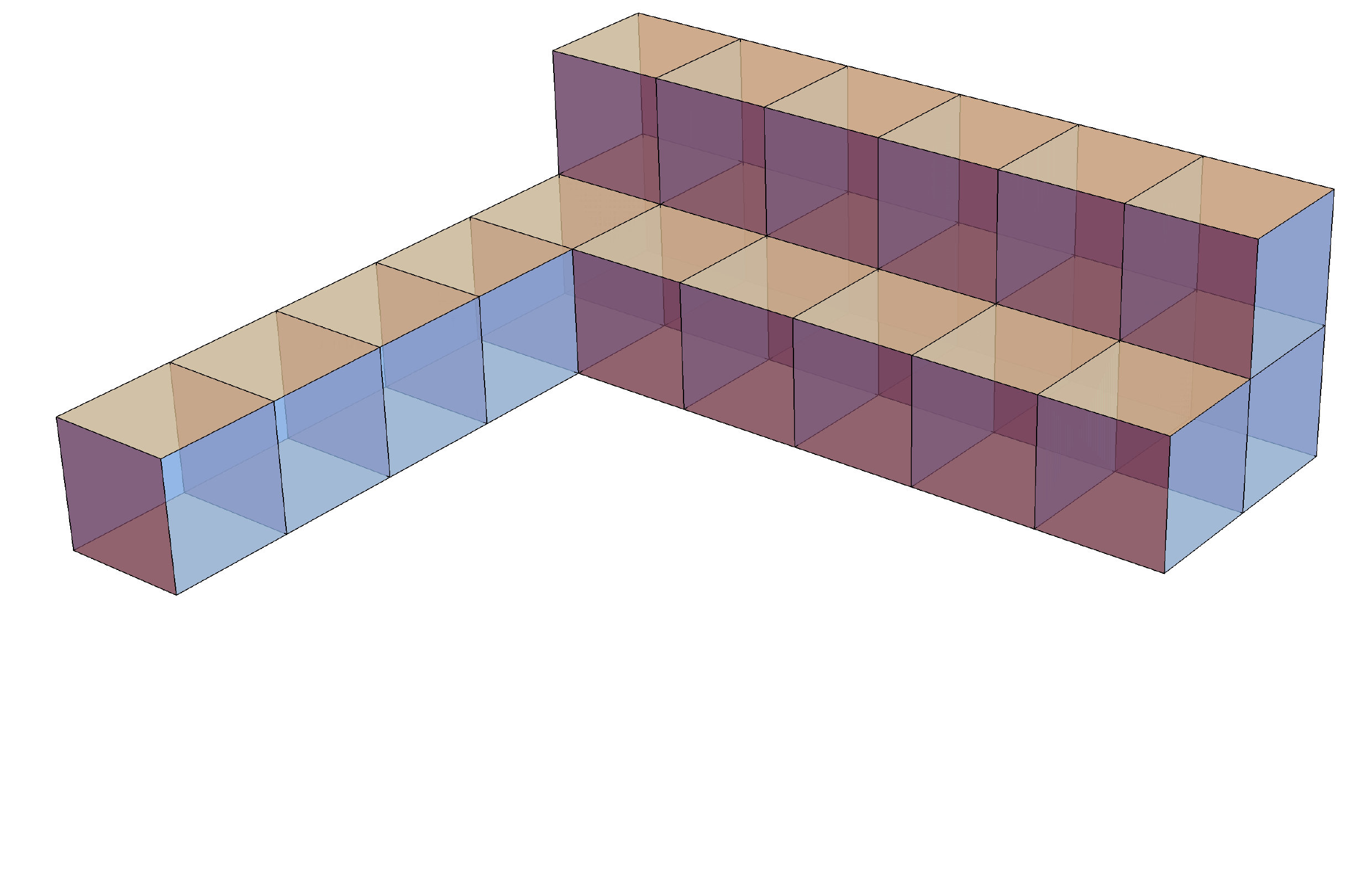}
\caption{}
\label{pic1}
    \end{subfigure}
    \begin{subfigure}[b]{0.44\textwidth}
        \includegraphics[width=\textwidth]{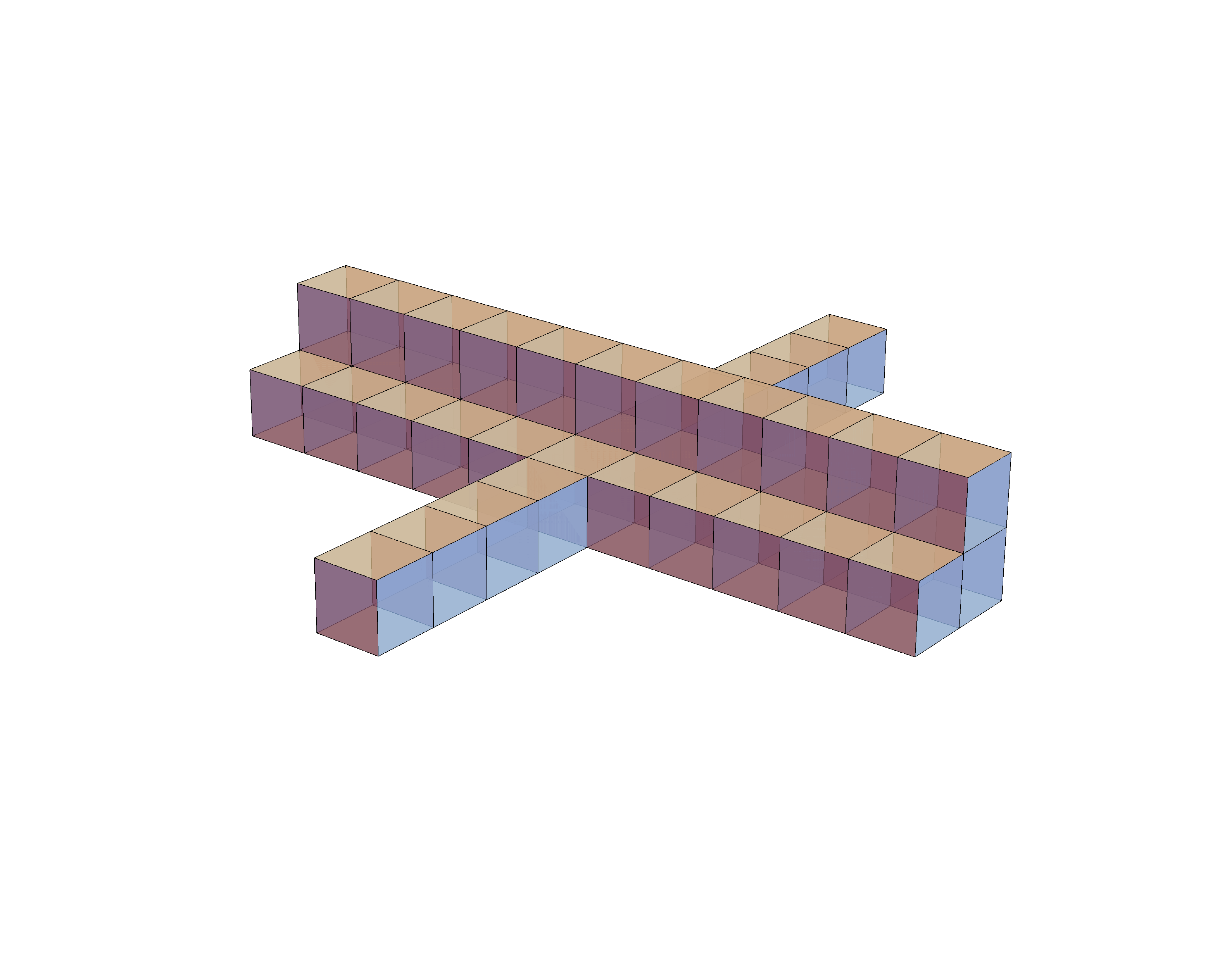}
        \caption{}
\label{pic2}
    \end{subfigure}
    \begin{subfigure}[b]{0.25\textwidth}
        \includegraphics[width=\textwidth]{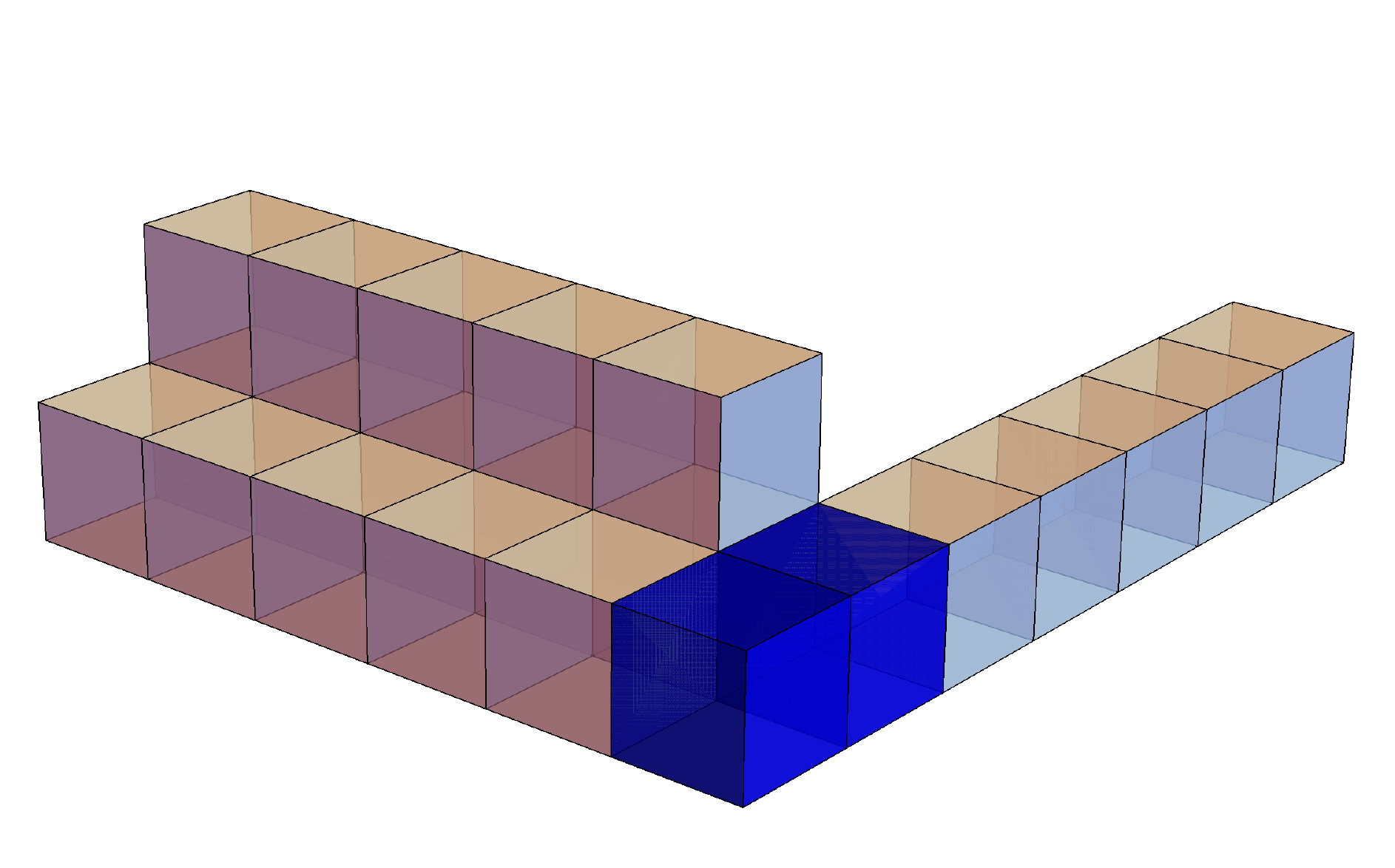}
        \caption{}
\label{pic3}
    \end{subfigure}
    \caption{Construction of the PT box-counting in the $(\protect \yng(1),\protect \yng(1,2),0)$ example. (a) Starting configuration associated to 3d partitions with two non-trivial asymptotics. (b) A configuration with asymptotic cylinders made infinite in both directions. (c) A configuration with removed and colored boxes.}
\end{figure}

\begin{figure}
    \centering
    \begin{subfigure}[b]{0.23\textwidth}
        \includegraphics[width=\textwidth]{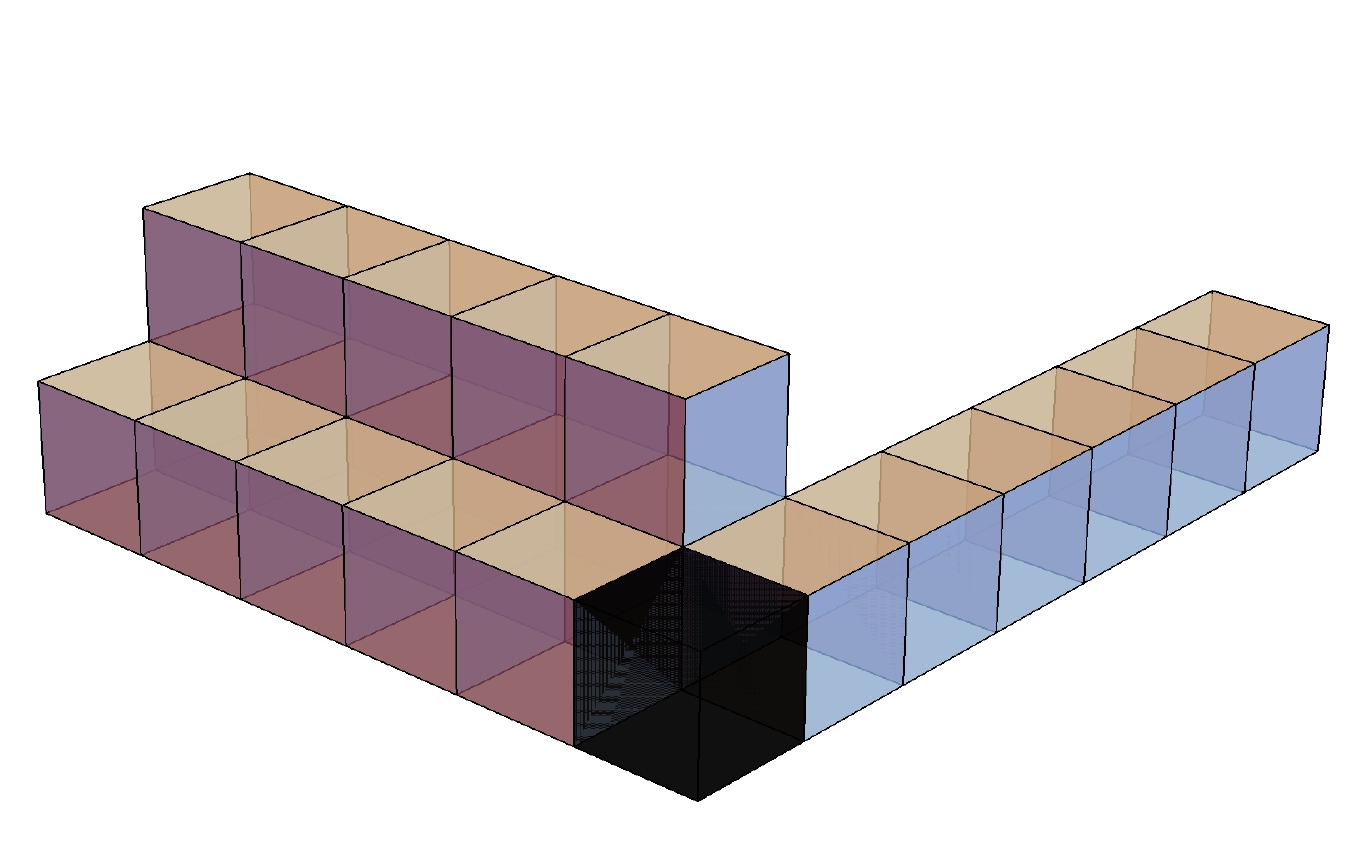}
    \end{subfigure}\quad
    \begin{subfigure}[b]{0.23\textwidth}
        \includegraphics[width=\textwidth]{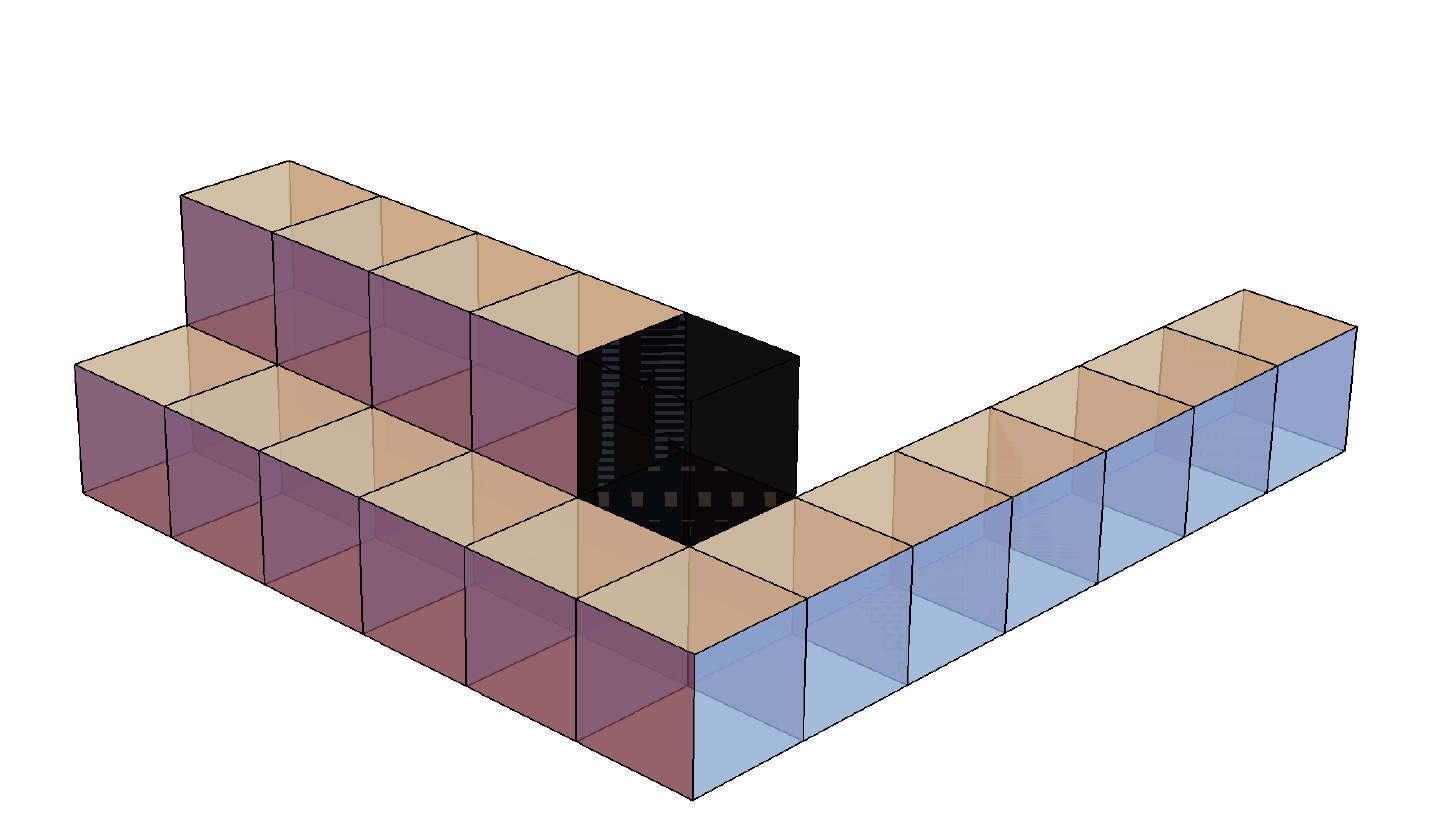}
    \end{subfigure}
 \caption{Two configurations with a single box in the $(\protect \yng(1),\protect \yng(1,2),0)$ example.}
\label{pic4}
\end{figure}

\begin{figure}
    \centering
    \begin{subfigure}[b]{0.23\textwidth}
        \includegraphics[width=\textwidth]{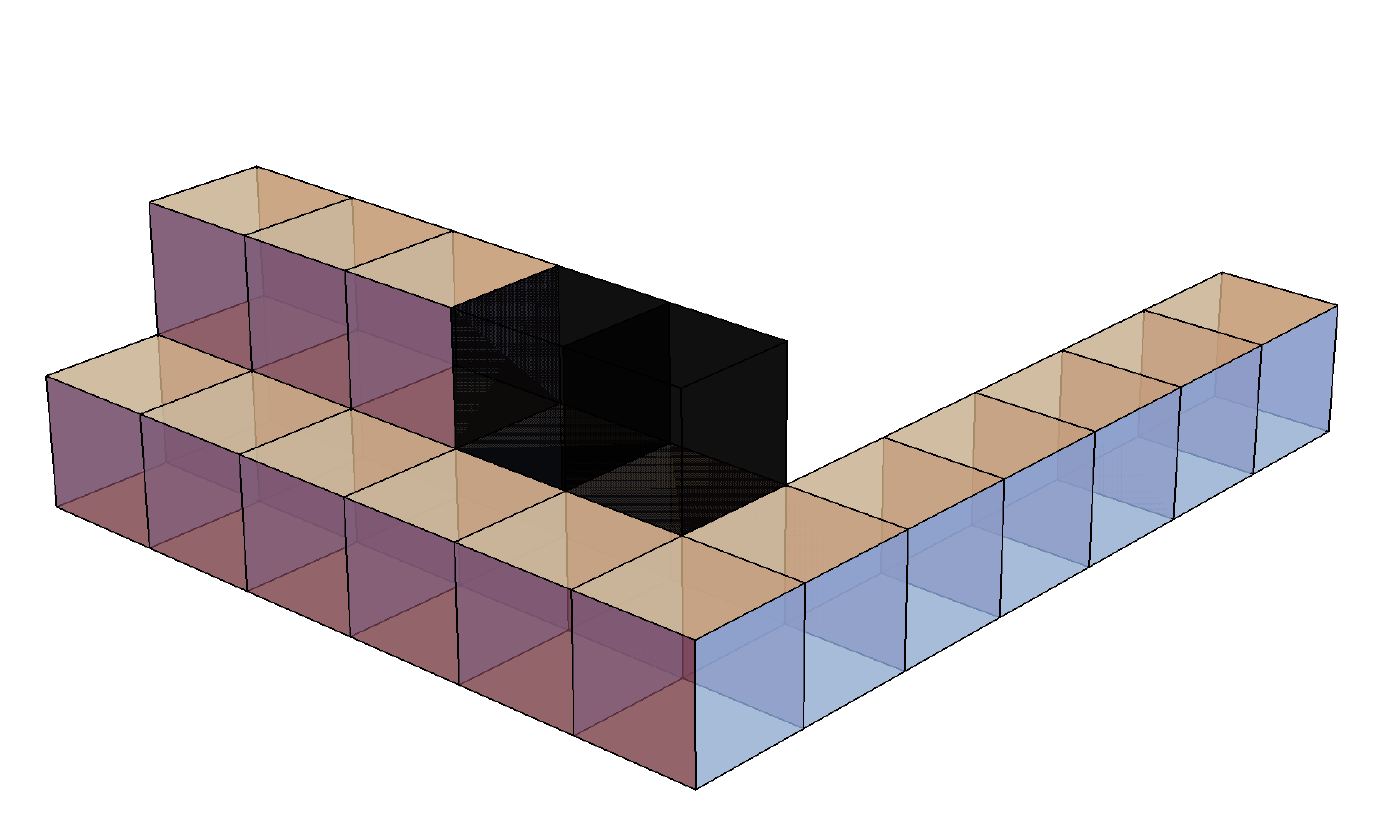}
    \end{subfigure}
    \begin{subfigure}[b]{0.23\textwidth}
        \includegraphics[width=\textwidth]{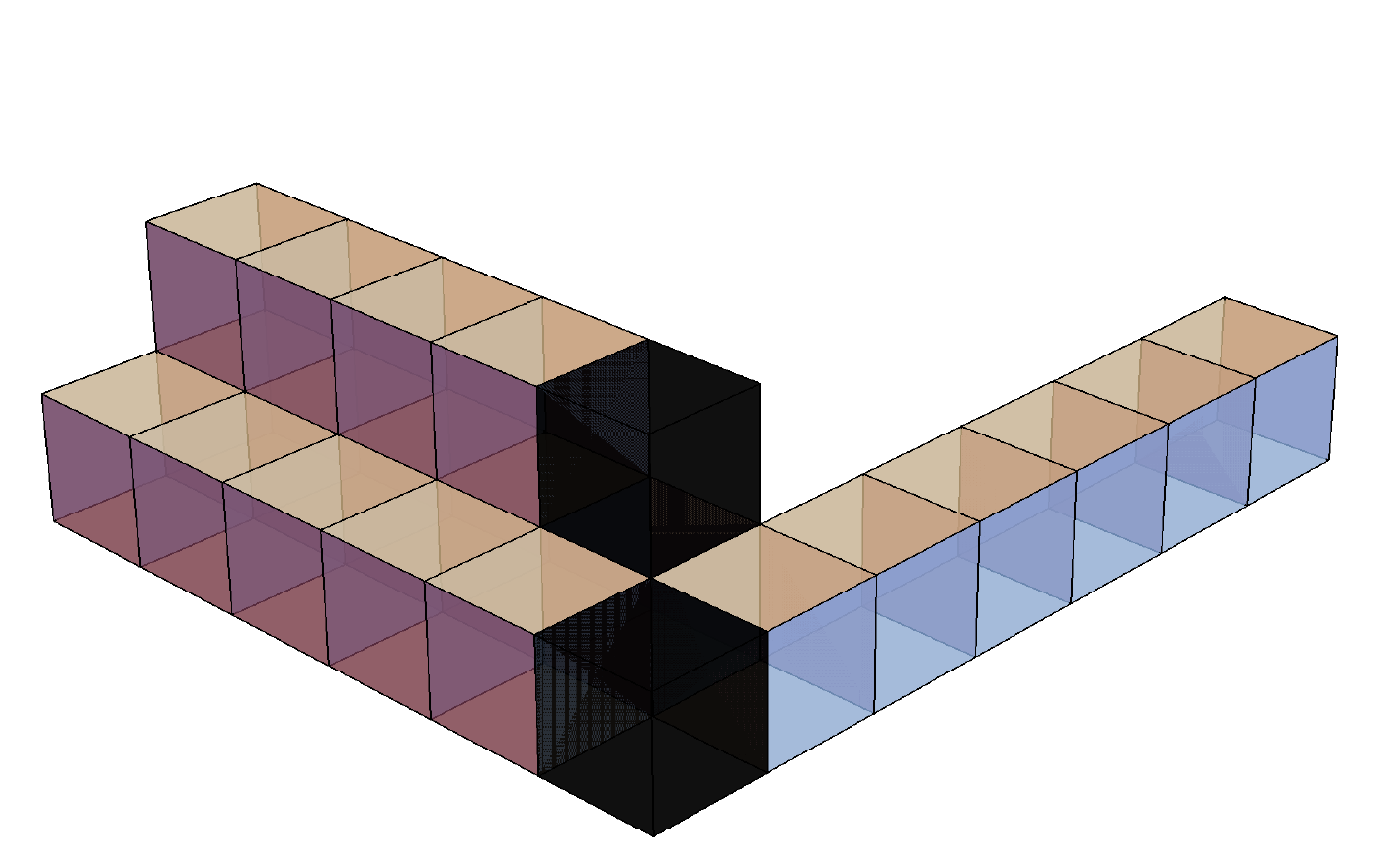}
    \end{subfigure}
    \begin{subfigure}[b]{0.23\textwidth}
        \includegraphics[width=\textwidth]{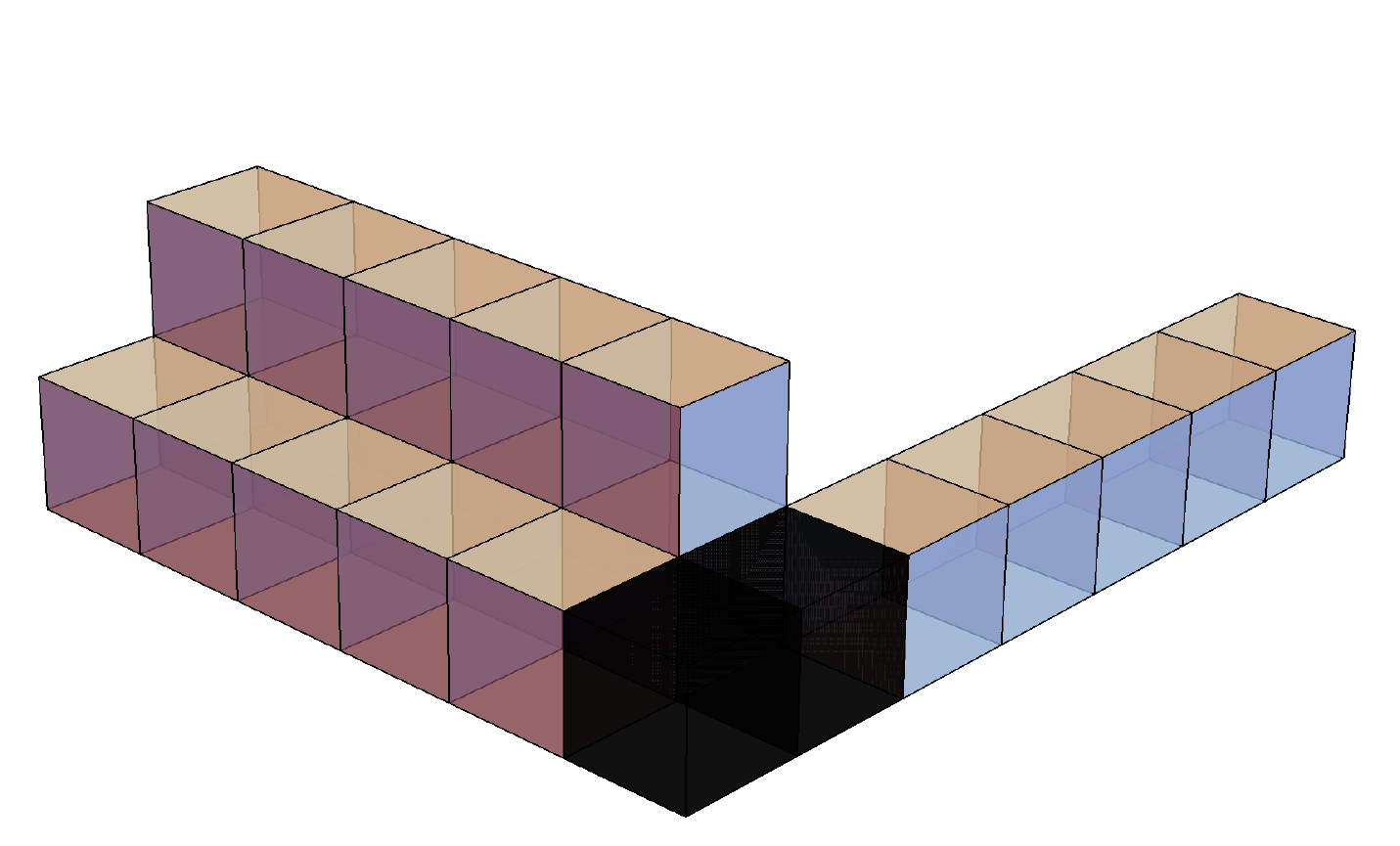}
    \end{subfigure}
    \begin{subfigure}[b]{0.23\textwidth}
        \includegraphics[width=\textwidth]{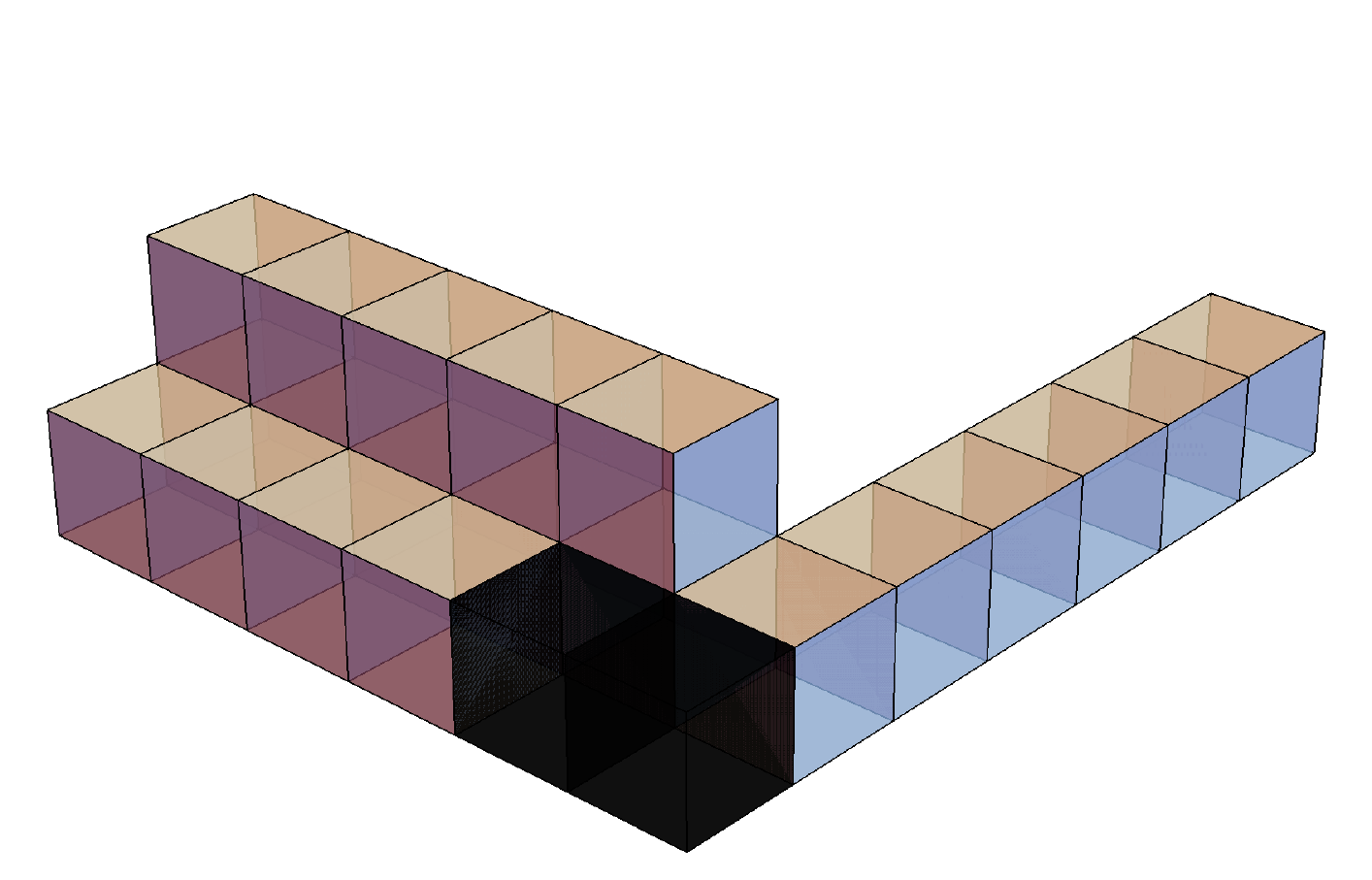}
    \end{subfigure}
 \caption{Four configurations with two boxes in the $(\protect \yng(1),\protect \yng(1,2),0)$ example.}
\label{pic6}
\end{figure}

Let us let us move to the slightly more complicated case case of $M_{\lambda,\mu,0}$. We are going to illustrate the box-counting on the example of $(\yng(1),\yng(1,2),0)$.  Our starting point is the vacuum configuration associated to the module $\mathcal{M}_{\lambda,\mu,0}$ depicted for our example in figure \ref{pic1}. All the other states of $\mathcal{M}_{\lambda,\mu,0}$ are associated to configurations of boxes that one can add on top of this vacuum configuration. As a first step towards the desired box counting of $M_{\lambda,\mu,0}$, let us extend both half-infinite cylinders of shapes $\lambda$ and $\mu$ in the negative direction along the $x_1$ and $x_2$ axes. Such a configuration for our example of $(\yng(1),\yng(1,2),0)$ is shown in figure \ref{pic2}. In the second step, let us remove all the boxes with coordinates $i,j,k\geq 0$ that do not live in the intersection of the two cylinders, i.e. boxes in $S_2$ in the notation above. Let us color those that lie in the intersection $S_2$ by a blue color. The corresponding configuration associated to our $(\yng(1),\yng(1,2),0)$ example is shown in figure \ref{pic3}. We can now think about the resulting diagram as a hollow structure inside which we can add boxes and start introducing boxes with the gravity pointing towards the positive direction $(1,1,1)$. The allowed configurations at level one and at level two are shown in figures \ref{pic4} and \ref{pic6}. Note that this box-counting is consistent with the proposal of the infinite wall of \cite{Gaberdiel:2018nbs} and its generalization from \cite{Harada:2018bkb}. As we are going to see next, the construction outlined here leads to a generalization to the general case of $M_{\lambda,\mu,\nu}$ modules.

\begin{figure}
    \centering
    \begin{subfigure}[b]{0.26\textwidth}
        \includegraphics[width=\textwidth]{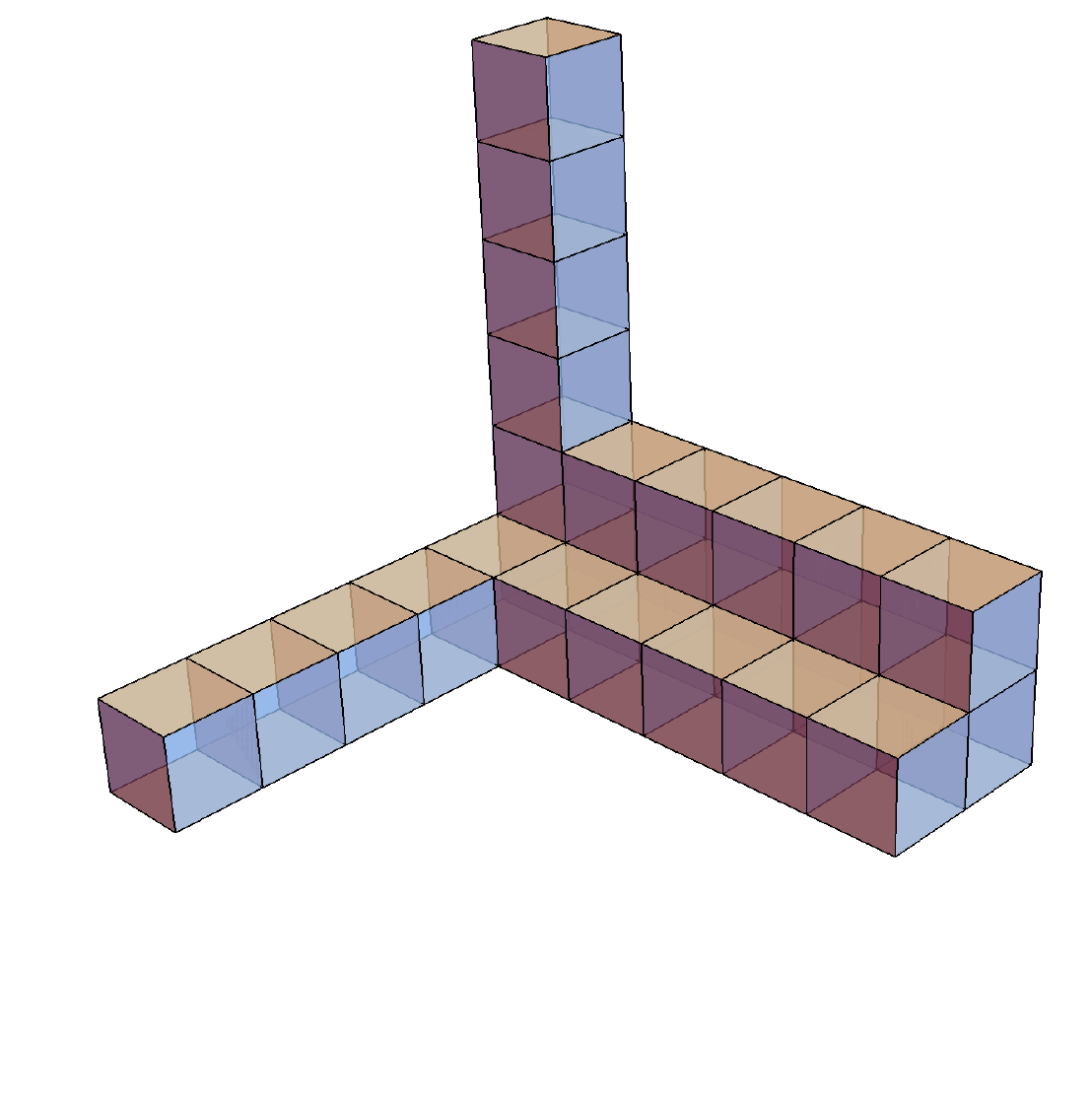}
        \caption{}
\label{pic10}
    \end{subfigure}
    \begin{subfigure}[b]{0.43\textwidth}
        \includegraphics[width=\textwidth]{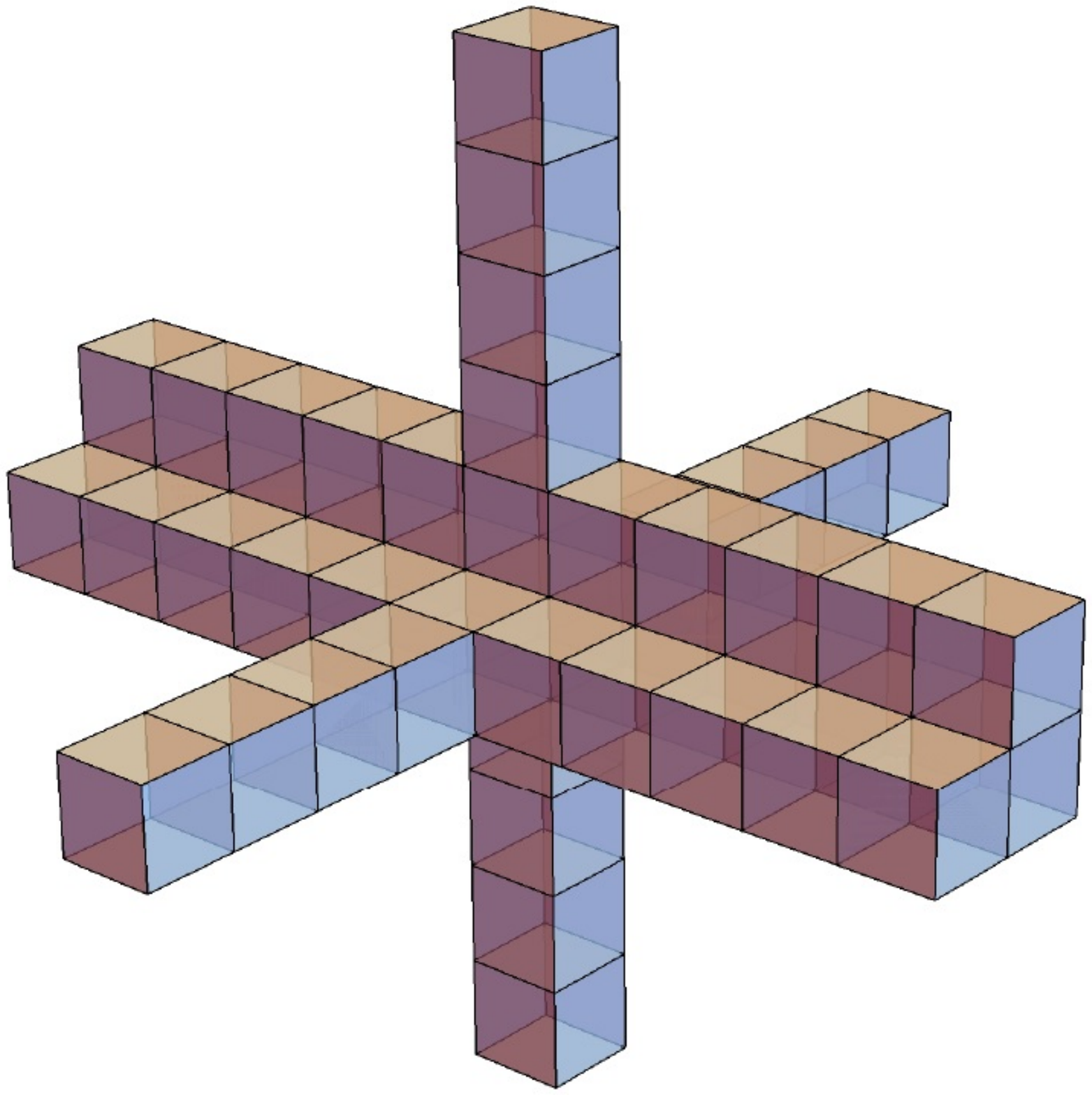}
        \caption{}
\label{pic11}
    \end{subfigure}
    \begin{subfigure}[b]{0.28\textwidth}
        \includegraphics[width=\textwidth]{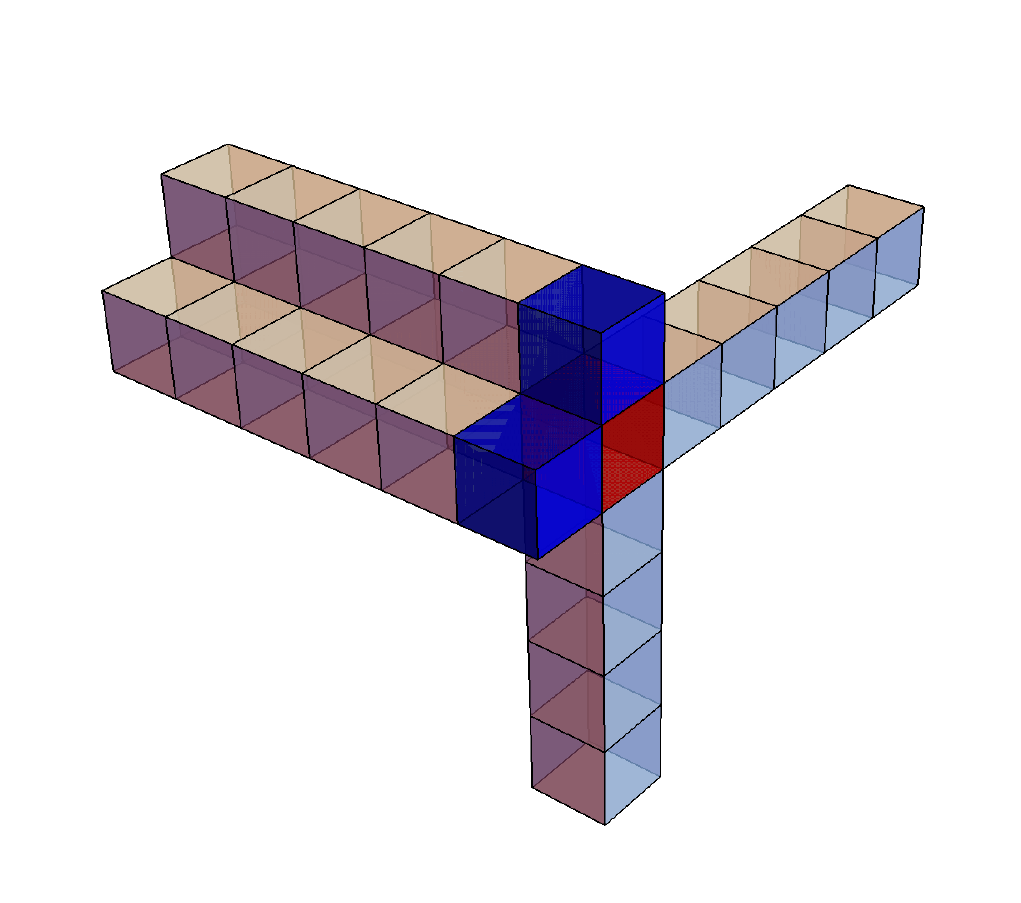}
        \caption{}
\label{pic12}
    \end{subfigure}
    \caption{Construction of the PT box-counting in the $(\protect \yng(1),\protect \yng(1,2),\protect \yng(1))$ example. (a) Starting configuration associated to 3d partitions with two non-trivial asymptotics. (b) A configuration with asymptotic cylinders made infinite in both directions. (c) A configuration with removed and colored boxes.}
\end{figure}

To find a generalization with all three asymptotics non-trivial $M_{\lambda,\mu,\nu}$, one has to introduce a couple of modifications associated to the boxes living in the intersection of all three cylinders, i.e. boxes in $S_3$. We are again going to illustrate the construction on the example of  $(\yng(1),\yng(1,2),\yng(1))$ with the standard vacuum configuration associated to $\mathcal{M}_{\lambda,\mu,\nu}$ depicted in \ref{pic10}. As above, the first step consists of extending the cylinders along the negative axes. This is illustrated for $(\yng(1),\yng(1,2),\yng(1))$  in figure \ref{pic11}.  Next, we remove the boxes in the positive quadrant that do not lie in the intersection of two or three cylinders. Boxes in $S_2$ that lie in the intersection of exactly two cylinders are labelled by the blue color as before and those in the intersection of all three cylinders, i.e. boxes in $S_3$, are labelled by the red color. The corresponding configuration for $(\yng(1),\yng(1,2),\yng(1))$ is shown in figure \ref{pic12}.

The boxes are again allowed to grow in the negative direction inside the hollow structure but new rules are necessary due to the presence of red boxes. In particular, one can associate three different kinds of boxes to the red position:
\begin{enumerate}
\item A \textbf{light box} that can support other light boxes.
\item A \textbf{heavy box} that can support other heavy boxes or boxes at uncolored locations, but allowed configurations are restricted by the rules below.
\item An \textbf{ultra-heavy box} that can support boxes in all directions. Compared to the previous two boxes, the ultra-heavy box contributes by weight two to the level.
\end{enumerate}
Let us now state the rules of adding boxes at red positions:
\begin{itemize}
\item All the boxes have to be supported from the positive direction except of the heavy box that does not have to be supported at blue locations that all lie in the intersection of two fixed cylinders along axes $i$ and $j$. Such a heavy box is labelled by direction $k$ such that $i\neq k\neq j$.
\item A heavy box can support heavy boxes and boxes at uncolored location if the resulting configuration admits a consistent labelling. In particular, all the heavy boxes can be labelled by direction 1, 2 or 3 in such a way that on top of a red box labelled by direction $i$, there can only be boxes at uncolored locations inside the cylinder along the axis $i$ or another heavy box labelled by direction $i$.
\item A family of adjacent red boxes can support light boxes if they admit at least two consistent labellings described in the previous point.
\end{itemize}
This box-counting is essentially equivalent to the one of \cite{Pandharipande:2007sq} that gives a more concise characterization of allowed configurations. Our ultra-heavy box is their unlabelled box and our light box and heavy box account for a possible degeneracy in the labelling. The distinction between the heavy and the light box makes the above rules slightly more complicated but it is going to play an essential role in the construction of the modules below.\footnote{Note also the observation of \cite{ptdt2,ptdt} who found an equivalent and more elegant characterization of PT configurations in terms of dimer models. It might be interesting to investigate whether modules constructed bellow have a more natural presentation in terms of dimer-model configurations, possibly along the lines of \cite{Li:2020rij}.}

Finally, let us list the situations under which a box at a red position can transform into a box of a different kind under the action of raising and lowering operators and which processes are not allowed:
\begin{itemize}
\item Both the light box and the heavy box can change into the ultra-heavy box or vice versa if the resulting configuration is allowed by the above rules.
\item When removing a box at uncolored location on top of the heavy box and the resulting configuration admits two or three consistent labellings, only the light box can be generated.
\item If an unsupported box is not supporting other boxes and becomes supported, only the light box is generated. In the opposite process, the heavy box can become unsupported but not the light one.
\item Whenever the resulting configuration contains heavy boxes at red locations admitting two or more consistent labellings (but only in situations distinct from the previous two points), both heavy boxes and light boxes can be generated, accounting for the corresponding degeneracy. 
\end{itemize}
The complicated-looking rules are illustrated on few examples in appendix \ref{app:box}. The rules listed above were determined by an analysis of the action of the conjugate embedding of $A$ on the highest-weight state of the well-known modules $\mathcal{M}_{\lambda,\mu,\nu}$ in the examples listed there. They are also consistent with the bootstrap analysis from the next section.  Though, at this point, they should still be considered conjectural. It is also likely that they admit a more concise formulation.

\subsection{Bootstraping anti-fundamental modules}

According to the above discussion, states of anti-fundamental modules $\mathcal{M}_{\bar{\lambda},\bar{\mu},\bar{\nu}}$ are labelled by a pair of a standard 3d partition and a PT box configuration. Let us now determine the action of generating functions $e(z),f(z),\psi (z)$ on states of $\mathcal{M}_{\bar{\lambda},\bar{\mu},\bar{\nu}}$ associated to such configurations. We are going to present a complete list of concrete rules that provide a definition of $\mathcal{M}_{\bar{\lambda},\bar{\mu},\bar{\nu}}$ in the presence of a single red box. We checked that such a proposal satisfies the relations of the affine Yangian in all allowed processes of creation and annihilation of boxes. After determining the  action in the presence of a single red box, we sketch its generalization to configurations containing more red boxes. A detailed analysis of processes allowing boxes at multiple red locations becomes rather lengthy and we leave it for future work.

Whenever we deal with boxes that are not at a red location or a location whose neighborhood contains a red location and whenever the resulting configuration does not contain a heavy box not supporting other boxes, the rules for adding and deleting box $a$ remain unchanged and we get the following terms in the action of $e(z)$ and $f(z)$:
\begin{eqnarray}\nonumber
e(z)|\Lambda\rangle &=&\sqrt{-\frac{\mbox{Res}_{z=x_a}\psi_{\Lambda}(z) }{\sigma_3}}\frac{1}{z-x_a}|\Lambda+a\rangle+\dots\\
f(z)|\Lambda\rangle &=&\sqrt{-\frac{\mbox{Res}_{z=x_a}\psi_{\Lambda}(z) }{\sigma_3}}\frac{1}{z-x_a}|\Lambda-a\rangle+\dots
\label{old}
\end{eqnarray}
The generating function $\psi (z)$ acting on the new state gets multiplied by the familiar factor of $\phi(z-x_a)$. The purpose of this section is to determine the action of $e(z),f(z),\psi (z)$ in processes that involve boxes at red locations or locations in their neighborhood. We are going to do that by writing an ansatz for the action of $e(z)$ and $f(z)$ and impose relations of the affine Yangian to solve for the coefficients of the ansatz. 

\subsubsection{Introducing a supported box at a red location}

Let us first find the rules for adding boxes at a supported red location.

Notice that the generating function $\psi(z)$ acting on a state associated to a configuration which allows two alternatives for the addition of a box at a red location contains a second-order pole. This can be already seen from the conjugation of (\ref{tribox}) for the simplest example of $\mathcal{M}_{\Box,\Box,\Box}$. In order to get a consistent action, generating functions $e(z)$ must also contain second-order poles when acting on the initial state. Analogously, $f(z)$ contains second-order poles in opposite processes. 

We label the allowed boxes at this position simply as $a_L$, $a_H$ and $a_U$ with $x_a=x_1\epsilon_1+x_2\epsilon_2+x_3\epsilon_3-\sigma_3\psi_0$. Let us write an ansatz for the terms in $e(z)$ that generate boxes  $a_L$ and $a_H$. We identify the state in front of the leading-order term with $|\Lambda+a_L\rangle$ and the one in front of the simple pole with $|\Lambda+a_H\rangle$, i.e.
\begin{eqnarray}
e(z)|\Lambda\rangle\sim \frac{E(\Lambda \rightarrow \Lambda+ a_L)}{(z-x_a)^2}|\Lambda+a_L\rangle+\frac{E(\Lambda\rightarrow \Lambda+a_H)}{z-x_a}|\Lambda+a_H\rangle
\end{eqnarray}
where $E(\Lambda\rightarrow \Lambda+a_L)$ and $E(\Lambda\rightarrow \Lambda+a_H)$ are the amplitudes for creating the light and the heavy box that need to be determined. From now on, the sign $\sim$ means that we are disregarding other possible terms on the right hand side that are not proportional to states written explicitly. 

The action of $\psi (z)$ on states $|\Lambda+a_L\rangle$ and $|\Lambda+a_H\rangle$ can be determined by employing the $\psi e$ relation. It can be easily verified that the solution is given by
\begin{eqnarray}\nonumber
\psi (z)|\Lambda+a_L\rangle&=&\phi (z-x_a)\psi_\Lambda(z)|\Lambda+a_L\rangle\\ \nonumber
\psi (z)|\Lambda+a_H\rangle&=&\phi (z-x_a)\psi_\Lambda(z) |\Lambda+a_H\rangle\\
&&-\frac{E(\Lambda \rightarrow \Lambda+ a_L)}{E(\Lambda \rightarrow \Lambda+ a_H)}\partial \phi (z-x_a) \psi_\Lambda (z) |\Lambda+a_L\rangle
\end{eqnarray}
where $\psi(z)|\Lambda\rangle =\psi_{\Lambda}(z)|\Lambda\rangle$. We can see that the presence of the second-order pole in the generating function $\psi (z)$ at some level leads to a non-diagonalizable action of $\psi (z)$ at higher levels. Boxes in the intersection of all three cylinders thus lead to nontrivial Jordan blocks for $\psi (z)$. Note also the appearance of the derivative of the function $\phi(z-x_a)$. This is responsible for introduction of new poles in $\psi(z)$ allowing a generation of configurations that would not be allowed otherwise. This also explains why more configurations can be created from $|\Lambda+a_H\rangle$ rather then $|\Lambda+a_L\rangle$ since the derivative factor is present only in the former case.

Let us now look at annihilation operators acting on states $|\Lambda+a_L\rangle$ and $|\Lambda+a_H\rangle$. Writing an ansatz of the form
\begin{eqnarray}\nonumber
f(z)|\Lambda+a_L\rangle &\sim&\frac{F(\Lambda + a_L \rightarrow \Lambda)}{z-x_a}|\Lambda\rangle \\ 
f(z)|\Lambda+a_H\rangle &\sim&\left (\frac{F^{(2)}(\Lambda+ a_H \rightarrow \Lambda)}{(z-x_a)^2}+\frac{F^{(1)}(\Lambda+a_H \rightarrow \Lambda)}{z-x_a}\right )|\Lambda\rangle
\end{eqnarray}
where amplitudes associated to the creation of the same state that appear as coefficients of singularity of different order are labelled by an index indicating the order. The $e f$ relation acting on the state $|\Lambda\rangle$ leads to conditions
\begin{eqnarray}\nonumber
E(\Lambda \rightarrow \Lambda + a_H)F^{(2)}(\Lambda+a_H \rightarrow \Lambda)&=&-\frac{1}{\sigma_3}\mbox{Res}_{z=x_a}(z-x_a)\psi_\Lambda (z)\\ \nonumber
E(\Lambda \rightarrow \Lambda + a_H)F^{(1)}(\Lambda+a_H \rightarrow \Lambda)&=&-\frac{1}{\sigma_3}\mbox{Res}_{z=x_a}\psi_\Lambda (z)\\
E(\Lambda \rightarrow \Lambda + a_L)F(\Lambda + a_L \rightarrow \Lambda)&=&-\frac{1}{\sigma_3}\mbox{Res}_{z=x_a}(z-x_a)\psi_\Lambda (z)
\end{eqnarray}
This can be easily checked by expanding the generating function $\psi_{\Lambda}(z)$ around $x_a$ and plugging into the $ef$ relation. It is also easy to check that the $\psi f$ relation is automatically satisfied. For later convenience, let us introduce the coefficients
\begin{eqnarray}
A_2=\mbox{Res}_{z=x_a}(z-x_a)\psi_\Lambda (z)\qquad A_1=\mbox{Res}_{z=x_a}\psi_\Lambda (z)
\end{eqnarray}
of the Laurent expansion of the generating function $\psi (z)$ acting on $|\Lambda\rangle$ at $z =x_a$.

We can now include the ultra-heavy box $|\Lambda+a_U\rangle$ in our discussion by writing an ansatz for the creation of the state
\begin{eqnarray}
e(z)|\Lambda+a_L\rangle &\sim& \frac{E(\Lambda+a_L\rightarrow \Lambda+a_U )}{z-x_a}|\Lambda+a_U\rangle \\ \nonumber
e(z)|\Lambda+a_H\rangle &\sim&\left ( \frac{E^{(2)}(\Lambda+a_H\rightarrow \Lambda+a_U )}{(z-x_a)^2}+\frac{E^{(1)}(\Lambda+a_H\rightarrow \Lambda+a_U )}{z-x_a}\right ) |\Lambda+a_U\rangle
\end{eqnarray}
Plugging into the $ee$ relation acting on the state $|\Lambda\rangle$ requires
\begin{eqnarray}\nonumber
E^{(2)}(\Lambda+a_H\rightarrow \Lambda+a_U )E(\Lambda \rightarrow \Lambda+a_H )&=&-E(\Lambda \rightarrow \Lambda+a_L )E(\Lambda+a_L\rightarrow \Lambda+a_U )\\ 
E^{(1)}(\Lambda+a_H\rightarrow \Lambda+a_U )&=&2\frac{\sigma_2}{\sigma_3}E^{(2)}(\Lambda +a_H \rightarrow \Lambda+a_U )
\end{eqnarray}

It is also easy to see that the proposal
\begin{eqnarray}
\psi (z)|\Lambda+a_U\rangle=\psi_{\Lambda}(z)\phi(z-x_a)^2|\Lambda+a_U\rangle
\end{eqnarray}
satisfies the $\psi e$ relation acting on $|\Lambda +a_H\rangle$ and $|\Lambda +a_L\rangle$ if we furthermore identify 
\begin{eqnarray}
E(\Lambda + a_L \rightarrow \Lambda + \Box_U) =-\frac{E(\Lambda \rightarrow \Lambda + a_H)}{E(\Lambda \rightarrow \Lambda + a_L)} E^{(2)}(\Lambda+a_H\rightarrow \Lambda+a_U )
\end{eqnarray}
We can see that the ultra-heavy box in a sense behaves as two boxes on top of each other. The last thing that needs to be determined is the action of $f(z)$ on $|\Lambda+a_U\rangle$. Writing the following ansatz
\begin{eqnarray}
f(z)|\Lambda+a_U\rangle &\sim&\frac{E(\Lambda+a_U\rightarrow \Lambda+a_H)}{z-x_a}|\Lambda+a_H\rangle\\ \nonumber
&&+\left ( \frac{E^{(2)}(\Lambda+a_U\rightarrow \Lambda+a_L)}{(z-x_a)^2}+\frac{E^{(1)}(\Lambda+a_U\rightarrow \Lambda+a_L)}{z-x_a}\right )|\Lambda+a_L\rangle
\end{eqnarray}
and plugging into the $ff$ relation, one gets constraints
\begin{eqnarray}
1&=&-\frac{F(\Lambda+a_L\rightarrow \Lambda )F^{(2)}(\Lambda+a_U \rightarrow \Lambda+a_L )}{F^{(2)}(\Lambda+a_H \rightarrow \Lambda )F(\Lambda+a_U \rightarrow \Lambda+a_H )}\\ \nonumber
1&=&-\frac{F(\Lambda+a_L\rightarrow \Lambda )F^{(1)}(\Lambda+a_U \rightarrow \Lambda+a_L )}{(F^{(1)}(\Lambda+a_H \rightarrow \Lambda  )+2\frac{\sigma_2}{\sigma_3}F^{(2)}(\Lambda+a_H \rightarrow \Lambda  ))F^{(1)}(\Lambda+a_U \rightarrow \Lambda+a_H )}
\end{eqnarray}
The only constraint coming from the $\psi f$ relation acting on the state $|\Lambda+a_U\rangle$ is
\begin{eqnarray}
F^{(2)}(\Lambda+a_U \rightarrow \Lambda+a_L)=-\frac{E(\Lambda \rightarrow \Lambda+a_L)}{E(\Lambda \rightarrow \Lambda+a_H)}F(\Lambda+a_U \rightarrow \Lambda+a_H)
\end{eqnarray}

Finally, action of the $ef$ relation on states $|\Lambda+a_L\rangle$ and $|\Lambda+a_H\rangle$ gives eight constraints
\begin{eqnarray}\nonumber
F^{(2)}(\Lambda+a_U \rightarrow \Lambda+a_L)E(\Lambda+a_L \rightarrow \Lambda+a_U)&=&\frac{A_2}{\sigma_3}\\\nonumber
F(\Lambda+a_L \rightarrow \Lambda)E(\Lambda \rightarrow \Lambda+a_L)&=&-\frac{A_2}{\sigma_3}\\ \nonumber
F^{(1)}(\Lambda+a_H \rightarrow \Lambda) E(\Lambda \rightarrow \Lambda+a_H)&=&-\frac{A_1}{\sigma_3}\\ \nonumber
F^{(2)}(\Lambda+a_H \rightarrow \Lambda)E(\Lambda \rightarrow \Lambda+a_H)&=&-\frac{A_2}{\sigma_3}\\ \nonumber
\frac{E^{(1)}(\Lambda +a_H\rightarrow \Lambda+a_U)E(\Lambda \rightarrow \Lambda+a_H)F(\Lambda +a_L\rightarrow \Lambda)}{E(\Lambda+a_L \rightarrow \Lambda+a_U)}&=&-2\frac{\sigma_2}{\sigma_3^2}A_2\\ \nonumber
\frac{E^{(2)}(\Lambda +a_H\rightarrow \Lambda+a_U)E(\Lambda \rightarrow \Lambda+a_H)F(\Lambda +a_L\rightarrow \Lambda)}{E(\Lambda+a_L \rightarrow \Lambda+a_U)}&=&\frac{A_2}{\sigma_3}\\ \nonumber
F^{(1)}(\Lambda+a_U \rightarrow \Lambda+a_L)E(\Lambda+a_L \rightarrow \Lambda+a_U)&=&\frac{\sigma_3A_1+2\sigma_2A_2}{\sigma_3^2}\\ 
\frac{F(\Lambda+a_L \rightarrow \Lambda)E(\Lambda \rightarrow \Lambda+a_H)}{E(\Lambda+a_L \rightarrow \Lambda+a_U)F(\Lambda+a_U \rightarrow \Lambda+a_H)}&=&1
\end{eqnarray}
The above-written constraints are mutually dependent. In total, we get a system of 8 independent constraints for 11 amplitudes. The remaining 3 coefficients are much harder to fix since they can be modified by renormalizing states $|\Lambda +a_L\rangle,|\Lambda +a_H\rangle,|\Lambda +a_U\rangle$. In \cite{Prochazka:2015deb}, this redundancy was fixed by requiring $F(\Lambda+a \rightarrow \Lambda)= E(\Lambda \rightarrow \Lambda+a)$ and checking in examples that the proposal is consistent. Imposing
\begin{eqnarray}\nonumber
F^{(1)}(\Lambda+a_H \rightarrow \Lambda)&=& E(\Lambda \rightarrow \Lambda+a_H)\\ \nonumber
F(\Lambda+a_L \rightarrow \Lambda)&=& E(\Lambda \rightarrow \Lambda+a_L)\\ 
F^{(2)}(\Lambda+a_U \rightarrow \Lambda+a_L )&=& E(\Lambda+a_L \rightarrow \Lambda+a_U)
\end{eqnarray}
in our refined case leads to the final proposal
\begin{eqnarray} \nonumber
e(z)|\Lambda\rangle&\sim&\sqrt{-\frac{A_2}{\sigma_3}}\frac{1}{(z-x_a)^2}|\Lambda+a_L\rangle+\sqrt{-\frac{A_1}{\sigma_3}}\frac{1}{z-x_a}|\Lambda+a_H\rangle\\ \nonumber
e(z)|\Lambda+a_L\rangle &\sim& \sqrt{\frac{A_2}{\sigma_3}}\frac{1}{z-x_a}|\Lambda+a_U\rangle \\ \nonumber
e(z)|\Lambda+a_H\rangle &\sim&\sqrt{\frac{A_2}{A_1}}\left (-\sqrt{-\frac{A_2}{\sigma_3}} \frac{1}{(z-x_a)^2}-2\frac{\sigma_2}{\sigma_3}\sqrt{\frac{A_2}{\sigma_3}}\frac{1}{z-x_a}\right ) |\Lambda+a_U\rangle\\
f(z)|\Lambda+a_L\rangle &\sim&\sqrt{-\frac{A_2}{\sigma_3}}\frac{1}{z-x_a}|\Lambda\rangle \\  \nonumber
f(z)|\Lambda+a_H\rangle &\sim&\left (\sqrt{\frac{A_2}{A_1}}\sqrt{-\frac{A_2}{\sigma_3}}\frac{1}{(z-x_a)^2}+\sqrt{-\frac{A_1}{\sigma_3}}\frac{1}{z-x_a}\right )|\Lambda\rangle \\ \nonumber
f(z)|\Lambda+a_U\rangle &\sim&-\sqrt{\frac{A_1}{\sigma_3}}\frac{1}{z-x_a}|\Lambda+a_H\rangle\\ \nonumber
&&+\left (\sqrt{\frac{A_2}{\sigma_3}} \frac{1}{(z-x_a)^2}+\left (\sqrt{\frac{A_1}{A_2}}\sqrt{-\frac{A_1}{\sigma_3}}+2\frac{\sigma_2}{\sigma_3}\sqrt{\frac{A_2}{\sigma_3}}\right )\frac{1}{z-x_a}\right )|\Lambda+a_L\rangle
\label{final}
\end{eqnarray}
This proposal satisfies all the above conditions and seems to be consistent in all the examples we have encountered. Note also that setting $A_2=0$, the expressions simplify considerably and admit a consistent restriction to $|\Lambda+a_L\rangle=|\Lambda+a_U\rangle=0$ leading to the standard proposal for the action of the creation and annihilation operators. The action of $\psi(z)$ is then given by
\begin{eqnarray}\nonumber
\psi(z)|\Lambda+a_L\rangle&=& \phi(z-x_a)\psi_{\Lambda}(z)|\Lambda+a_L\rangle\\ \nonumber
\psi(z)|\Lambda+a_H\rangle&=& \phi(z-x_a)\psi_{\Lambda}(z)|\Lambda+a_H\rangle-\sqrt{\frac{A_2}{A_1}}\partial\phi(z-x_a)\psi_{\Lambda}(z)|\Lambda+a_L\rangle\\
\psi(z)|\Lambda+a_U\rangle&=& \phi(z-x_a)^2\psi_{\Lambda}(z)|\Lambda+a_U\rangle
\end{eqnarray}

\subsubsection{Adding boxes in the presence of a heavy box}

Let us now determine how to add boxes away from the neighborhood of the red location if a red location is occupied by a supported box not supporting other boxes. If the red box supports other boxes or it is unsupported, the rules are of the standard form. 

If a red location supports an ultra-heavy box, the action of $\psi(z)$ is of the standard form and the addition and removal of boxes at other locations (that are not red) are governed by the same rule (\ref{old}). If the red location supports a box admitting two or three consistent labellings, it can support either light or heavy box. If the red location supports a light box, the action of $\psi(z)$ is of the standard form and the addition and removal of boxes at other locations (that are not red) are governed by the same rule (\ref{old}).  The situation becomes more complicated when the heavy box occupies the red location since addition and removal of boxes at location $x_{\tilde{a}}$ allows also the change of the heavy box into the light one. Since the generating function $\psi (z)$ acting on the initial state does not contain second-order poles at position $z=x_{\tilde{a}}$, we do not expect their presence even in the action of $e(z),f(z)$. We can thus write an ansatz
\begin{eqnarray}\nonumber
e(z)|\Lambda+a_H\rangle &\sim& \frac{E(\Lambda+a_H\rightarrow \Lambda+a_H+\tilde{a})}{z-x_{\tilde{a}}}|\Lambda+a_H+\tilde{a}\rangle\\ \nonumber
&&+ \frac{E(\Lambda+a_H\rightarrow \Lambda+a_L+\tilde{a})}{z-x_{\tilde{a}}}|\Lambda+a_L+\tilde{a}\rangle\\ \nonumber
f(z)|\Lambda+a_H+\tilde{a}\rangle &\sim& \frac{F(\Lambda+a_H+\tilde{a}\rightarrow \Lambda+a_H)}{z-x_{\tilde{a}}}|\Lambda+a_H\rangle\\
&&+ \frac{F(\Lambda+a_H+\tilde{a}\rightarrow \Lambda+a_L)}{z-x_{\tilde{a}}}|\Lambda+a_L\rangle
\end{eqnarray} 
Analogously to the detailed analysis in the previous section, the $ef$ relation together with an extra condition $E(\Lambda+a_H\rightarrow \Lambda+a_L+\tilde{a})=F(\Lambda+a_H+\tilde{a}\rightarrow \Lambda+a_L)$ fixes an explicit form of the action of the form
\begin{eqnarray}\nonumber
e(z)|\Lambda+a_H\rangle &\sim&\sqrt{-\frac{\mbox{Res}_{z=x_{\tilde{a}}}\phi(z-x_a)\psi_{\Lambda}(z)}{\sigma_3}} \frac{1}{z-x_{\tilde{a}}}|\Lambda+a_H+\tilde{a}\rangle\\ \nonumber
&&+ \sqrt{\frac{\mbox{Res}_{z=x_{\tilde{a}}}\sqrt{\frac{A_2}{A_1}}\partial\phi(z-x_a)\psi_{\Lambda}(z)}{\sigma_3}}\frac{1}{z-x_{\tilde{a}}}|\Lambda+a_L+\tilde{a}\rangle\\ \nonumber
f(z)|\Lambda+a_H+\tilde{a}\rangle &\sim& \sqrt{-\frac{\mbox{Res}_{z=x_{\tilde{a}}}\phi(z-x_a)\psi_{\Lambda}(z)}{\sigma_3}}\frac{1}{z-x_{\tilde{a}}}|\Lambda+a_H\rangle\\
&&+ \sqrt{\frac{\mbox{Res}_{z=x_{\tilde{a}}}\sqrt{\frac{A_2}{A_1}}\partial\phi(z-x_a)\psi_{\Lambda}(z)}{\sigma_3}}\frac{1}{z-x_{\tilde{a}}}|\Lambda+a_L\rangle
\end{eqnarray} 
One can check that this proposal satisfies other relations of the affine Yangian as well. An important detail to note is that the change of the heavy box into the light box is governed by the derivative $\partial \phi (z-x_a)$ responsible for the off-diagonal term in the action of $\psi(z)$.

\subsubsection{Adding boxes on top of the heavy and light box}

After determining the rules for creating and annihilating light, heavy and ultra-heavy boxes at a supported red location and rules for creating and annihilating boxes in their presence (but not on top of them), let us determine how to add boxes on top of the supported light and heavy box. Addition of boxes on top of the ultra-heavy box is governed by the standard rules. 

First, the creation of a box at an uncolored location is not allowed on top of the light supported box. The addition on top of the heavy box is governed by a simple-pole rule
\begin{eqnarray}
e(z)|\Lambda+a_H\rangle \sim\frac{E(\Lambda+a_H\rightarrow \Lambda+a_H+\tilde{a})}{z-x_{\tilde{a}}}|\Lambda+a_H+\tilde{a}\rangle 
\end{eqnarray}
where $\tilde{a}$ is the newly added box. In the opposite process of annihilating $\tilde{a}$, only $|\Lambda+a_L\rangle$ can be created and we write
\begin{eqnarray}
f(z)|\Lambda+a_H+\tilde{a}\rangle \sim\frac{F(\Lambda+a_H+\tilde{a}\rightarrow \Lambda+a_L)}{z-x_{\tilde{a}}}|\Lambda+a_L\rangle
\end{eqnarray}
It is easy to check that considering higher-order poles in these expressions would spoil relations of the affine Yangian. Let the generating function $\psi (z)$ acting on $|\Lambda \rangle$ have the following expansion \footnote{Note that it is generally not singular since the presence of the box $a$ is necessary for addition of the box $\tilde{a}$.} around $x_{\tilde{a}}$:
\begin{eqnarray}
\psi(z)|\Lambda \rangle = (A_0+ A_{-1}(z-x_{\tilde{a}})+A_{-2}(z-x_{\tilde{a}})^2+\dots)|\Lambda \rangle
\end{eqnarray}
From all the examples we have encountered, the leading coefficient in this expansion vanishes $A_0=0$, which implies that the function $\phi (z-x_{\tilde{a}})\psi(z)$ has no pole at $x_{\tilde{a}}$, justifying impossibility of creation of $|\Lambda+a_H+\tilde{a}\rangle $ from $|\Lambda+a_L\rangle$. Plugging into the $ef$ relation acting on the state $|\Lambda+a_H\rangle$, we get the following condition for the newly introduced amplitudes
\begin{eqnarray}\nonumber
&&E(\Lambda+a_H\rightarrow \Lambda+a_H+\tilde{a})F(\Lambda+a_H+\tilde{a}\rightarrow \Lambda+a_L)\\
&&=-\frac{1}{\sigma_3}\frac{E(\Lambda\rightarrow \Lambda+a_L)}{E(\Lambda\rightarrow \Lambda+a_H)}\mbox{Res}_{z=x_{\tilde{a}}}\partial \phi(z-x_{\tilde{a}})\psi_{\Lambda} (z)
\end{eqnarray}
In particular, the product is again governed by the poles of $\psi(z)$. Since $\phi(z-x_{\tilde{a}})\psi_{\Lambda} (z)$ does not have any pole at $x_{\tilde{a}}$ by itself, it is the derivative $\partial \phi(z-x_{\tilde{a}})\psi_{\Lambda} (z)$ that plays the crucial role. Imposing furthermore
\begin{eqnarray}
E(\Lambda+a_H\rightarrow \Lambda+a_H+\tilde{a})=F(\Lambda+a_H+\tilde{a}\rightarrow \Lambda+a_L)
\end{eqnarray}
leads to the final proposal for the rule of adding a box at uncolored location on top of the heavy box
\begin{eqnarray}\nonumber
e(z)|\Lambda+a_H\rangle &\sim&\frac{\sqrt{-\frac{1}{\sigma_3}\mbox{Res}_{z=x_{\tilde{a}}}\sqrt{\frac{A_2}{A_1}}\partial \phi(z-x_{\tilde{a}})\psi_{\Lambda} (z)]}}{z-x_{\tilde{a}}}|\Lambda+a_H+\tilde{a}\rangle \\
f(z)|\Lambda+a_H+\tilde{a}\rangle &\sim&\frac{\sqrt{-\frac{1}{\sigma_3}\mbox{Res}_{z=x_{\tilde{a}}}\sqrt{\frac{A_2}{A_1}}\partial \phi(z-x_{\tilde{a}})\psi_{\Lambda} (z)]}}{z-x_{\tilde{a}}}|\Lambda+a_L\rangle
\end{eqnarray}
We can see that the addition and removal of the boxes is again governed by the derivative term $\partial \phi(z-x_{\tilde{a}})\psi_{\Lambda} (z)$.

\subsubsection{Generating ultra-heavy box supporting other boxes}

Whenever we have boxes on top of the heavy box, we can change the heavy box to the ultra-heavy one and vice versa. This amplitude is governed by the standard rules with no modifications necessary.

\subsubsection{Transitions involving unsupported boxes}

Finally, an unsupported box can become supported and vice versa. Here, we have to distinguish two cases depending on weather the unsupported box supports other boxes or it does not. In the former case, the generating function $\psi(z)$ contains a single pole and the creation and the annihilation is governed by the standard rule. In the later case, the degeneration of the space associated to the red box makes the situation slightly more complicated and one can write the following ansatz
\begin{eqnarray}\nonumber
e(z)|\Lambda+ a_H \rangle \sim \frac{E(\Lambda+ a_L \rightarrow \Lambda+ \tilde{a}+ a_L)}{z-\epsilon_{\tilde{a}}}|\Lambda+ \tilde{a}+ a_L \rangle\\
f(z)|\Lambda+ \tilde{a}+ a_H\rangle \sim \frac{F(\Lambda+ \tilde{a}+ a_H \rightarrow  \Lambda+ a_L)}{z-\epsilon_{\tilde{a}}}|\Lambda+ a_H \rangle
\end{eqnarray} 
No second-order poles are expected since $\psi(z)$ itself does not contain higher-order poles in present situations. Note that the state $|\Lambda+ \tilde{a}+  a_H \rangle$ is not generated by creation operators and that the unsupported box $|\Lambda+ a_L \rangle$ cannot be generated from $|\Lambda+ \tilde{a}+ a_L \rangle$ by annihilation operators. This is consistent with the vanishing of the pole of $\psi (z)$ at the location where we are adding/removing a box. The $ef$ relation imposes conditions
\begin{eqnarray}\nonumber
\frac{E(\Lambda+ a_L \rightarrow \Lambda+ \tilde{a}+ a_L)}{E(\Lambda \rightarrow \Lambda+ \tilde{a})}&=&\frac{F(\Lambda+ a_L \rightarrow \Lambda)}{F(\Lambda+ \tilde{a}+ a_L \rightarrow \Lambda+ \tilde{a})}\\
\frac{E(\Lambda+ \tilde{a} \rightarrow \Lambda+ \tilde{a}+a_H)}{E(\Lambda\rightarrow a_L)}&=&\frac{F(\Lambda+ \tilde{a} \rightarrow \Lambda)}{F(\Lambda+ \tilde{a}+ a_H \rightarrow  \Lambda+ a_L)}
\end{eqnarray}
that allows a unique determination of the amplitudes in terms of the above.

\subsubsection{Adding boxes at multiple red locations on top of each other}

In the previous subsections, we have explicitly determined the action of $e(z),f(z),\psi(z)$ for anti-fundamental modules $\mathcal{M}_{\bar{\lambda},\bar{\mu},\bar{\nu}}$ containing a single red box. We have seen that $\psi(z)$ develops a Jordan block of size $2\times 2$ with the off-diagonal element proportional to the square-root of $\partial \phi (z-x_a)$, where $x_a$ is the position of the red box. In the presence of light and heavy boxes, the addition of boxes is governed by the position of poles of both $\phi (z-x_a)\psi_{\Lambda}(z)$ and $\partial \phi (z-x_a)\psi_{\Lambda}$, leading to a module with a quite non-trivial structure. The analysis in the presence of more red boxes becomes even more complex but we do not expect the rules to change in any significant way. In the next two subsections, we sketch how to deal with such situations. We start here by adding a box at a red location on top of the heavy and light box associated to a degenerate action of $\psi(z)$ such that the final configuration also admits two or three possible labellings. The resulting configuration contains a set of adjacent boxes at red locations that are all either light or heavy, leading again to a two-fold degeneracy in the action of $\psi(z)$.

Combining proposals from previous sections, we can write the following ansatz for the action of the creation operator
\begin{eqnarray}\nonumber
e(z)|\Lambda+a_L\rangle &\sim&\frac{E(\Lambda+a_L\rightarrow \Lambda+a_L+\tilde{a}_L )}{z-x_{\tilde{a}}}|\Lambda+a_L+\tilde{a}_L\rangle \\  \nonumber
e(z)|\Lambda+a_H\rangle &\sim&\frac{E(\Lambda+a_H \rightarrow \Lambda+a_L+\tilde{a}_L)}{(z-x_{\tilde{a}})^2}|\Lambda+a_L+\tilde{a}_L\rangle\\
&& +\frac{E(\Lambda+a_H \rightarrow \Lambda+a_H+\tilde{a}_H )}{z-x_{\tilde{a}}}|\Lambda+a_H+a_H\rangle 
\end{eqnarray} 
In particular, we can add only the light box on top of the heavy box whereas both possible configurations result from adding a box on top of the heavy box. The removal of boxes is then governed by
\begin{eqnarray}\nonumber
f(z)|\Lambda+a_L+\tilde{a}_L\rangle &\sim&\frac{F(\Lambda+a_L+\tilde{a}_L \rightarrow \Lambda+a_L)}{z-x_{\tilde{a}}}|\Lambda+a_L\rangle \\ \nonumber
f(z)|\Lambda+a_H+\tilde{a}_H\rangle &\sim& \frac{F(\Lambda+a_H+\tilde{a}_H\rightarrow \Lambda+a_H)}{z-x_{\tilde{a}}}|\Lambda+a_H\rangle\\
&&+\frac{F(\Lambda+a_H+\tilde{a}_H\rightarrow \Lambda+a_L)}{(z-x_{\tilde{a}})^2}|\Lambda+a_L\rangle
\end{eqnarray} 
The $\psi e$ relation requires the action of $\psi (z)$ to be of the following form
\begin{eqnarray}\nonumber
\psi(z)|\Lambda+a_L+\tilde{a}_L\rangle &=& \psi_{\Lambda}(z)\phi(z-x_{a})\phi(z-x_{\tilde{a}})  |\Lambda+a_L+\tilde{a}_L\rangle  \\ \nonumber
\psi(z)|\Lambda+a_H+\tilde{a}_H\rangle &=&  \psi_{\Lambda}(z)\phi(z-x_{a})\phi(z-x_{\tilde{a}})|\Lambda+a_H+\tilde{a}_H\rangle \\ \nonumber
&&-\bigg (  \frac{E(\Lambda+a_H \rightarrow \Lambda+a_H+\tilde{a}_L)}{E(\Lambda+a_H \rightarrow \Lambda+a_H+\tilde{a}_H)}\frac{\partial \phi (z-x_{\tilde{a}})}{\phi (z-x_{\tilde{a}})}\\ \nonumber
&&\ \ +\frac{E(\Lambda+a_L \rightarrow \Lambda+a_H+\tilde{a}_L) E(\Lambda \rightarrow \Lambda+a_L)}{E(\Lambda+a_H \rightarrow \Lambda+a_H+\tilde{a}_H)E(\Lambda\rightarrow \Lambda+a_H)}\frac{\partial \phi (z-x_{a})}{\phi (z-x_{a})}\bigg )\\ 
&& \times  \psi_{\Lambda}(z)\phi(z-x_{a})\phi(z-x_{\tilde{a}})\psi_{\Lambda}(z) |\Lambda+a_L+\tilde{a}_L\rangle
\label{ontop}
\end{eqnarray}
so the derivative of both $\phi (z-x_{a})$ and $\phi (z-x_{\tilde{a}})$ contribute to the off-diagonal term of the $2\times 2$ Jordan block. Generally, we expect derivative terms of all the light boxes to contribute if we stack more light boxes at red locations on top of each other. The coefficients in the above action of $e(z)$ adn $f(z)$ can be fixed analogously to the previous discussion and we leave the detailed analysis to reader. 

We expect that the lowest-lying box in the block of light and heavy boxes can transform into the ultra-heavy box. The block of light boxes should behave as a single light or heavy box when adding boxes at a different location with all the derivative  terms present in the action of $\psi(z)$ now contribution to the amplitude. Similarly, adding boxes on top of the block of heavy and light boxes is governed by analogous rules to a single light and heavy box with all the derivatives contributing. Transitions involving unsupported boxes should again remain the same with the block of light and heavy boxes behaving as a single light and heavy box.

\subsubsection{Supported boxes at multiple red location}

The above discussion should exhaust all the situation containing a single block of adjacent heavy and light boxes. We have not checked all the relations of the affine Yangian and we have not fixed explicitly all the coefficients in the ansatz for $e(z)$ and $f(z)$ in the presence of two or more boxes in the block block associated to a two-fold degeneration. We expect this to be straightforward though technically challenging and we leave it to the reader. Let us now briefly look at the generalization when more blocks of light and heavy boxes can be generated.

We start with a situation in the presence of a single heavy or light box at a supported red location $|\Lambda+a_L\rangle$  and $|\Lambda+a_H\rangle$. We can write the following ansatz for the addition of boxes at another supported red location that does not lie in the neighborhood of the initial box
\begin{eqnarray}\nonumber
e(z)|\Lambda+a_L\rangle &\sim& \frac{E(\Lambda+a_L\rightarrow \Lambda+a_L+\tilde{a}_L)}{(z-x_{\tilde{a}})^2}|\Lambda+a_L+\tilde{a}_L\rangle\\ \nonumber
&&+\frac{E(\Lambda+a_L\rightarrow \Lambda+a_L+\tilde{a}_H)}{z-x_{\tilde{a}}}|\Lambda+a_L+\tilde{a}_H\rangle\\ \nonumber
e(z)|\Lambda+a_H\rangle &\sim& \frac{E(\Lambda+a_H\rightarrow \Lambda+a_H+\tilde{a}_L)}{(z-x_{\tilde{a}})^2}|\Lambda+a_H+\tilde{a}_L\rangle\\ \nonumber
&&+\frac{E(\Lambda+a_H\rightarrow \Lambda+a_H+\tilde{a}_H)}{z-x_{\tilde{a}}}|\Lambda+a_H+\tilde{a}_H\rangle\\ \nonumber
&&+\frac{E(\Lambda+a_H\rightarrow \Lambda+a_L+\tilde{a}_L)}{(z-x_{\tilde{a}})^2}|\Lambda+a_L+\tilde{a}_L\rangle\\
&&+\frac{E(\Lambda+a_H\rightarrow \Lambda+a_L+\tilde{a}_H)}{z-x_{\tilde{a}}}|\Lambda+a_L+\tilde{a}_H\rangle
\end{eqnarray}
Note that we need to allow the original heavy box to change into the light box in order the relations of the affine Yangian to be satisfied. The corresponding amplitudes are expected to be proportional to the square root of the first-order and the second-order pole of $\psi_{\Lambda}(z)\phi(z-x_{a})$ at position $x_{\tilde{a}}$ for transitions not changing $a_H$ into $a_L$ and the first-order and the second-order pole of $\psi_{\Lambda}(z)\partial \phi(z-x_{a})$ at position $x_{\tilde{a}}$ for transitions that do change $a_H$ into $a_L$. 

Since each of the red locations now allow a two-fold degeneracy, the action of $\psi(z)$ contains a 4d Jordan blocks. In particular, we expect it to act on the vector
\begin{eqnarray}
|\Lambda+a+\tilde{a}\rangle = 
\begin{pmatrix}
|\Lambda+a_L+\tilde{a}_L\rangle\\
|\Lambda+a_L+\tilde{a}_H\rangle\\
|\Lambda+a_H+\tilde{a}_L\rangle\\
|\Lambda+a_H+\tilde{a}_H\rangle
\end{pmatrix}
\end{eqnarray}
as
\begin{eqnarray}
\psi(z)|\Lambda+a+\tilde{a}\rangle =
\psi_{\Lambda} (z)
\begin{pmatrix}
\phi\tilde{\phi}&-\tilde{E}\phi\partial \tilde{\phi}&-E\partial  \phi\tilde{\phi} & -E\tilde{E} \partial\phi \partial \tilde{\phi}\\
0 &\phi\tilde{\phi} & 0 &-E\partial \phi\tilde{\phi}\\
0 & 0 & \phi\tilde{\phi} &-\tilde{E}\phi \partial \tilde{\phi}\\
0 & 0 & 0 & \phi\tilde{\phi}
\end{pmatrix}|\Lambda+a+\tilde{a}\rangle.
\end{eqnarray}
where $E$ and $\tilde{E}$ are again ratios of the amplitudes for the generation of the light box and the heavy box and the tilded light box and the tilded heavy box. By $\phi$ and $\tilde{\phi}$ we abbreviate $\phi(z-x_{a})$ and $\phi(z-x_{\tilde{a}})$.

Analogously, generating boxes at yet another supported red location is going to contain the second-order pole associated to the added light box and the first-order pole associated to the added heavy box. In this process, the heavy boxes already present in the initial configurations can change into  light boxes with the amplitude proportional to the square root of the first-order and the second-order pole of
\begin{eqnarray}
\psi_{\Lambda}(z)\prod_{i\in L} \phi (z-x_{a_i})\prod_{i\in H} \partial \phi (z-x_{a_i})
\label{complete}
\end{eqnarray}
at the location of the added box. In this formula, the first product runs over all the light boxes and the second sum runs over all the heavy boxes added on top of the configuration $\Lambda$. If the boxes at red locations come in blocks of adjacent heavy and light boxes, the formula should generalize to
\begin{eqnarray}
\psi_{\Lambda}(z)\prod_{l\in L}\prod_{i\in l}  \phi (z-x_{a_i})\prod_{h\in H}\prod_{i\in h} \phi (z-x_{a_i}) \sum_{i\in h}\frac{\partial \phi (z-x_{a_i})}{\phi (z-x_{a_i})}
\label{complete2}
\end{eqnarray}
in accordance to (\ref{ontop}), where $L,H$ indexes different blocks of light and heavy boxes and $i$ labels boxes within a given block.

The $2^n$ dimensional Jordan block associated to $n$ blocks of adjacent boxes at red locations, each allowing two or three consistent labellings should be given by the tensor product of elementary building blocks
\begin{eqnarray}
\prod_i \phi(z-x_{a})\begin{pmatrix}
1& \sum_{i}E_i\frac{\partial \phi(z-x_{a_i})}{\phi(z-x_{a_i})}\\
0 &1
\end{pmatrix}
\end{eqnarray}
associated to each block multiplied, by the overall factor of $\psi_{\Lambda}(z)$.  The product and the sum runs over all the boxes in the given block and $E_i$ refers to the corresponding ratio of amplitudes associated to the creation of the given light and heavy box. The rules for removing boxes and all the other transitions investigated in previous sections should work in an analogous way, bearing in mind that heavy boxes can always change into light boxes with amplitude proportional to the square root of the residue of (\ref{complete2}). This completes our proposal for the definition of the anti-fundamental modules $\mathcal{M}_{\bar{\lambda},\bar{\mu},\bar{\nu}}$.

\subsection{Recap on the definition of $M_{\mu,\nu,\lambda}$ modules}

From the above discussion of the fundamental and the anti-fundamental modules, one can obtain the desired modules $M_{\lambda,\mu,\nu}$ of $A$ by either

\begin{enumerate}
\item Using the simple embedding $A$ as a shifted Yangian to act on the highest-weight state of the more combinatorially complicated anti-fundamental module $\mathcal{M}_{\bar{\lambda},\bar{\mu},\bar{\nu}}$ or
\item Using the more complicated conjugated embedding of $A$ inside the Yangian to act on the highest-weight state of the simpler fundamental module $\mathcal{M}_{\lambda,\mu,\nu}$.
\end{enumerate}

There exists yet another way to define modules $M_{\lambda,\mu,\nu}$ without the necessity to refer to modules of $Y$. Without going too much into details, let us comment on this possibility. The 1-shifted affine Yangian admits an equivalent definition\cite{Rapcak:2020ueh} in terms of the standard generating series \footnote{See also \cite{bfn} for the discussion of shifted finite Yangians.}

\begin{eqnarray}
e(z)=\sum_{i=0}^{\infty}\frac{e_i}{z^i}\qquad f(z)=\sum_{i=0}^{\infty}\frac{f_i}{z^i}\qquad  \psi(z)=1+\sigma_3\sum_{i=0}^{\infty}\frac{\psi_i}{z^{i+1}}
\end{eqnarray}
satisfying the $ee,eee,ff,fff,\psi e,\psi f$  relations in an unchanged form but with the relation $[e_i,f_j]=\psi_{i+j}$ now modified to $[e_i,f_j]=h_{i+j}$ for $h_i$ assembled into the generating function $H(z)$ related to $\psi (z)$ by

\begin{eqnarray}
H(z)=z\psi (z)
\end{eqnarray}

We can now define modules $M_{\lambda,\mu,\nu}$ in this formulation of $A$ by letting $\psi(z)$ and $f(z)$ to act as in $\mathcal{M}_{\bar{\lambda},\bar{\mu},\bar{\nu}}$ and restricting the action of $e(z)$ not to produce any configurations containing boxes at standard locations in $\mathcal{M}_{\bar{\lambda},\bar{\mu},\bar{\nu}}$. Given the bootstrap analysis above, this automatically guarantees relations $ee,eee,ff,fff,\psi e,\psi f$ to be satisfied. Moreover,   multiplying the generating function $\psi(z)$ of $\mathcal{M}_{\bar{\lambda},\bar{\mu},\bar{\nu}}$ by $z$ deletes the pole at the origin, consistently with dropping terms in the action of $e(z)$. We thus expect that the relation $ef$ should hold as well.

\subsection{General degenerate modules $\mathcal{M}_{\lambda,\bar{\lambda};\mu,\bar{\mu};\mu,\bar{\nu}}$}
It is natural to combine the combinatorial construction of fundamental and anti-fundamental modules 
by employing 3d partitions with $\lambda$, $\mu$, $\nu$ asymptotics at unshifted locations as well as 
configurations bound by $\bar{\lambda}$, $\bar{\mu}$, $\bar{\nu}$ at shifted locations. 

This should result in a broader class of Yangian modules labelled by six partitions $\mathcal{M}_{\lambda_1,\bar{\lambda}_2;\mu_1,\bar{\mu}_2;\mu_1,\bar{\nu}_2}$ with character\footnote{One should think about a single $\mu$ as labelling an irreducible representation of $\mathfrak{gl}(N)$ for large enough $N$ that can be constructed from tensor products of the fundamental representation. On the other hand, the pair $(\mu,\bar{\mu})$ labels any irreducible representation that appears in the fusion ring of both the fundamental and the anti-fundamental representation. Note also that this formula is the topological vertex for non-commutative DT invariants.}
\begin{eqnarray}
\hat{\chi}_{\mu_1,\bar{\mu}_2;\nu_1,\bar{\nu}_2;\lambda_1,\bar{\lambda}_2}=C_{\mu_1,\nu_1,\lambda_1}C_{\bar{\mu}_2,\bar{\nu}_2,\bar{\lambda}_2}\prod_{n=1}^{\infty} \frac{1}{(1-q^n)^{n}},
\end{eqnarray}
These modules should perhaps appear in the construction of general gauge-invariant junctions.

\subsection{Example: the simplest non-trivia anti-fundamental module}

In previous section, we constructed modules with basis vectors labelled by PT box configurations by writing an ansatz and fixing remaining coefficients by solving for the relations of the affine Yangian. Let us now illustrate how to construct these modules by utilizing the charge conjugation. We will only give two explicit examples, but we have checked numerically the PT relations on all the examples containing up to 5 asymptotic boxes up to level 3 numerically.

Let us start with $\mathcal{M}_{0,\bar{\yng(1)},0}$. We denote the level-one states of $\mathcal{M}_{0,\yng(1),0}$ as $|\epsilon_1\rangle$, $|\epsilon_3\rangle$. The rising operators act on the highest-weight state of the fundamental representation as
\begin{eqnarray}
e(z)|vac\rangle = \frac{1}{z-\epsilon_1}|\epsilon_1\rangle+\frac{1}{z-\epsilon_3}|\epsilon_3\rangle
\end{eqnarray}
i.e.
\begin{eqnarray}\nonumber
e_i|vac\rangle &=&\epsilon_1^i|\epsilon_1\rangle +\epsilon_3^i|\epsilon_3\rangle
\end{eqnarray}
Compared to the previous sections, we have renormalized our states to get rid of the square-root factors. We have
\begin{eqnarray}
\psi (z)|vac\rangle = \frac{(z+\sigma_3\psi_0)(z+\epsilon_2)}{(z-\epsilon_3)(z-\epsilon_1)}|vac\rangle
\end{eqnarray}
and in particular
\begin{eqnarray}\nonumber
\psi_1|vac\rangle &=&-\frac{1}{\epsilon_2}|vac\rangle\\
\psi_2|vac\rangle &=& (1-\epsilon_1\epsilon_3\psi_0)|vac\rangle
\end{eqnarray}

Composing with the conjugation automorphism (see formulas from appendix \ref{ainy}), we get
\begin{eqnarray}\nonumber
\tilde{e}_1|vac\rangle &=&\epsilon_1|\epsilon_1\rangle + \epsilon_3|\epsilon_3\rangle, \\ \nonumber
\tilde{e}_2|vac\rangle &=&-\epsilon_1^2|\epsilon_1\rangle - \epsilon_3^2|\epsilon_3\rangle -\sigma_3\psi_0 (\epsilon_1|\epsilon_1\rangle + \epsilon_3|\epsilon_3\rangle )-\epsilon_1 \epsilon_3 (|\epsilon_1\rangle +|\epsilon_3\rangle)\\
&=&(\epsilon_2-\sigma_3\psi_0)(\epsilon_1|\epsilon_1\rangle + \epsilon_3|\epsilon_3\rangle ).
\end{eqnarray}

On the other hand, action of rising operators on the highest-weight state of the anti-fundamental module $\mathcal{M}_{0,\bar{\yng(1)},0}$ is given by 
\begin{eqnarray}
e(z)|vac\rangle = \frac{1}{z}|0\rangle+\frac{1}{z+\sigma_3\psi_0-\epsilon_2}|\epsilon_2-\sigma_3\psi_0\rangle
\end{eqnarray}
in particular
\begin{eqnarray}\nonumber
e_0|vac\rangle &=&|0\rangle +|\epsilon_2-\sigma_3\psi_0\rangle, \\
e_i|vac\rangle &=&(\epsilon_2-\sigma_3\psi_0)^i|\epsilon_2-\sigma_3\psi_0\rangle\qquad \mbox{for}\ i>0.
\end{eqnarray}
This is consistent with the above expressions for the action of the conjugated generators $\tilde{e}_i$ for $i>0$ on the fundamental module. We have to identify
\begin{eqnarray}
|\epsilon_2-\sigma_3\psi_0\rangle= \frac{1}{\epsilon_2-\sigma_3\psi_0}(\epsilon_1|\epsilon_1\rangle + \epsilon_3|\epsilon_2\rangle)
\end{eqnarray}
and we can in particular see that only a single state is generated at level one by the action of $\tilde{e}_1$ and $\tilde{e}_2$ as expected. 

\subsection{Example: the simplest module with three non-trivial asymptotics}

The generating function of $\psi (z)$ acting on the highest-weight state of $\mathcal{M}_{\Box,\Box,\Box}$ is given by
\begin{eqnarray}
\frac{(z+\sigma_3\psi_0)z^2}{(z-\epsilon_1-\epsilon_2)(z-\epsilon_1-\epsilon_3)(z-\epsilon_2-\epsilon_3)}
\label{conex}
\end{eqnarray}
and in particular
\begin{eqnarray}
\psi_1|vac\rangle =-\left (\frac{1}{\epsilon_1}+\frac{1}{\epsilon_2}+\frac{1}{\epsilon_3}\right )|vac\rangle
\end{eqnarray}
We have
\begin{eqnarray}
e(z)|vac\rangle = \frac{1}{z-\epsilon_2-\epsilon_3}|\epsilon_2+\epsilon_3\rangle +\frac{1}{z-\epsilon_1-\epsilon_3}|\epsilon_1+\epsilon_3\rangle +\frac{1}{z-\epsilon_1-\epsilon_2}|\epsilon_1+\epsilon_2\rangle
\end{eqnarray}
and thus
\begin{eqnarray}
e_i|vac\rangle = (\epsilon_2+\epsilon_3)^i |\epsilon_2+\epsilon_3\rangle + (\epsilon_1+\epsilon_3)^i |\epsilon_1+\epsilon_3\rangle + (\epsilon_1+\epsilon_2)^i |\epsilon_1+\epsilon_2\rangle
\end{eqnarray}
The action of the conjugate generators is
\begin{eqnarray}\nonumber
\tilde{e}_1|vac\rangle &=& -\epsilon_1 |\epsilon_2+\epsilon_3\rangle-\epsilon_2|\epsilon_1+\epsilon_3\rangle-\epsilon_3|\epsilon_1+\epsilon_2\rangle \\ \nonumber
\tilde{e}_2|vac\rangle &=& -\epsilon_1^2 |\epsilon_2+\epsilon_3\rangle -\epsilon^2_2|\epsilon_1+\epsilon_3\rangle -\epsilon_3^2|\epsilon_1+\epsilon_2\rangle \\ \nonumber
&&-(\epsilon_1\epsilon_2+\epsilon_1\epsilon_3+\epsilon_2\epsilon_3)( |\epsilon_2+\epsilon_3\rangle+|\epsilon_1+\epsilon_3\rangle +|\epsilon_1+\epsilon_2\rangle)\\ \nonumber
&&+\sigma_3\psi_0(\epsilon_1 |\epsilon_2+\epsilon_3\rangle +\epsilon_2|\epsilon_1+\epsilon_3\rangle +\epsilon_3|\epsilon_1+\epsilon_2\rangle)\\ \nonumber
&=&-\epsilon_2\epsilon_3 |\epsilon_2+\epsilon_3\rangle -\epsilon_1\epsilon_3|\epsilon_1+\epsilon_3\rangle -\epsilon_1\epsilon_2|\epsilon_1+\epsilon_2\rangle\\ 
&&+\sigma_3\psi_0(\epsilon_1 |\epsilon_2+\epsilon_3\rangle +\epsilon_2|\epsilon_1+\epsilon_3\rangle +\epsilon_3|\epsilon_1+\epsilon_2\rangle
\end{eqnarray}

Let us determine the action of the $\psi_i$ generators on descendants. We have
\begin{eqnarray}\nonumber
\psi(z) |\epsilon_2+\epsilon_3\rangle=\frac{(z+\sigma_3\psi_0)z(z-\epsilon_2)(z-\epsilon_3)(z+2\epsilon_1)}{(z+\epsilon_1)(z+\epsilon_2)(z+\epsilon_3)(z-2\epsilon_2-\epsilon_3)(z-2\epsilon_3-\epsilon_2)} |\epsilon_2+\epsilon_3\rangle \\ \nonumber
\psi(z) |\epsilon_1+\epsilon_3\rangle =\frac{(z+\sigma_3\psi_0)z(z-\epsilon_1)(z-\epsilon_3)(z+2\epsilon_2)}{(z+\epsilon_1)(z+\epsilon_2)(z+\epsilon_3)(z-2\epsilon_1-\epsilon_3)(z-2\epsilon_3-\epsilon_1)} |\epsilon_1+\epsilon_3\rangle \\
\psi(z) |\epsilon_1+\epsilon_2\rangle =\frac{(z+\sigma_3\psi_0)z(z-\epsilon_1)(z-\epsilon_2)(z+2\epsilon_3)}{(z+\epsilon_1)(z+\epsilon_2)(z+\epsilon_3)(z-2\epsilon_1-\epsilon_2)(z-2\epsilon_2-\epsilon_1)} |\epsilon_1+\epsilon_2\rangle
\end{eqnarray}
so that
\begin{eqnarray}\nonumber
\psi_1 |\epsilon_i+\epsilon_j\rangle &=&-\frac{\sigma_2}{\sigma_3} |\epsilon_i+\epsilon_j\rangle\\ \nonumber
\psi_2 |\epsilon_i+\epsilon_j\rangle &=&\left (1-\sigma_2\psi_0\right ) |\epsilon_i+\epsilon_j\rangle\\ 
\psi_3 |\epsilon_i+\epsilon_j\rangle &=&\left (-6\frac{\sigma_3}{\epsilon_i\epsilon_j}+\frac{\sigma_2^2}{\sigma_3}+\sigma_3\psi_0\right ) |\epsilon_i+\epsilon_j\rangle
\end{eqnarray}
Furthermore, we have \footnote{Note that the full factor of $-\mbox{Res}\ \psi (z)/\sigma_3$ now appears in the action of $f(z)$ due to our normalization of level-one states.}
\begin{eqnarray}
f(z)|\epsilon_2+\epsilon_3\rangle =-\frac{\epsilon_1^2(1-\epsilon_2 \epsilon_3\psi_0)}{\epsilon_2\epsilon_3(\epsilon_1-\epsilon_3)(\epsilon_1-\epsilon_2)}\frac{1}{z+\epsilon_1}|vac\rangle
\end{eqnarray}
so that
\begin{eqnarray}
f_i|\epsilon_2+\epsilon_3\rangle=-\frac{\epsilon_1^2(1-\epsilon_2 \epsilon_3\psi_0)}{\epsilon_2\epsilon_3(\epsilon_1-\epsilon_3)(\epsilon_1-\epsilon_2)}(\epsilon_2+\epsilon_3)^n |vac\rangle
\end{eqnarray}
and analogously for $|\epsilon_1+\epsilon_3\rangle$ and $|\epsilon_1+\epsilon_2\rangle$. 

Using this data, we write the action of the conjugate generator
\begin{eqnarray}\nonumber
\tilde{\psi}_3|\epsilon_2+\epsilon_3\rangle&=&\left (-6\epsilon_1+\frac{2\sigma_2^2}{\sigma_3}+\sigma_3\sigma_2\psi_0^2\right )|\epsilon_2+\epsilon_3\rangle\\
&&+6\frac{\epsilon_1^2(\epsilon_1-\sigma_3\psi_0)}{(\epsilon_1-\epsilon_3)(\epsilon_1-\epsilon_2)}\left (|\epsilon_2+\epsilon_3\rangle+|\epsilon_1+\epsilon_3\rangle+|\epsilon_1+\epsilon_2\rangle \right )
\end{eqnarray}
and analogously for $|\epsilon_1+\epsilon_3\rangle$ and $|\epsilon_1+\epsilon_2\rangle$. Identifying
\begin{eqnarray}\nonumber
|0_L\rangle &=& -\frac{1}{\sigma_3\psi_0}\tilde{e}_2 |vac\rangle -\sigma_3\psi_0 \tilde{e}_1|vac\rangle\\
|0_H\rangle &=& -\frac{1}{\sigma_3^2\psi_0^2}\tilde{e}_2 |vac\rangle-\frac{2}{\sigma_3\psi_0}\tilde{e}_1 |vac\rangle
\end{eqnarray}
leads to the action
\begin{eqnarray}\nonumber
\tilde{\psi}_3|0_H\rangle &=&\sigma_3\psi_0(\sigma_2\psi_0-2)|0_H\rangle+6|0_L\rangle  \\
\tilde{\psi}_3|0_L\rangle &=&\sigma_3\psi_0(\sigma_2\psi_0-2)|0_L\rangle
\label{refex2}
\end{eqnarray}
and we can see that the action is not diagonalizable! Furthermore, we get for the action of the creation operators expressions
\begin{eqnarray}\nonumber
\tilde{e}_1|vac\rangle &=& -\sigma_3\psi_0 |0_H\rangle +|0_L\rangle\\
\tilde{e}_2|vac\rangle &=& \sigma_3^2\psi_0^2 |0_H\rangle -2\sigma_3\psi_0|0_L\rangle
\label{refex}
\end{eqnarray}
and we can see that the coefficients in front of the state $ |\tilde{0}_H\rangle$ are consistent with a simple-pole ansatz but the coefficients  in front of $|\tilde{0}_L\rangle$ require a modification.

Let us check that the above formulas are consistent with our proposal for the module $M_{\Box,\Box,\Box}$. The conjugate of the generating function (\ref{conex}) is
\begin{eqnarray}
\tilde{\psi}(z)=\frac{(z-\epsilon_1+\sigma_3\psi_0)(z-\epsilon_2+\sigma_3\psi_0)(z-\epsilon_3+\sigma_3\psi_0)}{z(z+\sigma_3\psi_0)^2}
\end{eqnarray}
Level-one states of the conjugated module $\mathcal{M}_{\bar{\Box},\bar{\Box},\bar{\Box}}$ are generated as
\begin{eqnarray}
e(z)|vac\rangle =\frac{1}{z}|0\rangle+\frac{1}{(z+\sigma_3\psi_0)^2}|0_L\rangle+\frac{1}{z+\sigma_3\psi_0}|0_H\rangle
\end{eqnarray}
in particular
\begin{eqnarray}\nonumber
e_1|vac\rangle &=&-\sigma_3\psi_0|0_H\rangle + |0_L\rangle\\
e_2|vac\rangle &=& \sigma_3^2\psi_0^2|0_H\rangle-2\sigma_3\psi_0|0_L\rangle
\end{eqnarray}
recovering the above expression (\ref{refex}).

Similarly, our proposal for the action of $\psi(z)$ gives
\begin{eqnarray}\nonumber
\psi (z)|0_H\rangle &=& \tilde{\psi}(z)\phi (z+\sigma_3\psi_0)|A\rangle-\tilde{\psi} (z)\partial \phi (z+\sigma_3\psi_0)|0_L\rangle\\ 
\psi (z)|0_L\rangle &=& \tilde{\psi}(z)\phi (z+\sigma_3\psi_0)|0_L\rangle
\end{eqnarray}
in particular
\begin{eqnarray}\nonumber
\psi (z)|0_H\rangle &=&\left  (1+\sigma_3\left (\frac{\psi_0}{z}+\frac{\sigma_2}{\sigma_3}\frac{1}{z^2}+\frac{1-\sigma_2\psi_0}{z^3}-\frac{\sigma_3\psi_0(2-\sigma_2\psi_0)}{z^4}+\dots\right )\right )|0_H\rangle\\ 
&&+\left (\frac{6}{z^4}+\dots \right)|0_L\rangle\\ \nonumber
\psi (z)|0_L\rangle &=&\left  (1+\sigma_3\left (\frac{\psi_0}{z}+\frac{\sigma_2}{\sigma_3}\frac{1}{z^2}+\frac{1-\sigma_2\psi_0}{z^3}-\frac{\sigma_3\psi_0(2-\sigma_2\psi_0)}{z^4}+\dots\right )\right )|0_L\rangle 
\end{eqnarray}
again recovering the action of $\psi_3$ form (\ref{refex2}).

\section{Conclusions and future directions}\label{sec:future}
We have provided a self-consistent characterization of gauge-invariant junctions and fusion of defects within 
a $\mathbb{R} \times \mathbb{C}$ subspace of the twisted M-theory setup with $\mathbb{C}_{\epsilon_1} \times \mathbb{C}_{\epsilon_2} \times \mathbb{C}_{\epsilon_3}$ internal geometry. This characterization led to a variety of results concerning 
the M2 brane algebra $A$, the M5 brane VOA $\cW_\infty$, their interplay and truncations. It also provided a novel interpretation
of many well-known facts about these algebras. 

Several of our results are somewhat conjectural in nature and could be made rigorous. It would be particularly nice to have a clear characterization of Koszul duality for holomorphic-topological systems. It would also be nice to have a geometric construction of modules introduced in this paper.

Our work admits several natural generalizations
\begin{itemize}
\item We can look at a different space-time configuration. There are conjectural proposals for the M2 brane algebras associated to transverse geometries given by ADE singularities, $\mathbb{C} \times \mathbb{C}^*$ or $\mathbb{C}^* \times \mathbb{C}^*$
\cite{Gaiotto:2019wcc} which could be treated in a similar manner. The simplest example would involve replacing the rational Calogero Hamiltonian with the trigonometric one. 
\item We can change the internal geometry, say to $\frac{\mathbb{C}_{\epsilon_1} \times \mathbb{C}_{\epsilon_2}}{\mathbb{Z}_k} \times \mathbb{C}_{\epsilon_3}$ or other toric manifolds. This will give constructions associated to VOAs such as 
the matrix analogue of the $\cW_{\infty}$.
\end{itemize}

\acknowledgments
We thank M. Aganagic, K. Costello, H. Jenne, J. Oh, T. Proch\'{a}zka, G. Zhao for  discussions. M.R. is grateful to the string-theory group at the Czech Academy of Sciences for its hospitality during the difficult Covid-19 times. The research of D.G. is supported in part by a grant from the
Krembil foundation by the Perimeter Institute for Theoretical Physics. Research at Perimeter Institute is supported in part by the Government of Canada through the Department of Innovation, Science and Economic Development Canada and by the Province of Ontario through the Ministry of Economic Development, Job Creation and Trade. The research of M.R. was supported by NSF grant 1521446, NSF grant 1820912, the Berkeley Center for Theoretical Physics and the Simons Foundation.

\appendix

\section{Defining relations of $W_{\infty}$} \label{app:winfty}
\label{appendixa}

\subsection{OPEs of $U$-generators}
\label{uope}

The $U_i$ generators coming from the composition of Miura operators $\mathcal{L}_{0,0,1}^{0,0,1}$ have the following OPEs 
\begin{align}
U_1(z)U_1(w)\sim&-\frac{\epsilon_1\epsilon_2N}{(z-w)^2}\cr
U_1(z)U_2(w)\sim&-\frac{\epsilon_1\epsilon_2\epsilon_3N(N-1)}{(z-w)^3}-\frac{\epsilon_1\epsilon_2(N-1)U_1(w)}{(z-w)^3}\cr
U_1(z)U_3(w)\sim&-\frac{\epsilon_1\epsilon_2\epsilon_3^2N(N-1)(N-2)}{(z-w)^4}-\frac{\epsilon_1\epsilon_2\epsilon_3(N-1)(N-2) U_1(w)}{(z-w)^3}\cr
&-\frac{\epsilon_1\epsilon_2(N-2)U_2(w)}{(z-w)^2}\cr
U_2(z)U_2(w)\sim&  \frac{N(N-1)(\epsilon_1^2\epsilon_2^2+2\epsilon_1\epsilon_2\epsilon_3^2(2N-1))}{2(z-w)^4}\cr
&+\frac{2\epsilon_1\epsilon_2U_2-\epsilon_1\epsilon_2(N-1)U_1U_1-\epsilon_1\epsilon_2\epsilon_3 N(N-1)\partial U_1}{(z-w)^2}\cr
&+\frac{2\epsilon_1\epsilon_2\partial U_2-2\epsilon_1\epsilon_2(N-1)\partial U_1 U_1-\epsilon_1\epsilon_2\epsilon_3 N(N-1)\partial^2 U_1}{2(z-w)}
\end{align}
These can be identified with formulas from \cite{Prochazka:2014gqa} by setting $\epsilon_1\epsilon_2=-1$ and $\epsilon_3=\alpha_0$.

\subsection{From $U$-basis to the primary basis}
\label{utop}

The $U$-basis consists of quasi-primary fields $U_i$ appearing as coefficients in the Miura pseudo-differential operator.  Another commonly used basis is the primary basis generated by the stress-energy tensor $W_2$ together with an infinite series of primary fields $W_1,W_3,W_4,\dots$ with the subscript labeling the conformal weight of the corresponding field. The transformation between the two bases for the first few generators reads
\begin{eqnarray}\nonumber
W_1&=&-\frac{1}{\epsilon_1\epsilon_2}U_1,\\ \nonumber
W_2&=&\frac{1}{\epsilon_1\epsilon_2}U_2-\frac{1}{2\epsilon_1\epsilon_2}U_1^2-\frac{(N-1)\epsilon_3}{2\epsilon_1\epsilon_2}\partial U_1,\\ \nonumber
W_3&=&-\frac{1}{\epsilon_1\epsilon_2}\bigg (U_3-\frac{N-2}{N}U_1U_2+\frac{(N-1)(N-2)}{3N^2}U_1^3-\frac{(N-2)\epsilon_3}{2}\partial U_2\\ 
&&\quad+\frac{(N-1)(N-2)\epsilon_3}{2N}\partial U_1 U_1 +\frac{(N-1)(N-2)\epsilon_3^2}{12}\partial^2 U_1\bigg ).
\end{eqnarray}

Writing explicitly OPEs of first few primary generators 
\begin{eqnarray}\nonumber
W_1(z)W_1(w)&\sim& \frac{\psi_0}{(z-w)^2}\\ \nonumber
W_1(z)W_3(w)&\sim&0\\
W_2(z)W_2(w)&\sim& -\frac{\sigma_2\psi_0+\sigma_3^2\psi_0^3}{2}\frac{1}{(z-w)^4}+\frac{W_2(w)}{(z-w)^2}+\frac{\partial W_2(w)}{z-w}
\end{eqnarray}
where we identified $\sigma_3\psi_0=-\epsilon_3N$ for $N$ being the parameter $N$ of the $U$-basis coming from $\mathcal{L}_{0,0,1}^{0,0,1}$. The field $W_3$ is normalized such that the leading order pole is
\begin{eqnarray}
\frac{(1+\epsilon_1\epsilon_2\psi_0)(1+\epsilon_1\epsilon_3\psi_0)(1+\epsilon_2\epsilon_3\psi_0)(2+\epsilon_1\epsilon_2\psi_0)(2+\epsilon_1\epsilon_3\psi_0)(2+\epsilon_2\epsilon_3\psi_0)}{6\psi_0}
\end{eqnarray}
We see that OPEs of primary generators can be written in a triality-invariant way. 

The algebra $\mathcal{W}_{\infty}$ admits a simple charge-conjugation automorphism reversing the sign of the primary generators of odd spin $W_n \leftrightarrow (-1)^n W_n$. In terms of the $U$-generators, the charge-conjugation reads
\begin{eqnarray}\nonumber
U_1&\rightarrow&-U_1\\ \nonumber
U_2&\rightarrow&U_2-(N-1)\epsilon_3 \partial U_1\\ 
U_3&\rightarrow&-U_3+(N-2)\epsilon_3\partial U_2-\frac{(N-1)(N-2)}{2}\epsilon_3^2\partial^2 U_1
\label{utrans}
\end{eqnarray}

\subsection{Relations of affine Yangian}
\label{yrelations}

The generators $e_i,f_i,\psi_i$ of the affine Yangian satisfy defining relations \cite{Tsymbaliuk:2014fvq}
\begin{eqnarray}\nonumber
\psi_{i+j}&=&{[}e_i,f_j{]}\qquad {[}\psi_i,\psi_j{]}=0\\ \nonumber
0&=&{[}e_{i+3},e_j{]}-3{[}e_{i+2},e_{j+1}{]}+3{[}e_{i+1},e_{j+2}{]}-{[}e_{i},e_{j+3}{]}\\ \nonumber
&&+\sigma_2 {[}e_{i+1},e_j{]}-\sigma_2{[}e_{i},e_{j+1}{]}-\sigma_3 \{ e_{i},e_j\}\\ \nonumber
0&=&{[}f_{i+3},f_j{]}-3{[}f_{i+2},f_{j+1}{]}+3{[}f_{i+1},f_{j+2}{]}-{[}f_{i},f_{j+3}{]}\\ \nonumber
&&+\sigma_2 {[}f_{i+1},f_j{]}-\sigma_2{[}f_{i},f_{j+1}{]}+\sigma_3 \{ f_{i},f_j\}\\ \nonumber
0&=&{[}\psi_{i+3},e_j{]}-3{[}\psi_{i+2},e_{j+1}{]}+3{[}\psi_{i+1},e_{j+2}{]}-{[}\psi_{i},e_{j+3}{]}\\ \nonumber
&&+\sigma_2 {[}\psi_{i+1},e_j{]}-\sigma_2{[}\psi_{i},e_{j+1}{]}-\sigma_3 \{ \psi_{i},e_j\}\\ \nonumber
0&=&{[}\psi_{i+3},f_j{]}-3{[}\psi_{i+2},f_{j+1}{]}+3{[}\psi_{i+1},f_{j+2}{]}-{[}\psi_{i},f_{j+3}{]}\\ 
&&+\sigma_2 {[}\psi_{i+1},f_j{]}-\sigma_2{[}\psi_{i},f_{j+1}{]}+\sigma_3 \{ \psi_{i},f_j\}
\label{yrelations2}
\end{eqnarray}
together with 
\begin{eqnarray}\nonumber
&{[}\psi_0,e_i{]}={[}\psi_0,f_i{]}={[}\psi_1,e_i{]}={[}\psi_1,f_i{]}=0&\\
&{[}\psi_2,e_i{]}=2e_i\qquad {[}\psi_2,f_i{]}=-2f_i&
\label{yrelations3}
\end{eqnarray}
and
\begin{eqnarray}\nonumber
\mbox{Sym}_{i,j,k}{[}e_i,{[}e_j,e_k{]}{]}&=&0\\
\mbox{Sym}_{i,j,k}{[}f_i,{[}f_j,f_k{]}{]}&=&0
\label{yrelations3}
\end{eqnarray}
where $\mbox{Sym}_{i,j,k}$ denotes symmetrization over indices $i,j,k$.

\subsection{Relations in terms of generating functions}
\label{relationsfunctions}

Defining generating functions
\begin{eqnarray}
e(z)=\sum_{i=0}^{\infty}\frac{e_i}{z^{1+i}}\qquad f(z)=\sum_{i=0}^{\infty}\frac{f_i}{z^{1+i}}\qquad \psi (z)=1+\sigma_3\sum_{i=0}^{\infty}\frac{\psi_i}{z^{1+i}},
\end{eqnarray}
the relations (\ref{yrelations2}) can be compactly encoded by
\begin{align}
e(z)e(w)&\sim \phi (z-w)e(w)e(z),\cr
f(z)f(w)&\sim \phi^{-1} (z-w)f(w)f(z),\cr
\psi(z)e(w)&\sim \phi (z-w)e(w)\psi(z),\cr
\psi(z)f(w)&\sim \phi^{-1} (z-w)f(w)\psi(z),
\label{relationsfunctions1}
\end{align}
together with
\begin{eqnarray}
e(z)f(w)-f(w)e(z)\sim -\frac{1}{\sigma_3}\frac{\psi (z)-\psi (w)}{z-w}
\end{eqnarray}
where $\phi(z)$ is the standard factor
\begin{eqnarray}
\phi(z)=\frac{(z+\epsilon_1)(z+\epsilon_2)(z+\epsilon_3)}{(z-\epsilon_1)(z-\epsilon_2)(z-\epsilon_3)}
\end{eqnarray}
Relations (\ref{yrelations2}) and (\ref{yrelations3}) can be then deduced from (\ref{relationsfunctions1}) by multiplying by denominators (the denominator of the $\phi$ factor and the $z-w$ factor), expanding both sides in powers of $z$ and $w$ around infinity and restricting to coefficients of $z^{-n}w^{-m}$ for $m,n>0$ in the first two relations and $m>0$ with $n$ unconstrained in the rest. Similarly, the Serre relations (\ref{yrelations3}) are encoded by
\begin{eqnarray}
\sum_{\pi\in S_3}(u_{\pi (1)}-2u_{\pi (2)}+u_{\pi (3)})e(u_{\pi (1)})e(u_{\pi (2)})e(u_{\pi (3)})=0
\end{eqnarray}
where $\pi$ is an element of the symmetric group $S_3$. 

The relations written in terms of generating functions listed in this section are used in bootstraping the PT modules in the body of the paper.

\subsection{From $W_\infty$ to affine Yangian}
\label{wtoy}

The algebra $W_\infty$ is known to be isomorphic to the affine Yangian from appendix \ref{yrelations}. In particular the affine Yangian contains the Heisenberg algebra at level $\psi_0$ generated by 
\begin{eqnarray}
W_{1,0}=\psi_1\quad W_{1,-1}=e_0\quad W_{1,1}=-f_0\quad W_{1,-2}=[e_1,e_0]\quad W_{1,2}=-[f_1,f_0]
\end{eqnarray}
or generally
\begin{eqnarray}
W_{1,-n}=\frac{1}{(m-1)!}\mbox{ad}_{e_1}^{m-1} e_0\qquad W_{1,n}=-\frac{1}{(m-1)!}\mbox{ad}_{f_1}^{m-1} f_0,
\end{eqnarray}
where $\mbox{ad}_{e_1} e_0$ denotes an adjoint action of $e_1$ on $e_0$. The affine Yangian is known to be generated \footnote{Knowing $e_0$ and $f_0$, one can recursively determine $e_i,f_i$ from relations $[\psi_3,e_{i-1}]$ and $[\psi_3,f_{i-1}]$  and then $\psi_{i}$ from the commutator of $e_0$ with $f_i$.} by $W_{1,n}$ for $n\in \mathbb{Z}$ together with $\psi_3$ that can be in turn identified with the non-trivial combination
\begin{eqnarray}
\psi_3&=&V_0-\sigma_3\psi_0W_{2,0}-\frac{1}{2}\sigma_3W_{1,0}^2 -\frac{3}{2}\sigma_3\sum_{m=-\infty}^{\infty}|m|:W_{1,-m}W_{1,m}:
\label{psi}
\end{eqnarray}
where $V_0$ is the zero mode of the quasi-primary field
\begin{align}
V=&-\frac{1}{\epsilon_1\epsilon_2}\bigg (U_3-:U_1U_2:+\frac{1}{3}:U_1U_1U_1: +\frac{\epsilon_3}{2}(N-2)\partial U_2\cr
&\quad -\frac{\epsilon_3}{2}(N-1) :\partial U_1 U_1:-\frac{\epsilon_3^2(N-1)(N-2)}{12}\partial^2 U_1\bigg )\cr
=&W_3+\frac{2}{\psi_0}:W_1W_2:-\frac{2}{3}\frac{1}{\psi_0^2}:W_1W_1W_1: 
\label{V}
\end{align}
The field $V$ can be uniquely specified by its OPEs with the current $W_1$ and the stress-energy tensor $W_2$, namely 
\begin{align}
W_1(z)V(w)&=\frac{2W_2(w)}{(z-w)^2}\cr
W_2(z)V(w)&=-\frac{\left (\sigma_2+\sigma_3^2\psi_0^3\right )W_1(w)}{(z-w)^4}+\frac{3V(w)}{(z-w)^2}+\frac{\partial V(w)}{z-w}
\label{vope}
\end{align}

Other useful generators can be then identified by commuting the above, e.g. the global conformal generators are
\begin{eqnarray}
W_{2,-1}=e_1\qquad W_{2,0}=\frac{1}{2}\psi_2 \qquad W_{2,1}=-f_1
\end{eqnarray}
We also have
\begin{eqnarray}\nonumber
e_2&=& V_1-\frac{1}{2}\sigma_3\psi_0W_{2,1}+\frac{\sigma_3}{2}\sum_{m=-\infty}^{\infty}\left |l-\frac{1}{2}\right |:W_{1,-m}W_{m+1}:\\
f_2 &=& -V_1+\frac{1}{2}\sigma_3\psi_0W_{2,1}+\frac{\sigma_3}{2}\sum_{m=-\infty}^{\infty}\left |l-\frac{1}{2}\right |:W_{1,1-m}W_{m}:
\end{eqnarray}
and important commutators
\begin{eqnarray}\nonumber
{[}e_1,e_2{]}&=&V_{-2}+\frac{\sigma_3}{2}\sum_{m=-\infty}^{\infty}|m|W_{1,-m-1}W_{1,m-1}\\
{[}f_1,f_2{]}&=&V_{2}+\frac{\sigma_3}{2}\sum_{m=-\infty}^{\infty}|m|W_{1,-m+1}W_{1,m+1}
\end{eqnarray}

\subsection{Charge conjugation in Yangian basis}
\label{conjugation}

Recall the charge-conjugation automorphism from the body of the paper
\begin{eqnarray}\nonumber
&\tilde{e}_0=-e_0,\qquad \tilde{e}_1=e_1\\
&\tilde{f}_0=-f_0,\qquad \tilde{f}_1=f_1\\ \nonumber
&\tilde{\psi}_0=\psi_0,\qquad \tilde{\psi}_1=-\psi_1,\qquad \tilde{\psi}_2=\psi_2
\end{eqnarray}
together with
\begin{equation}
\tilde{\psi}_3=-\psi_3-\sigma_3(\psi_0\psi_2-\psi_1^2)+6\sigma_3\sum_{m>0}mW_{1,-m}W_{1,m}
\label{psi3conj}
\end{equation}
Let us collect explicit formulas for the charge conjugation of other generators. For example, we have
\begin{eqnarray}\nonumber
\tilde{e}_{2}&=&-e_2-\sigma_3\psi_0 e_1 +\sigma_3\sum_{m\geq 0}(2m+1)W_{1,-m-1}W_{1,m}\\
\tilde{f}_{2}&=&-f_2-\sigma_3\psi_0 f_1 +\sigma_3\sum_{m\geq 0}(2m+1)W_{1,-m}W_{1,m+1}
\end{eqnarray}
and for the commutators
\begin{eqnarray}\nonumber
{[}\tilde{e}_2,\tilde{e}_1{]}&=&-{[}e_2,e_1{]}+\sigma_3\sum_{m=-\infty}^{\infty}|m| :W_{1,-m-1}W_{1,m-1}:\\
{[}\tilde{f}_2,\tilde{f}_1{]}&=&-{[}f_2,f_1{]}+\sigma_3\sum_{m=-\infty}^{\infty}|m|:W_{1,-m+1}W_{1,m+1}:.
\end{eqnarray}
Note the non-trivial tails present in charge-conjugation formulas in Yangian basis compared to the simple expressions in the primary basis of $\mathcal{W}_{\infty}$.

\subsection{Two embeddings of $A$ inside $Y$}
\label{ainy}

In the main text, we identified
\begin{eqnarray}\nonumber
T_{0,n}&\rightarrow&W_{1,n}=\frac{1}{(n-1)!}\mbox{ad}_{f_1}^{n-1}f_0\\
T_{2,0}&\rightarrow&V_{-2}+\sigma_3\sum_{n=0}^{\infty}(n+1)W_{1,-n-2}W_{1,n}=-[e_2,e_1]
\label{emb}
\end{eqnarray}
Other useful generators are given by
\begin{eqnarray}\nonumber 
T_{1,n}&\rightarrow&-W_{2,n-1}-\frac{\sigma_3\psi_0}{2}n W_{1,n-1}\\ \nonumber
T_{2,n}&\rightarrow&V_{n-2}+\frac{\sigma_3}{2}\psi_0nW_{2,n-2}+\frac{n(n-1)}{12}(\sigma_2+4\sigma_3^2\psi_0^2)W_{1,n-2}\\ 
&&+\frac{\sigma_3}{2}\sum_{m=0}^{\infty}(m+1)\left ((m+2)W_{1,n-m-2}W_{1,m}-mW_{1,-m-2}W_{1,n+m}\right )
\end{eqnarray}
Using these explicit formulas, one can explicitly check that the relations of $A$ are satisfied. It is also straightforward to check that this carves out a subalgebra of $Y$ generated by $\psi_i,f_i$ for $i\geq 0$ but $e_i$ only for $i>0$. In particular, the commutator of $[e_2,e_1]$ with $W_{1,3}$ allows an identification of $f_1$ and the mutual commutator of $[e_2,e_1]$ with $f_1$ produces $e_2$. Commutator of $e_2$ with $f_1$ gives the generator $\psi_3$ that can be then used to find all $f_i$ for $i\geq 0$ and $e_i$ for $i>0$ with the use of the $[\psi_i,e_j]$ relation. We thus get a 1-shifted affine Yangian of $\mathfrak{u}(1)$, first defined in \cite{Kodera:2016faj}.

The charge-conjugation automorphism acting on (\ref{emb}) produces an alternative embedding of $A$ inside $Y$ given by
\begin{eqnarray}\nonumber
T_{0,n}&\rightarrow&-W_{1,n}=- \frac{1}{(n-1)!}\mbox{ad}_{f_1}^{n-1}f_0\\ \nonumber
T_{2,0}&\rightarrow&-V_{-2}+\sigma_3\sum_{n=0}^{\infty}(n+1)W_{1,-n-2}W_{1,n}\\
&\rightarrow&[e_2,e_1]+2\sigma_3\sum_{n=0}^{\infty}(n+1)W_{1,-n-2}W_{1,n}
\end{eqnarray}
Note that compared to (\ref{emb}), the second embedding has much more complicated form in terms of the Yangian generators, non-trivially mixing the rising generators $e_i$ with the lowering ones $f_i$. This non-trivial form of the charge conjugation in the Yangian basis is the main reason why fundamental and anti-fundamental modules look so differently when defined in terms of $e(z),f(z),\psi(z)$.

\section{Box-counting for $A$-modules}
\label{app:box}

 \Yboxdim{6pt}

In this appendix, we present a collection of examples showing how to build configurations of boxes for $A$-modules at low levels. We present examples with all three asymptotics non-trivial and with the total number of asymptotic boxes at most 5. We checked all the processes listed bellow by analyzing the action of the conjugated embedding of $A$ in $Y$ on the highest-weight state of fundamental modules $\mathcal{M}_{\lambda,\mu,\nu}$. 

\subsection{Example 1:}

\begin{figure}[h]
    \centering
        \includegraphics[width=3.8cm]{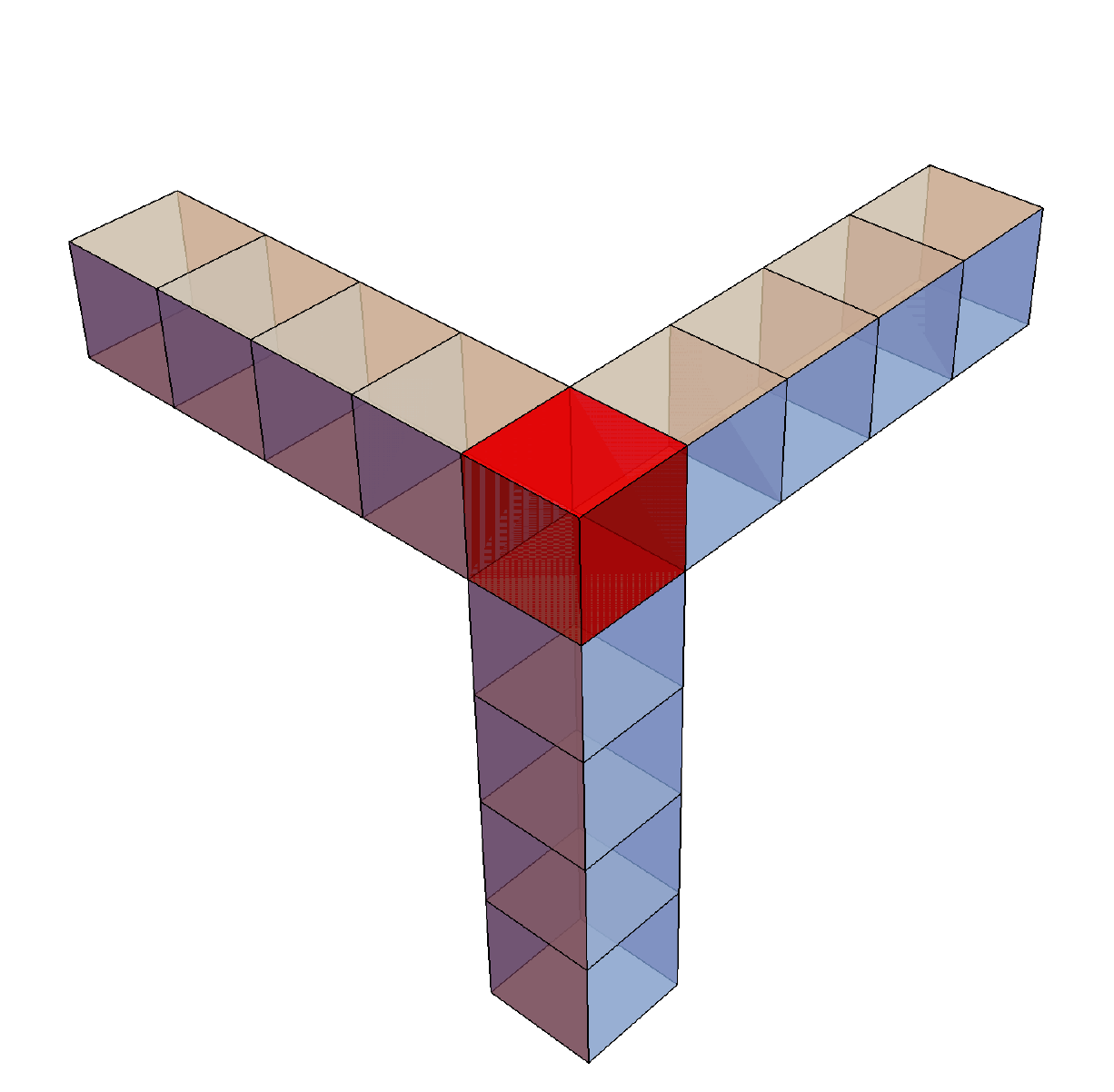}
 \caption{The box counting setup associated to $C_{\protect \yng(1),\protect \yng(1),\protect \yng(1)}$. }
\label{pic15}
\end{figure}

The first example comes from considering a single box along all three asymptotics, a configuration depicted in \ref{pic15}, with character
\begin{eqnarray}
C_{\yng(1),\yng(1),\yng(1)} =\frac{1-q+q^2-q^3+q^4}{(1-q)^3} =  1+2q+4q^2+6q^3+\dots.
\end{eqnarray}

At level one, we can obviously introduce a two-dimensional space associated to the red box
\begin{eqnarray}
|vac\rangle \rightarrow |0_H\rangle + |0_L\rangle
\end{eqnarray}
since the red box allows any labelling.

The light box at level one can only transform into the ultra-heavy box. On the other hand, we can stack boxes on top of the heavy box as
\begin{eqnarray}\nonumber
|0_H\rangle &\rightarrow& |0_U\rangle + |0_H,\epsilon_1\rangle + |0_H,\epsilon_2\rangle + |0_H,\epsilon_3\rangle\\
|0_L\rangle &\rightarrow& |0_U\rangle
\end{eqnarray}
One needs to check the consistency of the configuration by determining a labelling associated to the heavy box. For each $|0_H,\epsilon_i\rangle$, one has a single uncolored box in direction $i$ and one can simply label the box at the origin by this direction. For simplicity, throughout this appendix, we disregard the shift by $-\sigma_3\psi_0$ in the position of boxes and furthermore reverses its sign. 

At level two, one can stuck any boxes on top of the ultra-heavy box. Furthermore, one can either change the light box in $|0_H,\epsilon_i\rangle$ into the ultra-heavy box or add an extra box in the direction of $i$. In total, one gets six states at level three
\begin{eqnarray}\nonumber 
|0_U\rangle &\rightarrow&|0_U,\epsilon_1\rangle + |0_U,\epsilon_2\rangle + |0_U,\epsilon_3\rangle\\ \nonumber
|0_H\epsilon_1\rangle &\rightarrow&|0_H,\epsilon_1,2\epsilon_1\rangle +|0_U,\epsilon_1\rangle\\ \nonumber
|0_H,\epsilon_2\rangle &\rightarrow&|0_H,\epsilon_2,2\epsilon_2\rangle +|0_U,\epsilon_2\rangle\\
|0_H,\epsilon_3\rangle &\rightarrow&|0_H,\epsilon_3,2\epsilon_3\rangle +|0_U,\epsilon_3\rangle
\end{eqnarray}
Note that one cannot add a box at position $\epsilon_j$ for $i\neq j$ on top of the configuration $|0_H,\epsilon_i\rangle$ since such a configuration would not allow a consistent labelling of the red box due to the presence of an uncolored box on top of the heavy box in two different directions. 

\subsection{Example 2: }

\begin{figure}[h]
    \centering
        \includegraphics[width=3.8cm]{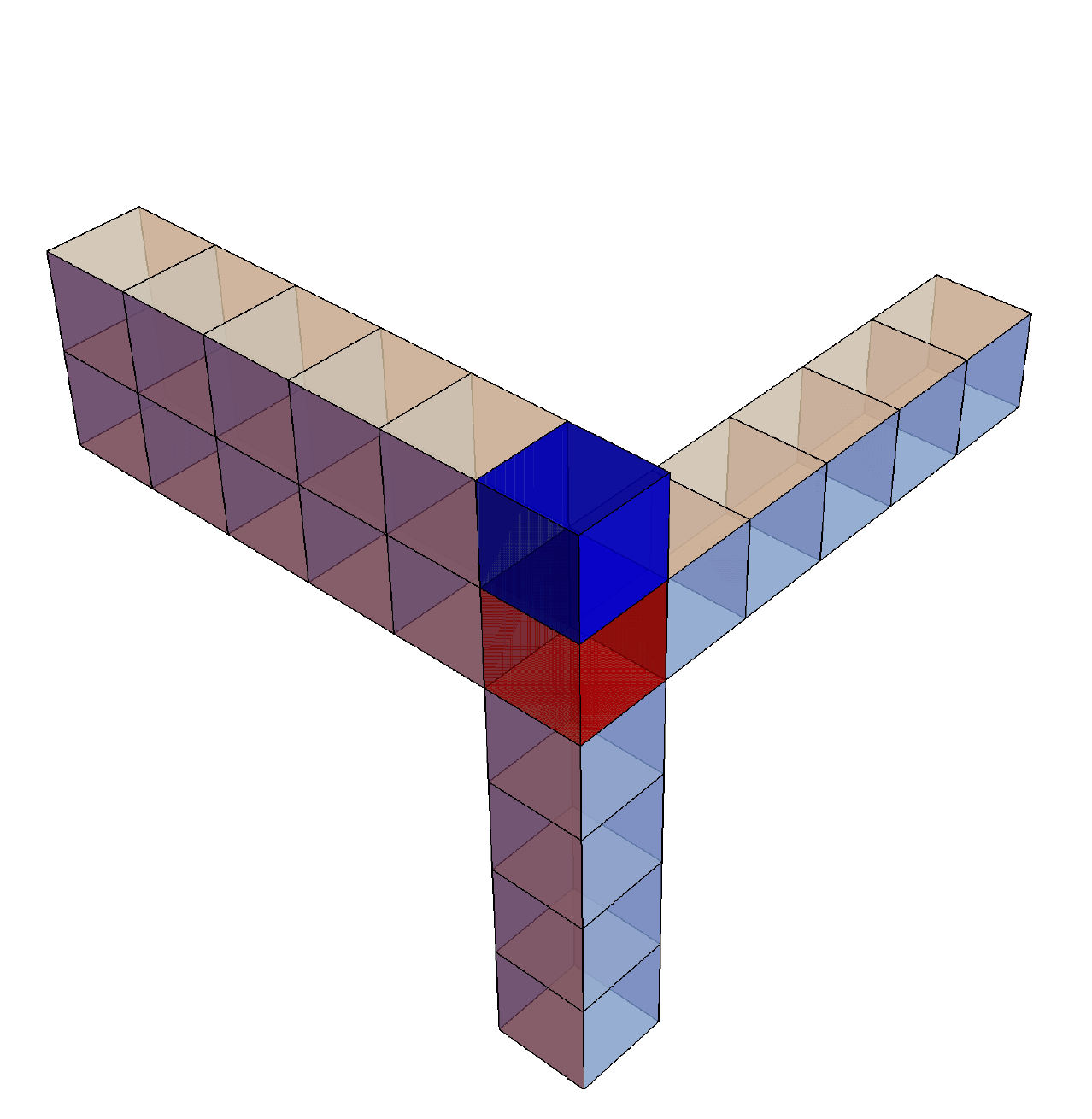}
 \caption{The box counting setup associated to $C_{\protect \yng(1), \protect \yng(2),\protect \yng(1)}$.}
\label{pic16}
\end{figure}

The second example is the only case containing the total of four asymptotic boxes, a configuration depicted in figure \ref{pic16}, with character
\begin{eqnarray}
C_{\yng(1),\yng(2),\yng(1)} =\frac{1-q+q^3-q^5+q^6}{(1-q)^3(1-q^2)}  =  1+2q+4q^2+7q^3+\dots
\end{eqnarray}

At level one, we can obviously generate the box at position $-\epsilon_3$ but that is not the only possibility. One can also generate a heavy box at the red position since it is not supported only from one side by a missing blue box and the configuration can thus be consistently labelled by direction 1 that is distinct from the orientation of the two cylinders whose intersection contains the blue box. We thus get
\begin{eqnarray}
|vac\rangle \rightarrow  |-\epsilon_3\rangle + |0_H\rangle 
\end{eqnarray}

At level one, we get the following possibilities
\begin{eqnarray}\nonumber
|-\epsilon_3\rangle &\rightarrow& |-\epsilon_3,-\epsilon_3+\epsilon_2\rangle +|-\epsilon_3,0_H\rangle +|-\epsilon_3,0_L\rangle \\
|0_H\rangle &\rightarrow&  |0_H, \epsilon_1\rangle + |-\epsilon_3,0_L\rangle
\end{eqnarray}
In particular, we can generate a two-dimensional vector space associated to the red location on top of $|-\epsilon_3\rangle$ since the red box is supported from all directions and admits any labelling. On the other hand, one can put a box on top of the unsupported box $|0_H\rangle$ only in  direction 1 since the missing blue box already requires the heavy box to be labelled by direction 1. Note that the configuration $|-\epsilon_3,0_L\rangle$ can be generated from $|0_H\rangle$ even though  the generating function $\psi(z)$ does not contain a pole at position $-\epsilon_3$. On the other hand,  $|0_H\rangle$ cannot be produced by an action of $f(z)$ on $|-\epsilon_3,0_L\rangle$  consistently with the $ef$ relation and the vanishing of the residue at $-\epsilon_3$. The action of $f(z)$ on $|-\epsilon_3,0_H\rangle$ is required to produce $|0_H\rangle$.

At level three, we can produce the following seven states
\begin{eqnarray}\nonumber
|-\epsilon_3,-\epsilon_3+\epsilon_2\rangle &\rightarrow&|-\epsilon_3,-\epsilon_3+\epsilon_2,0_L\rangle+ |-\epsilon_3,-\epsilon_3+\epsilon_2,0_H\rangle \\ \nonumber
&&+|-\epsilon_3,-\epsilon_3+\epsilon_2,-\epsilon_3+2\epsilon_2\rangle  \\ \nonumber
|-\epsilon_3,0_H\rangle &\rightarrow&  |-\epsilon_3,0_U\rangle + |-\epsilon_3,-\epsilon_3+\epsilon_2,0_L\rangle + |-\epsilon_3,-\epsilon_3+\epsilon_2,0_H\rangle\\ \nonumber
&& + |-\epsilon_3,0_H,\epsilon_1\rangle +|-\epsilon_3,0_H,\epsilon_3\rangle \\ \nonumber
|-\epsilon_3,0_L\rangle &\rightarrow& |-\epsilon_3,0_U\rangle + |-\epsilon_3,-\epsilon_3+\epsilon_2,0_L\rangle  \\ 
 |0_H, \epsilon_1\rangle&\rightarrow& |-\epsilon_3,0_H, \epsilon_1\rangle + |0_H, \epsilon_1,2\epsilon_1\rangle
\end{eqnarray}
Since the red box is supported from all sides in the $|-\epsilon_3,-\epsilon_3+\epsilon_2\rangle$ configuration, one can generate either a box at $-\epsilon_3+2\epsilon_2$ or a two-dimensional subspace at the red position. The heavy box of $|-\epsilon_3,0_H\rangle$ can either change to the ultra-heavy one or we can stuck boxes at positions $\epsilon_1$ or $\epsilon_3$ on top of it. Both configurations are consistent since there is a single uncolored box on top of the red box in direction $1$ or $3$ respectively and the heavy box can be labelled by these directions. One can also obviously introduce a box at position  $-\epsilon_3+\epsilon_2$ on top of $|-\epsilon_3,0_H\rangle $ producing both $|-\epsilon_3,-\epsilon_3+\epsilon_2,0_L\rangle $ and $|-\epsilon_3,-\epsilon_3+\epsilon_2,0_H\rangle$ since $|-\epsilon_3,-\epsilon_3+\epsilon_2,0_L\rangle $ can be always generated by the action of $\psi (z)$ on $|\epsilon_3,-\epsilon_3+\epsilon_2,0_H\rangle$ . On the other hand, we cannot stuck boxes on top of the light box and we can only add a box at position $-\epsilon_3+\epsilon_2$ on top of $|-\epsilon_3,0_L\rangle$ or we can change the light box into the ultra-heavy one. Finally, on top of $|-\epsilon_3,0_H\rangle$, only the box at position $2\epsilon_1$ can be generated since adding a box at $\epsilon_3$ would not allow a consistent labelling of the red box. Another obvious option is a generation of the box at position $-\epsilon_3$. Compared to the previous level, the action of $\psi (z)$ on $|0_H, \epsilon_1\rangle$ does contain a pole at $-\epsilon_3$ and the transition $|-\epsilon_3,0_H, \epsilon_1\rangle  \rightarrow |0_H, \epsilon_1\rangle$ is now allowed.

\subsection{Example 3: }

\begin{figure}[h]
    \centering
        \includegraphics[width=3.8cm]{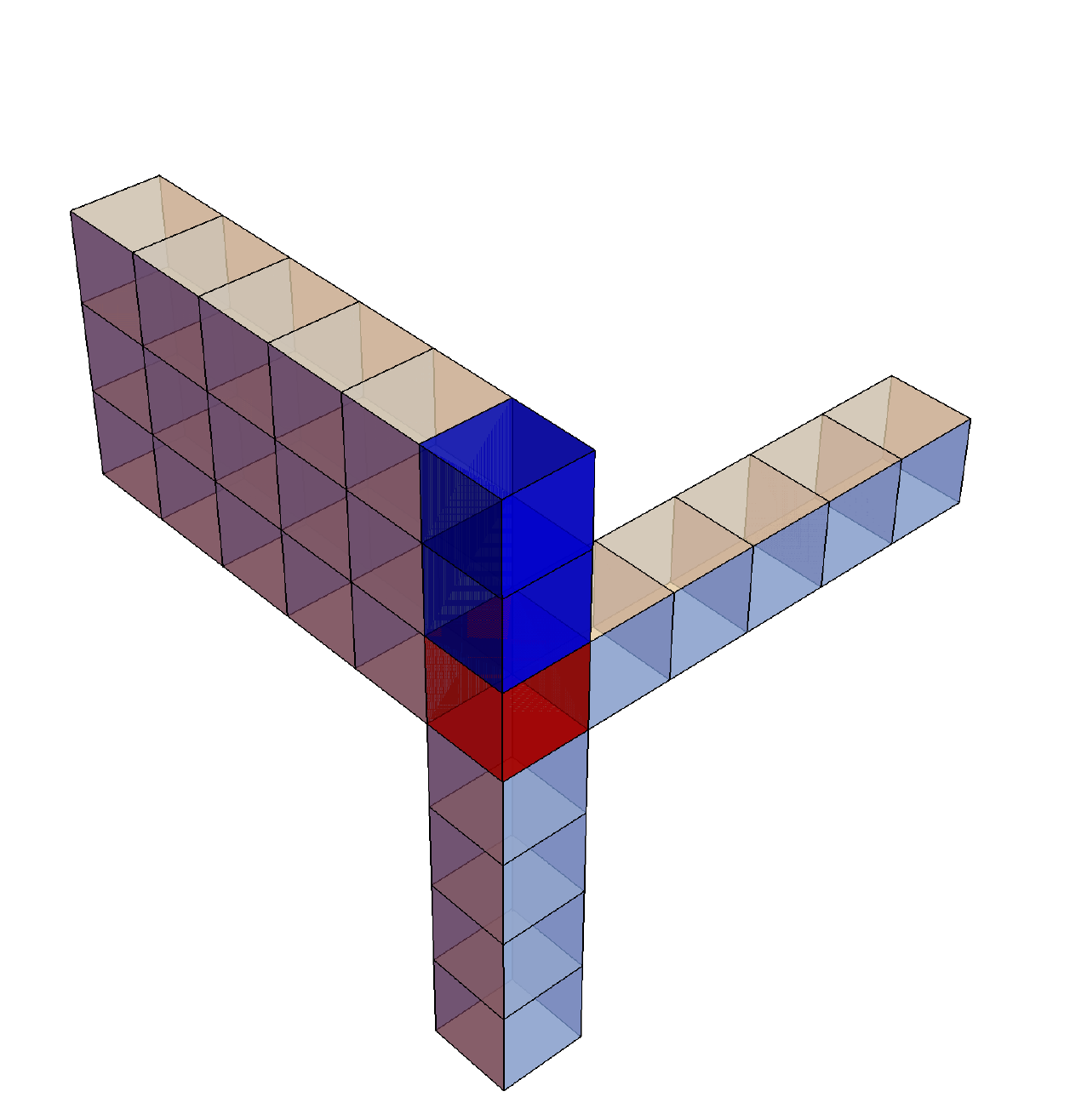}
 \caption{The box counting setup associated to $C_{\protect \yng(1), \protect \yng(3),\protect \yng(1)}$.}
\label{pic17}
\end{figure}

The next example is the first example containing in total five asymptotic boxes, shown in figure \ref{pic17} and with character
\begin{eqnarray}
\chi_{\yng(1),\yng(3),\yng(1)} =\frac{1-q+q^4-q^7+q^8}{(1-q)^3(1-q^2)(1-q^3)}  =  1+2q+4q^2+7q^3+\dots
\end{eqnarray}

Analogously to the previous example, at level one, can can generate either a box at the blue position $-2\epsilon_3$ or a heavy box at the origin
\begin{eqnarray}
|vac\rangle \rightarrow  |-2\epsilon_3\rangle + |0_H\rangle 
\end{eqnarray}

At level two, one can either stuck boxes on either of the two positions on top of $|-2\epsilon_3\rangle$ or one can introduce an unsupported heavy box at the origin. On top of $|0_H\rangle$, one can only grow boxes in direction 1 since the red box is already labelled by this direction. (The red box is unsupported only by a single blue box that lies in the intersection of cylinders of orientations 2 and 3.) One can also obviously introduce a box at location $-2\epsilon_3$. In total, we get at level two
\begin{eqnarray}\nonumber
|-2\epsilon_3\rangle &\rightarrow& |-2\epsilon_3,-2\epsilon_3+\epsilon_2\rangle +|-2\epsilon_3,-\epsilon_3\rangle  +|-2\epsilon_3,0_H\rangle \\
|0_H\rangle &\rightarrow& |-2\epsilon_3,0_H\rangle  + |0_H, \epsilon_1\rangle
\end{eqnarray}
Note also that compared to the second example, the state $|0_H\rangle $ can is produced by an annihilation of the box at $-2\epsilon_3$ in $|-2\epsilon_3,0_H\rangle $.

At next level, we get the following allowed transitions
\begin{eqnarray}\nonumber
|-2\epsilon_3,-2\epsilon_3+\epsilon_2\rangle &\rightarrow& |-2\epsilon_3,-2\epsilon_3+\epsilon_2,-\epsilon_3\rangle + |-2\epsilon_3,-2\epsilon_3+\epsilon_2,0_H\rangle   \\ \nonumber
&&+ |-2\epsilon_3,-2\epsilon_3+\epsilon_2,-2\epsilon_3+2\epsilon_2\rangle,\\ \nonumber
|-2\epsilon_3,-\epsilon_3\rangle &\rightarrow&  |-2\epsilon_3,-2\epsilon_3+\epsilon_2,-\epsilon_3\rangle +|-2\epsilon_3,-\epsilon_3,0_H\rangle +|-2\epsilon_3,-\epsilon_3,0_L\rangle,  \\ \nonumber
|-2\epsilon_3,0_H\rangle&\rightarrow&|-2\epsilon_3,-2\epsilon_3+\epsilon_2,0_H\rangle +|-2\epsilon_3,-\epsilon_3,0_L\rangle +|-2\epsilon_3,0_H,\epsilon_1\rangle,  \\ 
 |0_H, \epsilon_1\rangle &\rightarrow& |-2\epsilon_3,0_H, \epsilon_1\rangle+|0_H, \epsilon_1,2\epsilon_1\rangle.
\end{eqnarray}
The generation of boxes on top of $|-2\epsilon_3,-2\epsilon_3+\epsilon_2\rangle$ is analogous to previous levels. Since the red box of the configuration  $|-2\epsilon_3,-\epsilon_3\rangle$ is supported from all sides, one can introduce a two-dimensional space of the heavy and the light box at the origin or one can introduce the box at $-2\epsilon_3+\epsilon_2$. On top of $|-2\epsilon_3,0_H\rangle$, one can stuck boxes only in the direction 1 since $0_H$ is an unsupported box or one can add boxes under the box at the origin, in particular at positions $ -2\epsilon_3+\epsilon_2$ and $ -\epsilon_3$. The second option changes the heavy box into the light one. Analogously to example 2, the generating function $\psi(z) $ acting on the state $|-2\epsilon_3,0_H\rangle$ does not contain a pole at position $-\epsilon_3$ and the configuration $|-2\epsilon_3,0_H\rangle$ cannot be generated by an annihilation of a box of $|-2\epsilon_3,-\epsilon_3,0_L\rangle$. Finally, one can grow only boxes in direction 1 on top of $ |0_H, \epsilon_1\rangle$ since otherwise the red box would not allow a consistent labelling. One can also grow boxes under the red box. 

\subsection{Example 4: }

The example from figure \ref{pic18} has the character of the form
\begin{eqnarray}\nonumber
\chi_{\yng(1),\yng(1,2),\yng(1)} &=&\frac{1-2q+3q^2-3q^3+3q^4-3q^5+3q^6-2q^7+q^8}{(1-q)^4(1-q^3)} \\
 &=&  1+2q+5q^2+10q^3+\dots
\end{eqnarray}

\begin{figure}
    \centering
        \includegraphics[width=3.8cm]{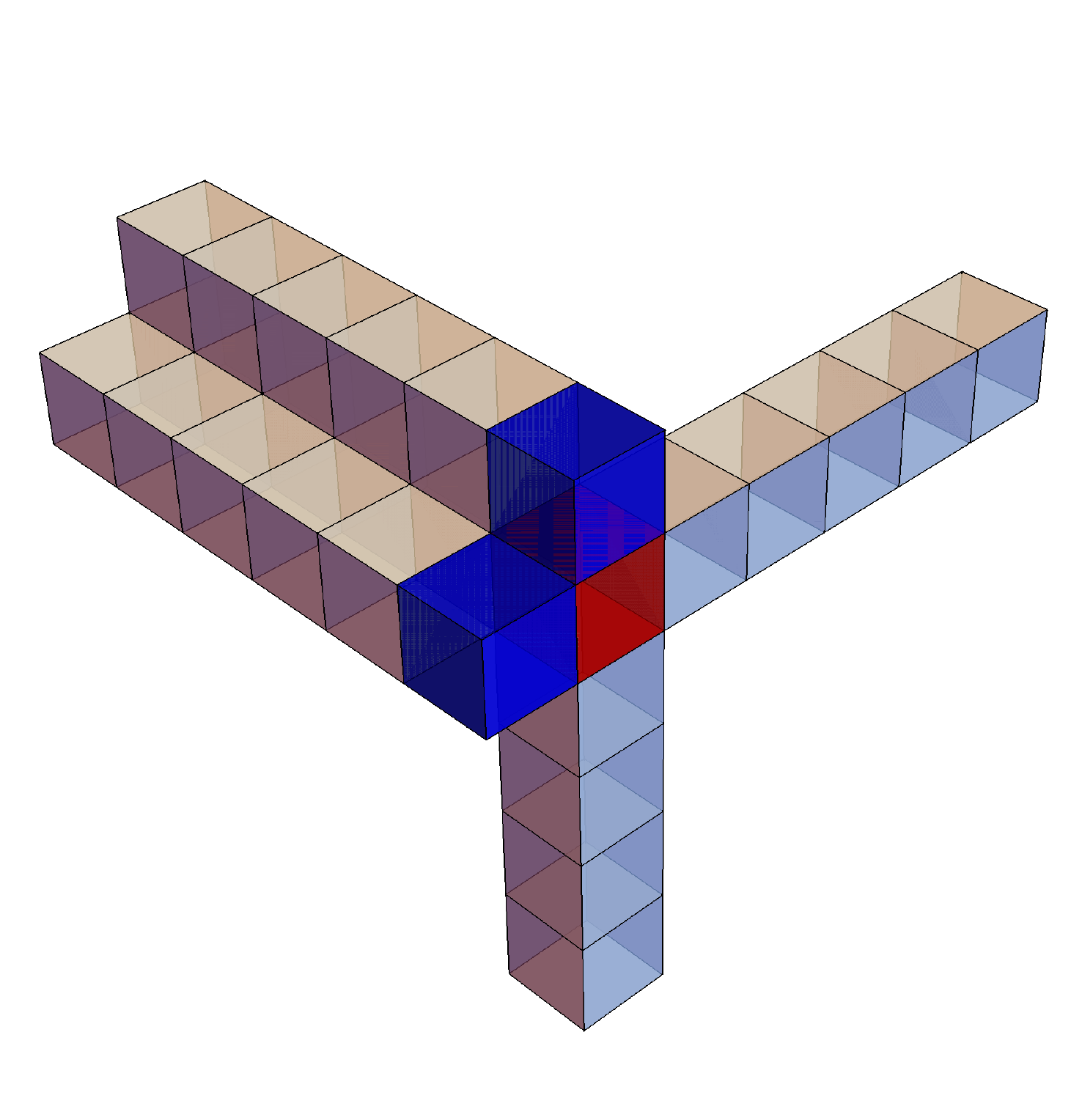}
 \caption{The box counting setup associated to $C_{\protect \yng(1), \protect \yng(1,2),\protect \yng(1)}$.}
\label{pic18}
\end{figure}

At level one, we can generate boxes at either of the two blue locations
\begin{eqnarray}
|vac\rangle \rightarrow |-\epsilon_3\rangle + |-\epsilon_1\rangle
\end{eqnarray}
Note that one cannot introduce a light box at the red position since it would not allow a consistent labelling. The missing box at position $-\epsilon_3$ lying at the intersection of the cylinders with orientations 3 and 2 require the labelling to be 1 but the missing blue box at position $-\epsilon_1$ lying at the intersection of cylinders with orientation 1 and 2 requires it to be 3.

At level two, we get five possible states
\begin{eqnarray}\nonumber
|-\epsilon_3\rangle &\rightarrow& |-\epsilon_3,-\epsilon_3+\epsilon_2\rangle+ |-\epsilon_3,-\epsilon_1\rangle+|-\epsilon_3,0_H\rangle\\
|-\epsilon_1\rangle &\rightarrow&  |-\epsilon_1,-\epsilon_1+\epsilon_2\rangle+ |-\epsilon_3,-\epsilon_1\rangle+|-\epsilon_1,0_H\rangle
\end{eqnarray}
In particular, one can generate two configurations containing an unsupported heavy box at the red position since it is not supported only at a single blue location and one can label the corresponding red box either by 3 in the case of $|-\epsilon_1,0_H\rangle$ or by 1 in the case of $|-\epsilon_3,0_H\rangle$.

At level two, we have the following transitions
\begin{eqnarray}\nonumber
 |-\epsilon_3,-\epsilon_3+\epsilon_2\rangle&\rightarrow&  |-\epsilon_3,-\epsilon_1,-\epsilon_3+\epsilon_2\rangle + |-\epsilon_3,-\epsilon_3+\epsilon_2,0_H\rangle \\ \nonumber
&&+|-\epsilon_3,-\epsilon_3+\epsilon_2,-\epsilon_3+2\epsilon_2\rangle \\ \nonumber
 |-\epsilon_1,-\epsilon_1+\epsilon_2\rangle &\rightarrow& |-\epsilon_3,-\epsilon_1,-\epsilon_1+\epsilon_2\rangle + |-\epsilon_1,-\epsilon_1+\epsilon_2,0_H\rangle\\ \nonumber
&& +|-\epsilon_1,-\epsilon_1+\epsilon_2,-\epsilon_1+2\epsilon_2\rangle \\ \nonumber
|-\epsilon_3,-\epsilon_1\rangle &\rightarrow& |-\epsilon_3,-\epsilon_1,-\epsilon_3+\epsilon_2\rangle+ |-\epsilon_3,-\epsilon_1,-\epsilon_1+\epsilon_2\rangle\\ \nonumber
&&+|-\epsilon_3,-\epsilon_1,0_L\rangle+|-\epsilon_3,-\epsilon_1,0_H\rangle\\ \nonumber
|-\epsilon_3,0_H\rangle &\rightarrow&|-\epsilon_3,-\epsilon_1,0_L\rangle  + |-\epsilon_3,-\epsilon_3+\epsilon_2,0_H\rangle +|-\epsilon_3,0_H,\epsilon_3\rangle \\ 
|-\epsilon_1,0_H\rangle &\rightarrow&|-\epsilon_3,-\epsilon_1,0_L\rangle + |-\epsilon_1,-\epsilon_1+\epsilon_2,0_H\rangle +|-\epsilon_1,0_H,\epsilon_1\rangle
\end{eqnarray}
On top of $ |-\epsilon_3,-\epsilon_3+\epsilon_2\rangle$ and $|-\epsilon_1,-\epsilon_1+\epsilon_2\rangle$, one can either continue adding boxes along direction 2, add a box at the other blue location or generate an unsupported box at the red location analogously to $|-\epsilon_3,0_H\rangle$ and  $|-\epsilon_1,0_H\rangle$. On top of $|-\epsilon_3,-\epsilon_1\rangle$, one can either continue adding boxes in the second direction or one can generate a two-dimensional space at the red location since the red box is now supported from all directions. Since $|-\epsilon_3,0_H\rangle $ and $|-\epsilon_1,0_H\rangle $ contain an unsupported box that allows labelling by either 1 or 3, one can only grow boxes in these two directions on top of the heavy box. Another option is growing boxes on top of the blue boxes or introducing a box at the other blue location. In the later case, the heavy box becomes supported and turns into the light box.

\subsection{Example 5: }

\begin{figure}[h]
    \centering
        \includegraphics[width=3.8cm]{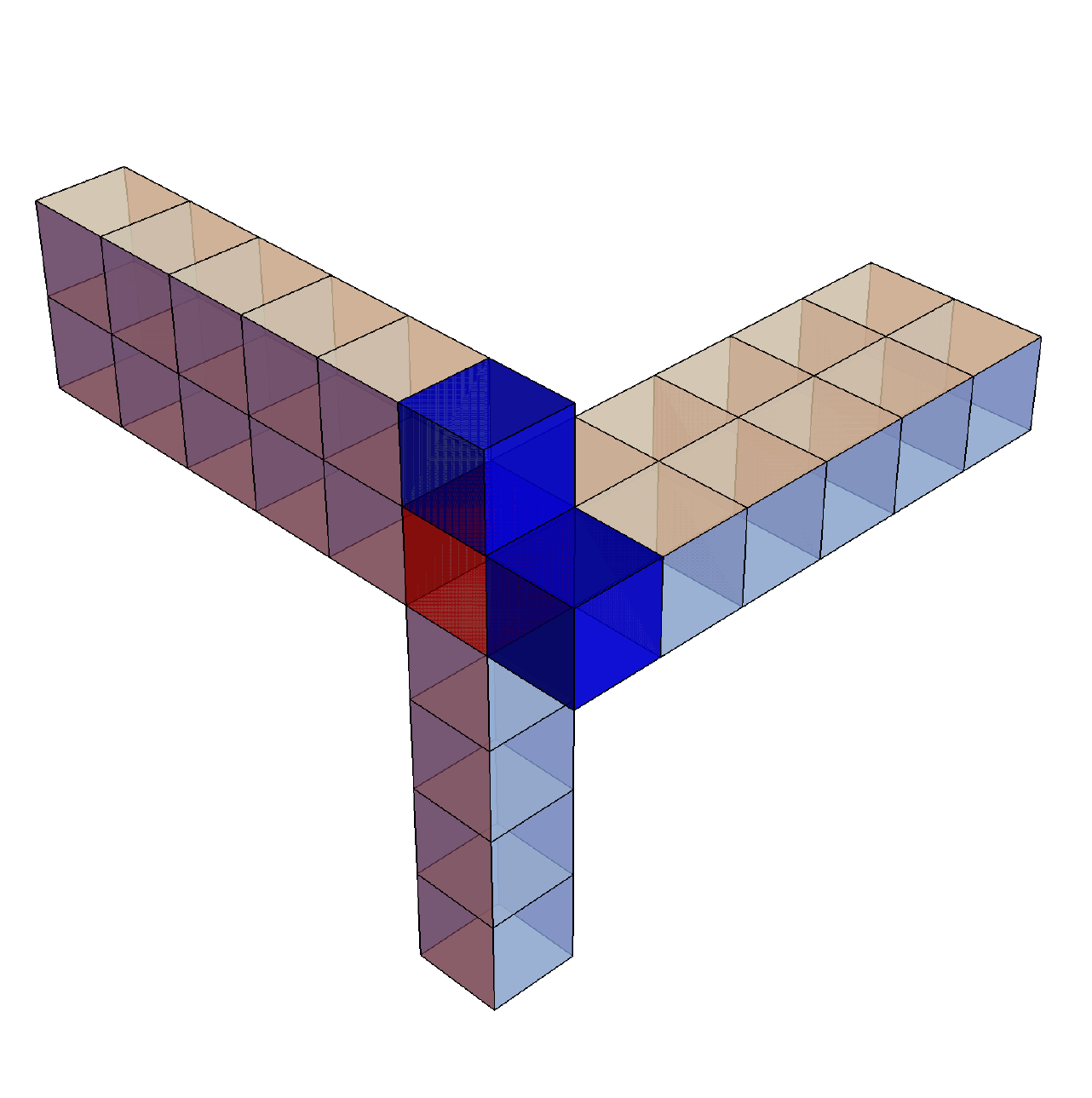}
 \caption{The box counting setup associated to $C_{\protect \yng(2), \protect \yng(2),\protect \yng(1)}$.}
\label{pic20}
\end{figure}

Our next example with character
\begin{eqnarray}\nonumber
C_{\yng(2),\yng(2),\yng(1)}&=&\frac{1-q+q^3-q^4+q^5-q^7+q^8}{(1-q)^3(1-q^2)^2} \\
& =&  1+2q+5q^2+9q^3+16q^4+26q^5+\dots
\end{eqnarray}
is depicted in figure \ref{pic20}.

At level one, we can introduce boxes at either of the two blue locations
\begin{eqnarray}
|vac\rangle \rightarrow |-\epsilon_3\rangle + |-\epsilon_2\rangle
\end{eqnarray}
We cannot introduce a box at the red location since it is not supported by blue boxes that lie at the intersection of only two cylinders.

At level two, one can freely grow boxes on top of the present blue box for either $ |-\epsilon_3\rangle$ or $ |-\epsilon_2\rangle$ or one can introduce a box at the other blue location or one can add a heavy box at the red location
\begin{eqnarray}\nonumber
|-\epsilon_3\rangle &\rightarrow& |-\epsilon_3,-\epsilon_3+\epsilon_2\rangle+|-\epsilon_3,-\epsilon_2\rangle +|-\epsilon_3,0_H\rangle  \\
|-\epsilon_2\rangle &\rightarrow& |-\epsilon_2,-\epsilon_2+\epsilon_1\rangle +|-\epsilon_3,-\epsilon_2\rangle+|-\epsilon_2,0_H\rangle 
\end{eqnarray}
The configurations containing a box at the origin is consistent since the light box is not supported only at one direction. 

At level two, we have the following allowed creations
\begin{eqnarray}\nonumber
|-\epsilon_3,-\epsilon_3+\epsilon_2\rangle&\rightarrow& |-\epsilon_3,-\epsilon_3+\epsilon_2,-\epsilon_3+2\epsilon_2\rangle  +|-\epsilon_3,-\epsilon_2,-\epsilon_3+\epsilon_2\rangle \\ \nonumber
&&+|-\epsilon_3,-\epsilon_3+\epsilon_2,0_H\rangle\\ \nonumber
 |-\epsilon_2,-\epsilon_2+\epsilon_1\rangle &\rightarrow& |-\epsilon_2,-\epsilon_2+\epsilon_1,-\epsilon_2+2\epsilon_1\rangle + |-\epsilon_3,-\epsilon_2,-\epsilon_2+\epsilon_1\rangle  \\ \nonumber
&&+  |-\epsilon_2,-\epsilon_2+\epsilon_1,0_H\rangle\\ \nonumber
|-\epsilon_3,-\epsilon_2\rangle  &\rightarrow& |-\epsilon_3,-\epsilon_2,-\epsilon_3+\epsilon_2\rangle +|-\epsilon_3,-\epsilon_2,-\epsilon_2+\epsilon_1\rangle \\ \nonumber
&&+|-\epsilon_3,-\epsilon_2,0_L\rangle  +|-\epsilon_3,-\epsilon_2,0_H\rangle \\ \nonumber
|-\epsilon_3,0_H\rangle &\rightarrow&|-\epsilon_3,0_H,\epsilon_3\rangle + |-\epsilon_3,-\epsilon_3+\epsilon_2,0_H\rangle + |-\epsilon_3,-\epsilon_2,0_L\rangle \\ 
|-\epsilon_2,0_H\rangle  &\rightarrow&|-\epsilon_2,-\epsilon_2+\epsilon_1,0_H\rangle + |-\epsilon_3,-\epsilon_2,0_L\rangle
\end{eqnarray}
Creation of boxes on top of $|-\epsilon_3,-\epsilon_3+\epsilon_2\rangle$ and $|-\epsilon_2,-\epsilon_2+\epsilon_1\rangle$ is analogous to the previous level. Since the red box in configuration $|-\epsilon_3,-\epsilon_2\rangle$ is supported from all sides, one can introduce, apart from the boxes at positions $-\epsilon_2+\epsilon_1$ and $-\epsilon_3+\epsilon_2$, a two-dimensional space associated to the red box. On top of the configuration $|-\epsilon_3,0_H\rangle$, one can either stuck a box on top of the blue box at position $-\epsilon_3+\epsilon_2$, generate a box at the other blue location or introduce a box on top of the red box at position $\epsilon_3$. Note that this is the only direction in which we can add a box on top of the unsupported red box since it is not being supported by a blue location lying at the intersection of cylinders 1 and 2. On the other hand $|-\epsilon_2,0_H\rangle$ allows only adding a box at the other blue location or a box on top of the present blue box. Adding a box at position $\epsilon_3$ is prohibited due to the absence of the box at $-\epsilon_3$.

\subsection{Example 6: }

The next example with character
\begin{eqnarray}\nonumber
\chi_{\yng(2),\yng(1,1),\yng(1)} &=&\frac{1-q-q^2+2q^3-q^5+q^6-q^8+q^9}{(1-q)^3(1-q^2)^2} \\
 &=&  1+2q+4q^2+7q^3+12q^4+19q^5+\dots
\end{eqnarray}
is depicted in figure \ref{pic20}.

\begin{figure}[h]
    \centering
        \includegraphics[width=3.8cm]{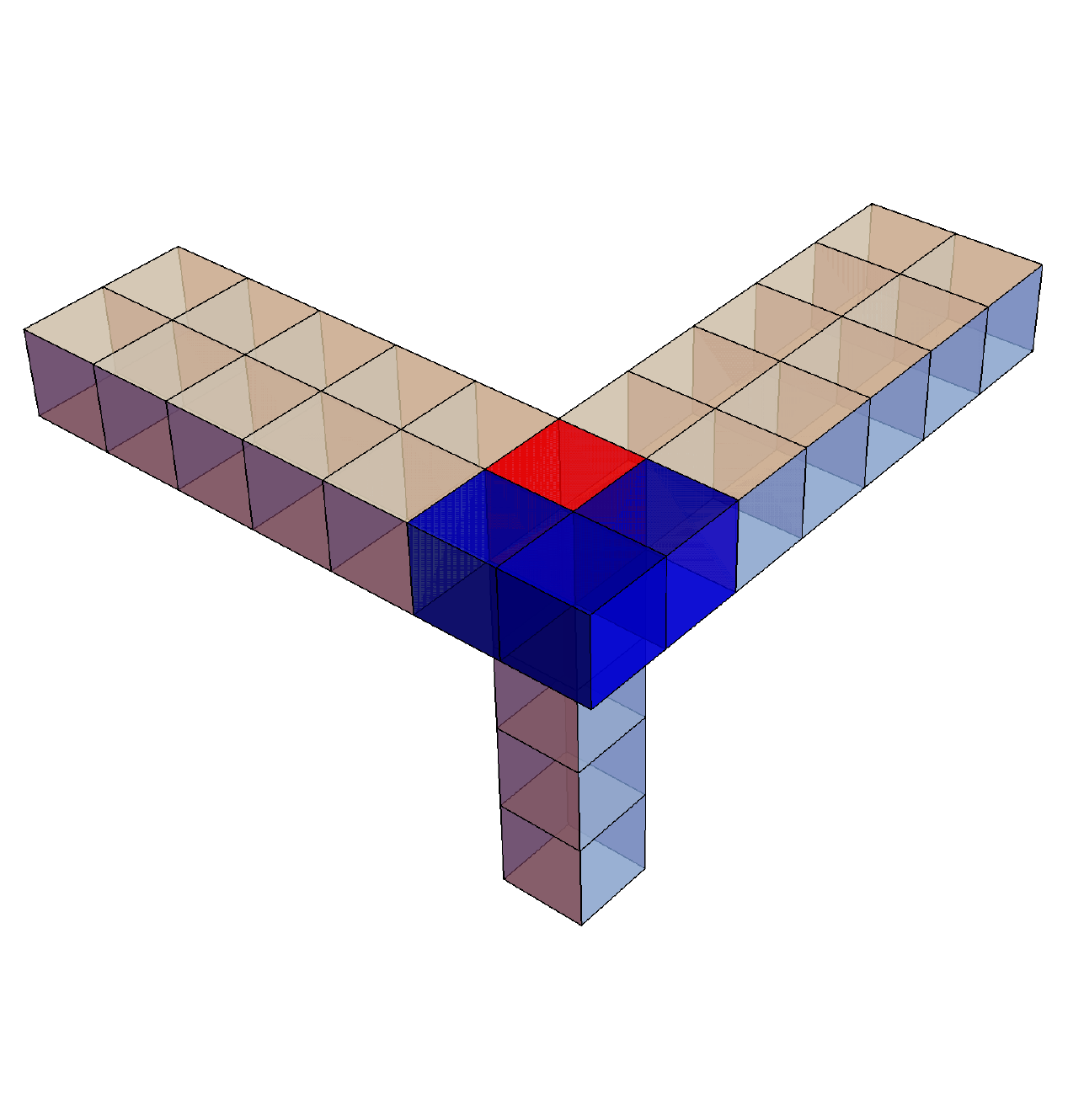}
 \caption{The box counting setup associated to $C_{\protect \yng(2), \protect \yng(1,1),\protect \yng(1)}$.}
\label{pic18}
\end{figure}

At level one, we can obviously add a box at the blue location $-\epsilon_1-\epsilon_2$ but we can also introduce a light box at the origin
\begin{eqnarray}
|vac\rangle \rightarrow |-\epsilon_1-\epsilon_2\rangle + |0_H\rangle
\end{eqnarray}
even though it is unsupported from two directions. The reason is that all the blue boxes lie at the intersection of the same cylinders along directions 1 and 2. 

At level two, we have the following allowed configurations
\begin{eqnarray}\nonumber
|-\epsilon_1-\epsilon_2\rangle&\rightarrow& |-\epsilon_1-\epsilon_2,0_H\rangle+ |-\epsilon_1-\epsilon_2,-\epsilon_1\rangle + |-\epsilon_1-\epsilon_2,-\epsilon_2\rangle \\
|0_H\rangle &\rightarrow&  |-\epsilon_1-\epsilon_2,0_H\rangle +|0_H,\epsilon_3\rangle
\end{eqnarray}
We can either add a box on top of the blue box at $-\epsilon_1-\epsilon_2$ or introduce an unsupported box at the red position. One can also grow boxes in direction 3 on top of $|0_H\rangle $ since all the blue boxes lie at the intersection of cylinders with orientations 1 and 2. 

At level two, the following creations are possible
\begin{eqnarray}\nonumber
 |-\epsilon_1-\epsilon_2,0_H\rangle&\rightarrow& |-\epsilon_1-\epsilon_2,-\epsilon_1,0_H\rangle+  |-\epsilon_1-\epsilon_2,-\epsilon_2,0_H\rangle +  |-\epsilon_1-\epsilon_2,0_H,\epsilon_3\rangle\\ \nonumber
  |-\epsilon_1-\epsilon_2,-\epsilon_1\rangle &\rightarrow&   |-\epsilon_1-\epsilon_2,-\epsilon_1,-\epsilon_2\rangle + |-\epsilon_1-\epsilon_2,-\epsilon_1,-\epsilon_1+\epsilon_2\rangle    \\ \nonumber
&& +|-\epsilon_1-\epsilon_2,-\epsilon_1,0_H\rangle\\ \nonumber
 |-\epsilon_1-\epsilon_2,-\epsilon_2\rangle &\rightarrow&   |-\epsilon_1-\epsilon_2,-\epsilon_1,-\epsilon_2\rangle +  |-\epsilon_1-\epsilon_2,-\epsilon_2,-\epsilon_2+\epsilon_1\rangle \\ \nonumber
&&+ |-\epsilon_1-\epsilon_2,-\epsilon_2,0_H\rangle\\ 
|0_H,\epsilon_3\rangle &\rightarrow& |0_H,\epsilon_3,2\epsilon_3\rangle + |-\epsilon_1-\epsilon_2,0_H,\epsilon_3\rangle 
\end{eqnarray}
For both $|-\epsilon_1-\epsilon_2,0_H\rangle$ and $|0_L,\epsilon_3\rangle $, one can either continue adding boxes along the direction $\epsilon_3$ or add boxes at available blue locations. For $|-\epsilon_1-\epsilon_2,-\epsilon_1\rangle$ and $|-\epsilon_1-\epsilon_2,-\epsilon_2\rangle $, one can either add boxes on top of the already present blue boxes or introduce an unsupported box at the red location. 

\subsection{Example 7: }

\begin{figure}[h]
    \centering
        \includegraphics[width=3.8cm]{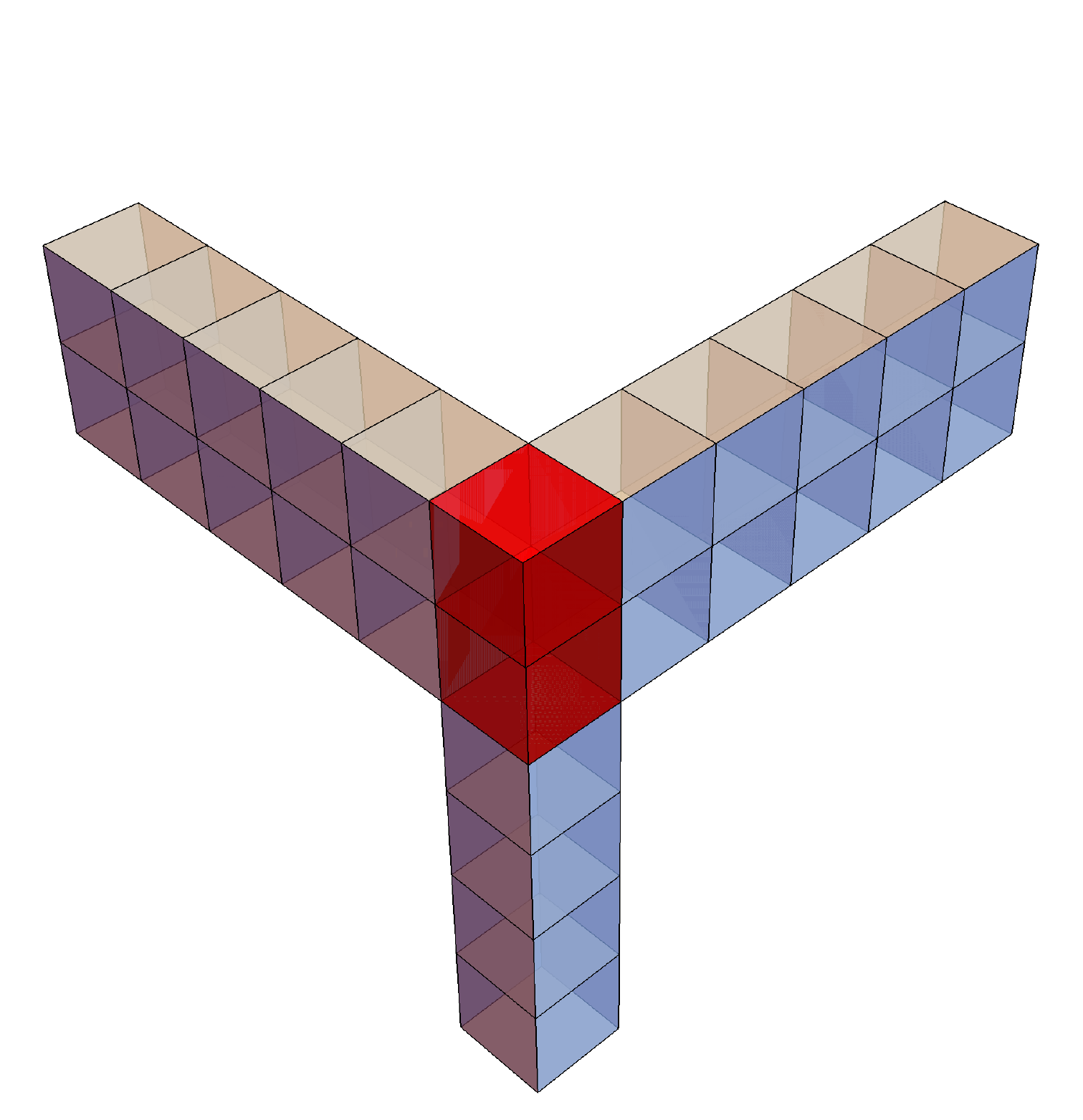}
 \caption{The box counting setup associated to $C_{\protect \yng(1,1), \protect \yng(2),\protect \yng(1)}$.}
\label{pic19}
\end{figure}

Our last example containing two red boxes is shown in figure \ref{pic19} and the corresponding character is 
\begin{eqnarray}\nonumber
\chi_{\yng(1,1),\yng(2),\yng(1)} &=&\frac{1-q+q^3-q^4+2q^6-q^7-q^8+q^9}{(1-q)^3(1-q^2)^2} \\
& =&  1+2q+5q^2+9q^3+16q^4+25q^5+\dots
\end{eqnarray}

As in our first example, we can introduce a two-dimensional space of the heavy and the light box at level one
\begin{eqnarray}
|vac\rangle \rightarrow |(-\epsilon_3)_L\rangle +|(-\epsilon_3)_H\rangle 
\end{eqnarray}
but now at the location $-\epsilon_3$. Note that we cannot add a box at the origin since it would be unsupported by a box at the a location.

The light box can again transform into the ultra-heavy box or one can introduce another light box on top of it. Note that the resulting configuration allows three possible labellings, where both red boxes carry the same label and one can have a block of two heavy or to light boxes. The heavy box allows more configurations to be generated
\begin{eqnarray}\nonumber
 |(-\epsilon_3)_H\rangle &\rightarrow&  |(-\epsilon_3)_U\rangle + |(-\epsilon_3)_H,-\epsilon_3+\epsilon_1\rangle \\ \nonumber
&&+ |(-\epsilon_3)_H,-\epsilon_3+\epsilon_2\rangle +|(-\epsilon_3)_L,0_L\rangle +|(-\epsilon_3)_H,0_H\rangle  \\
|(-\epsilon_3)_L\rangle  &\rightarrow& |(-\epsilon_3)_U\rangle+|(-\epsilon_3)_L,0_L\rangle
\end{eqnarray}
In particular, one can add boxes at either of the uncolored locations $-\epsilon_3+\epsilon_1$ and $-\epsilon_3+\epsilon_2$, transform the heavy box into the ultraheavy one or add a two-dimensional space associated to a block of heavy and light boxes.

At the next level, we have the following processes
\begin{eqnarray}\nonumber
|(-\epsilon_3)_L,0_L \rangle &\rightarrow& |(-\epsilon_3)_U,0_L \rangle \\ \nonumber
|(-\epsilon_3)_H,0_H\rangle  &\rightarrow&|(-\epsilon_3)_H,0_H,\epsilon_3\rangle +|(-\epsilon_3)_U,0_L\rangle +|(-\epsilon_3)_U,0_H\rangle \\ \nonumber
&&+|(-\epsilon_3)_H,0_H,-\epsilon_3+\epsilon_2\rangle +|(-\epsilon_3)_H,0_H,-\epsilon_3+\epsilon_1\rangle   \\ \nonumber
 |(-\epsilon_3)_H,-\epsilon_3+\epsilon_2\rangle &\rightarrow& |(-\epsilon_3)_U,-\epsilon_3+\epsilon_2\rangle  + |(-\epsilon_3)_H,0_H,-\epsilon_3+\epsilon_2\rangle \\ \nonumber
&&+   |(-\epsilon_3)_H,-\epsilon_3+\epsilon_2,-\epsilon_3+2\epsilon_2\rangle\\ \nonumber
|(-\epsilon_3)_H,-\epsilon_3+\epsilon_1\rangle &\rightarrow&|(-\epsilon_3)_U,-\epsilon_3+\epsilon_1\rangle +|(-\epsilon_3)_H,0_H,-\epsilon_3+\epsilon_1\rangle \\ \nonumber
&&+|(-\epsilon_3)_H,-\epsilon_3+\epsilon_1,-\epsilon_3+2\epsilon_1\rangle\\ \nonumber
 |(-\epsilon_3)_U\rangle  &\rightarrow& |(-\epsilon_3)_U,-\epsilon_3+\epsilon_1\rangle+ |(-\epsilon_3)_U,-\epsilon_3+\epsilon_2\rangle\\ 
&&+|(-\epsilon_3)_U,0_L\rangle +|(-\epsilon_3)_U,0_H\rangle
\end{eqnarray}
Let us go through them in a greater detail. First, the bottom box of the light block $|(-\epsilon_3)_L,0_L \rangle $ can change into the ultra-heavy but that is the only possibility since we cannot add boxes on top of the light box at uncolored location. In the heavy block $|(-\epsilon_3)_H,0_H\rangle$, the heavy box at $-\epsilon_3$ can transform into the ultra-heavy box. The creation of $|(-\epsilon_3)_U,0_L\rangle$ is possible as well since this state is generated by the action of $\psi (z)$ on $|(-\epsilon_3)_U,0_H\rangle$. The transformation of the heavy box at the origin to the ultra-heavy box is not allowed since a ultra-heavy box has to be supported only by another ultra-heavy box at the red location.  It is also possible to add boxes at locations $-\epsilon_3+\epsilon_2$, $-\epsilon_3+\epsilon_1$ and $\epsilon_3$. In the first two cases, the labels associated to both of the red boxes are 2 and 1 respectively whereas in the last case the labels are 3.  Note that boxes on top of the red box at position $-\epsilon_3$ in $|(-\epsilon_3)_H,0_H,-\epsilon_3+\epsilon_1\rangle$contain the box at uncolored location in direction 1 and the heavy red box at the origin that can be labelled by 1. Creation of boxes on top of $|(-\epsilon_3)_H,-\epsilon_3+\epsilon_1\rangle$ and $|(-\epsilon_3)_H,-\epsilon_3+\epsilon_2\rangle$ are analogous so let us discuss only the first case only. One can obviously change the heavy box into the ultra-heavy one or add a box at position $-\epsilon_3+2\epsilon_1$ with the heavy box admitting labeling by direction 1. One can also generate a heavy box at the origin of the same label 1. The generation of the light box is not allowed since the box at position $-\epsilon_3+\epsilon_1$ requires the red boxes to be labelled by direction 1. Finally, on top of the ultra-heavy box of $|(-\epsilon_3)_U$, we can add boxes in either direction including the two-dimensional space at the red position at 0.

\bibliographystyle{JHEP}

\bibliography{mono}

\end{document}